\documentclass[a4paper,fleqn,usenatbib]{mnras}

\usepackage{newtxtext,newtxmath}
\usepackage{pifont}

\usepackage[T1]{fontenc}
\usepackage{ae,aecompl}

\usepackage{graphicx}	
\usepackage{amsmath}	
\usepackage{amssymb}	
\usepackage{setspace}
\usepackage{ragged2e}
\usepackage{stackengine}
\usepackage{float}
\usepackage{subcaption} 
\usepackage{threeparttable}
\usepackage{natbib}

\def\wig#1{\mathrel{\hbox{\hbox to 0pt{%
          \lower.6ex\hbox{$\sim$}\hss}\raise.4ex\hbox{$#1$}}}}
\def\Ro{{\rm Ro}}
\def\Rom{{\rm Ro_{crit}^{m}}}
\def\RoM{{\rm Ro_{crit}^{M}}}
\def\tauc{\tau_{\rm c}}
\def\Msun{{\rm M_\odot}}

\newcommand{\cmark}{\textcolor{green}{\ding{51}}}%
\newcommand{\xmark}{\textcolor{red}{\ding{55}}}%

\title[Spin-Orbit Evolution of Stars and Exoplanets]{A semi-empirical model for magnetic braking of solar-type stars}

\author[L. S. Ardestani et al.]{
Leila Sadeghi Ardestani,$^{1}$\thanks{E-mail: lsadeghi@ipm.ir}
Tristan Guillot,$^{2}$
Pierre Morel $^{2}$
\\
$^{1}$School of Astronomy, Institute for Research in Fundamental Sciences (IPM), P. O. Box 19395-5531, Tehran, Iran\\
$^{2}$Universit\'{e} C\^ote d'Azur, Observatoire de la C\^ote d'Azur, CNRS UMR 7293, 06304 Nice Cedex 4, France\\
}

\date{Published: 14 August 2017}

\pubyear{2016}

\begin{document}
\label{firstpage}
\pagerange{\pageref{firstpage}--\pageref{lastpage}}
\maketitle

\begin{abstract}
We develop new angular momentum evolution models for stars with masses of $0.5$ to $1.6~\rm M_\odot$ and from the pre-main-sequence (\rm PMS) through the end of their main-sequence  (\rm MS) lifetime. The parametric models include magnetic braking based on numerical simulations of magnetised stellar winds, mass loss rate prescription, core-envelope decoupling as well as disk locking phenomena.
 We have also accounted for recent developments in modelling dramatically weakened magnetic braking in stars more evolved than the Sun. We fit the free parameters in our model by comparing model predictions to rotational distributions of a number of stellar clusters as well as individual field stars. Our model reasonably successfully reproduces the rotational behaviour of stars during the \rm PMS phase to the zero-age main-sequence (\rm ZAMS) spin up, sudden \rm ZAMS spin down, and convergence of the rotation rates afterwards. We find that including core-envelope decoupling improves our models especially for low-mass stars at younger ages. In addition, by accounting for the almost complete suppression of magnetic braking at slow spin periods, we provide better fits to observations of stellar rotations compared to previous models.
\end{abstract}

\begin{keywords}
Spin-orbit Evolution - Angular Momentum Transfer - Magnetic Braking 
\end{keywords}



\section{Introduction}

Angular momentum evolution in low-mass stars is a result of a complex interplay between initial conditions during star formation, the stellar structure evolution, and the behaviour of stellar winds and magnetic fields. During the life of a low-mass star on the main-sequence (MS), angular momentum is lost to the magnetised wind (magnetic braking). By removing angular momentum, winds cause their host stars to spin down with time \citep[e.g.,][]{Schatzman1962,Kawaler1988}. Since rotation is the most important parameter that determines the strength of a star's magnetic dynamo, this spin down leads to a decrease in the magnetic activity of low-mass stars as they age \citep[e.g.,][]{Skumanich1972,Vidotto2014}.

The study of angular momentum evolution has benefited a lot from recent measurements of rotation rates in star-forming regions \citep[e.g.,][]{Irwin2011}, young open clusters \citep[e.g.,][]{Hartman2010} and the field stars \citep[e.g.,][]{McQuillan2014,Leao2015} covering an age range from $1\,\rm Myr$ to about $10\,\rm Gyr$. These results provide a detailed view of how surface rotational velocity changes as stars evolve from the PMS, through the ZAMS to the late MS. And a number of models have been developed to account for these observations \citep[e.g.,][]{Irwin2007, Gallet&Bouvier2015,Johnstone2015}.	 These models generally incorporate a wind braking law, disk locking and core-envelope decoupling into their parametric models and in order to find an explanation for the observations.

In order to satisfy observational constraints, most of these models have to incorporate three major physical processes: star-disk interaction during the PMS, redistribution of angular momentum in the stellar interior, and angular momentum loss due to stellar winds. Star-disk interaction prevents the star spinning up during the PMS, although it is contracting at a fast rate. Recent theoretical advances have highlighted the impact of accretion/ejection phenomena on the angular momentum evolution of young suns \citep[e.g.,][]{Matt2012,Zanni2013}. The amount of angular momentum stored in the stellar interior throughout its evolution is usually unknown because observations have so far only revealed surface rotation, except for the Sun \citep[e.g.,][]{Hiremath2013} and for a few evolved giants \citep[e.g.,][]{Deheuvels2012}.

It has recently been possible to gain a qualitative estimate of  angular momentum loss rates due to magnetic braking with 2D and 3D numerical simulations of realistic magnetised stellar winds  \citep[e.g.,][]{Vidotto2011,Matt2012,Vidotto2014}. Much of the difficulty to predict stellar wind torque arises from the uncertainty in our knowledge of the magnetic and stellar wind properties (both observational and theoretical). There is still much work needed in order to understand how wind properties depend upon stellar mass, rotation rate, and time.

The aim of the present study is to develop new angular momentum evolution models for a wide range of stars from F to M-type and from a few $\rm Myr$ to the end of their main-sequence life and incorporate some of the most recent advances.

The structure of this article is as follows: in Section~\ref{sec:Parametrization}, we re-derive the parametrised model of \cite{Kawaler1988}, update it with available information and present our model for rotational evolution. In Section~\ref{sec:Observational-constraints}, we introduce our observational sample of over 1500 stars in 13 clusters and a number of individual field stars to derive observational constraints on the rotational evolution of stars. In Section~\ref{sec:Parameters}, we calibrate the free parameters in our rotational evolution model using the observational constraints allowing us to find values that generate best fit models. Finally in Section~\ref{sec:results}, we compare the results of our rotational evolution model to observations.

\section{Parameterisation of magnetic braking}
\label{sec:Parametrization}
The angular momentum evolution of isolated low-mass stars is controlled by the balance between three main physical mechanisms: angular momentum removal by magnetised stellar winds (here after "magnetic braking"), the star-disk interaction, and the angular momentum transfer within the stellar interior. In this section we discuss the corresponding model assumptions.

\subsection{General formalism}
As first proposed by \cite{Schatzman1962}, following the pioneering work of \cite{Parker1958}, stars spin down because a coupling between their magnetic field and their ionised wind transfers angular momentum to the wind at the expense of that in the star. The stellar angular momentum $J$ lost by stellar wind in unit time can be estimated as a function of the stellar mass loss rate $\dot{M}$ and rotation frequency $\Omega$: 
\begin{equation}
{d J\over d t}={2\over 3} \dot{M} R^2 \Omega \left(r_{\rm J} \over R\right)^2,
\label{general-formalism}
\end{equation}
where $R$ is the stellar radius and $r_{\rm J}$ is an equivalent radius that is to be determined. It can be thought as the distance after which wind material is no longer in bound to the star.\footnote{Although it has been common to use the Alfv\'en radius instead of $r_{\rm J}$ in this expression, we believe that it is important to separate these two quantities which are physically distinct.} An important parameter entering this estimation is the  Alfv\'en velocity in the radial direction,  $v_{\rm A}\equiv B_r(r)/\sqrt{\mu_0\rho(r)}$, and the corresponding Alfv\'en radius $r_{\rm A}$ at which $v_{\rm A}(r_{\rm A})=v(r_{\rm A})$, $v$ being the wind speed, and $B_r$ the radial component of the magnetic field. 

The radius $r_{\rm J}$ depends mostly on the strength and geometry of the stellar magnetic field and the launching of the stellar wind. For example, \cite{Mestel1986} proposes that for a radial magnetic field $B_r=B_0 (r/R)^{-2}$ and $r_{\rm J}=r_{\rm A}$, while for a dipolar magnetic field $B_r=B_0(r/R)^{-3}$ and $r_{\rm J}=\sqrt{r_{\rm A} R}$.

In order to express $r_{\rm A}$ as a function of $\dot{M}$, $M$ and $B_0$ we make two assumptions: (i) the wind velocity at $r_{\rm A}$ is a fraction of the escape speed at that location,
\begin{equation}
v_{\rm A}=K\sqrt{2GM\over r_{\rm A}},
\end{equation}
where $K\wig{<}1$, and (ii) the wind can be considered isotropic,
\begin{equation}
\dot{M}=4\pi r^2 \rho(r) v(r).
\end{equation}
Solving for $\rho(r_{\rm A})$ it is easy for us to show that 
\begin{equation}
r_{\rm A}^3={\left[r_{\rm A}^2 B_r(r_{\rm A})\right]^4 \over 2 K^2 G M \dot{M}^2}.
\end{equation}
Given $B_r(r_{\rm A})$, this last equation is used to find $r_{\rm A}$. 

With the above expressions for $B_r(r_{\rm A})$ and $r_{\rm J}$, we obtain
\begin{equation}
{r_{\rm J}\over R}=\left[B_0^4 R^5  \over 2 K^2 G M \dot{M}^2\right]^{n/6}=\left[{B_0^2 R^2\over K\dot{M}v_{\rm{esc}}}\right]^{n/3}
\label{eq:rJ}
\end{equation}
with $n=2$ for a radial field and $n=3/7$ for a dipolar field and $v_{\rm{esc}}$ is the surface escape velocity.  The study of \cite{Matt2008} shows that equation~\eqref{eq:rJ} has indeed the correct functional form for relatively slowly rotating stars, with the two parameters $K_{\rm 1}$ and $n$ which may be fitted to simulations. However, for more rapidly rotating stars the centrifugal acceleration changes the functional form slightly \citep{Matt2012} to
\begin{equation}
\label{eq:new-rJ}
{r_{\rm J}\over R}=K_{\rm 1} K_{\rm 3}\left[{B_0^2R^2\over \dot{M}(K_{\rm 2}^2 v_{\rm{esc}}^2+\Omega^2 R^2)^{1/2}}\right]^{m}.
\end{equation}
where \cite{Matt2012} find $K_{\rm 1}=1.3$, $K_{\rm 2}=0.0506$, $m=0.2177$. And all quantities are in cgs units. $K_{\rm 3}=1$ is a coefficient that we will modify later. Equation \ref{eq:new-rJ} is equivalent  to \ref{eq:rJ} in the limit that $\Omega^2R^2 \ll K_{\rm 2}^2 v^2_{\rm {esc}} $, $m=n/3$ and $K_{\rm 1}=\left(K_{\rm 2}/ K \right)^{2m}$. We can thus rewrite equation \eqref{general-formalism} as

\begin{equation}
{dJ\over dt}={2\over 3}\left(K_{\rm 1} K_{\rm 3}\right)^2\dot{M}^{(1-{2m})} R^{(2+{4m})}B_0^{{4m}}{\Omega\over(K_{\rm 2}^2 v_{\rm{esc}}^2+\Omega^2 R^2)^m}.
\label{new-general-formalism}
\end{equation}

\subsection{Magnetic field parameterisation}
\label{sec:Magnetic field parameterization}

\subsubsection{Empirical evidence}
\label{Empirical evidence}
The question of how to estimate the magnetic field strength is a difficult one. It has been known since \cite{Kraft1967} that rapidly rotating stars have higher levels of magnetic activity than slowly rotating stars. At a given stellar mass, the magnetic field strength has approximately a power law dependence on rotation rate, such that $B_0\propto\Omega^a$ for slow rotators and saturates at fast rotation. Following the work of \cite{Saar1996} it is usually assumed that the surface average field strength is a strong function of the spin frequency. \cite{Wright2011} find  $B_0\propto\Omega^{2.3}$.  Furthermore, using the results of recent Zeeman-Doppler Imaging (ZDI) studies of over 70~stars, \cite{Vidotto2014} showed that the large-scale field strength averaged over the stellar surface, $\overline{B}_V$, scales with rotations as $\overline{B}_V\propto\Omega^{1.32}$. 

On the other hand, studies of the magnetic field in M-dwarfs indicate that the large-scale magnetic field is inversely proportional to the Rossby number,
\begin{equation}
\Ro=\frac{2\pi}{\Omega\tauc},
\end{equation}
where $\tauc$ is the convective turnover timescale which appears to be a natural scaling factor for the stellar activity and magnetic field. For example, \cite{Noyes1984} showed that the mean level of chromospheric emission is inversely related to the rotation period. Also, \cite{Baliunas1996} found a significant correlation between the magnetic and rotational moments for the Mount Wilson sample. More recently, using a sample of $66$~stars in the range of $0.1$ to $1.34\,\rm M_\odot$, for which the large-scale surface magnetic field has been mapped using the Zeeman Doppler Imaging (ZDI) technique, \cite{See2017} have shown that the magnetic field strength and mass loss show less scatter when plotted against Rossby number rather than rotation period.

We therefore assume that magnetic field scales with the Rossby number as $B_0\propto \Ro^{-a}$. \cite{See2017} and \cite{Folsom2016} provide predictions on the dependency of magnetic field strength on Rossby number using ZDI. In their figure 5, \cite{See2017} have plotted magnetic field strength against Rossby number. Using the bisector ordinary least-squares method and after performing a cut for the critical Rossby number ($\Rom$)  at $0.1$ they found $a=1.65\pm 0.14$. Using linear regression we were also able to fit their sample with $a= 0.8\pm 0.20$. In addition, using a different sample and the $\chi^2$ method, \cite{Folsom2016} have approximated $a=1.0\pm 0.1$.  We consider $a$ as a free parameter to be determined.

\subsubsection{The convective turnover timescale}
\label{convective}
The convective turnover timescale is defined as the ratio of the pressure scale height (or the mixing length) to the convective velocity near the base of the outer convective zone. Here, we calculate it using the CESAM stellar evolution code \citep{2008Ap&SS.316...61M}, as described in Appendix~\ref{sec:tconv}. 

\begin{figure}
\centerline{\includegraphics[width=95mm]{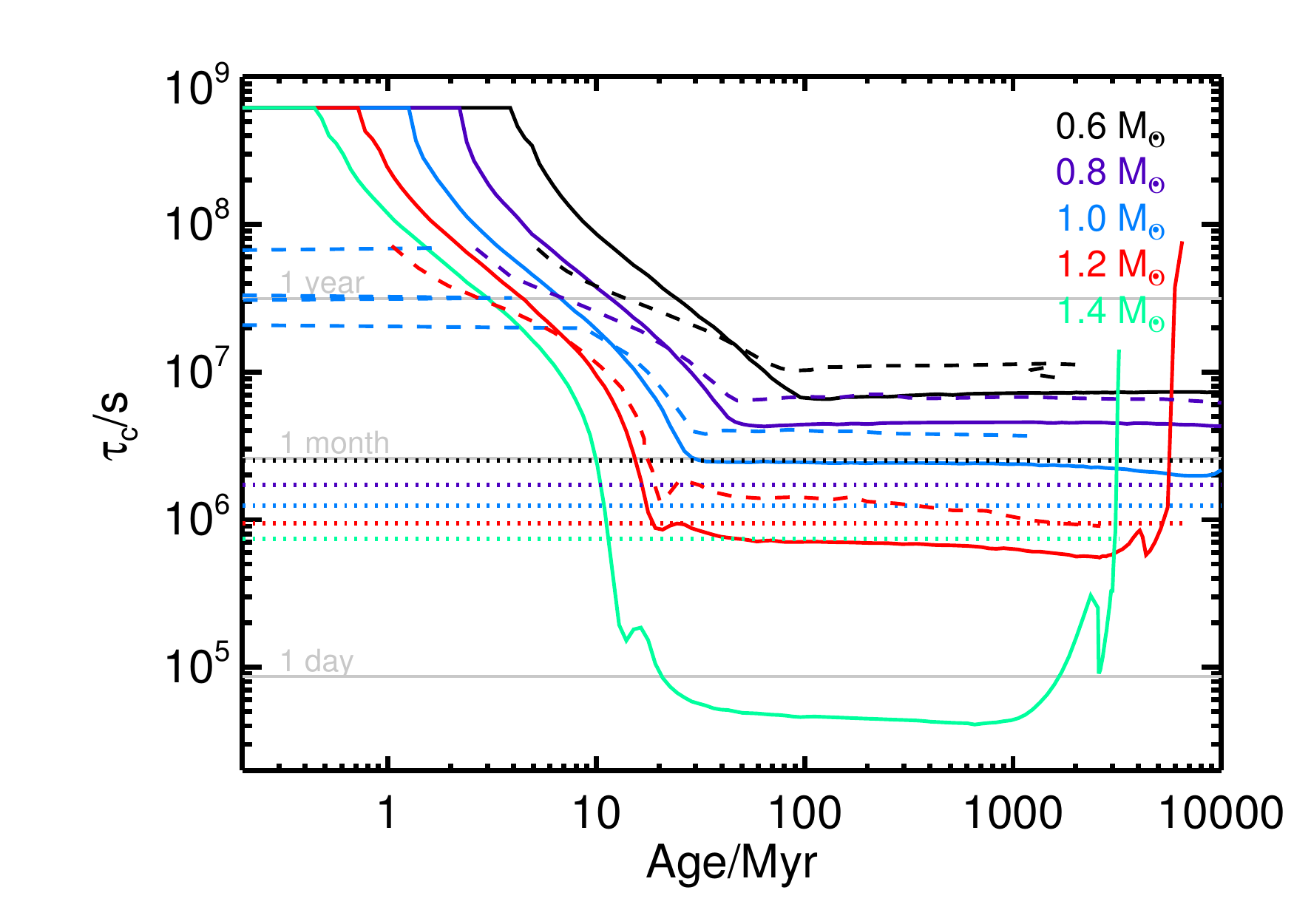}}
\caption{Comparison between the convective turnover timescale calculated in this work (solid) from those of \protect\cite{Landin+2010} (dashed) and \protect\cite{Wright2011} (dotted), for stellar masses between 0.6 and 1.4$\,\rm M_\odot$, as labelled.}
\label{tauc}
\end{figure}

Fig.~\ref{tauc} shows how $\tauc$ evolves as a function of time for stars of different masses. It is initially large (of order 20) and constant when the stars are fully convective. It then decreases on the pre-main-sequence to become almost constant on the main-sequence. It is smaller for higher mass stars, with a significant drop above $1.2\,\rm M_\odot$ which coincides with the gradual suppression of the external convective zone for stars of these masses and higher. The fact that our $\tau_c$ differs slightly from those of \citep{Landin+2010} is due to a slightly different location for the calculation of $\tauc$. We use the pressure scale height and they use the mixing length of 1.5 times the scale height. The differential behaviour between stars of different masses is very similar, as it should. 

In fig.~\ref{tauc}, we also compare our results to those estimated empirically by \cite{Wright2011}. By necessity, due to the large number of stars required, the latter estimates apply only to main-sequence stars. It is striking that although the $\tau_c$ is similar within an order-of-magnitude, the range, as a function of stellar mass, is much smaller in the empirical determination than in our calculations or those of \cite{Landin+2010}.

 In this work, we choose to use $\rm \tau_{c}$ based on our self-consistent CESAM calculations. We note that the comparison with  the Rossby number in the literature is not necessarily straightforward because of the different definitions and/or calculations used. However, with the exception of the pre-main-sequence evolution phase, the different definitions and calculations lead to $\Ro$ which is approximately proportional between different methods. Hence we can use proportionality relations between $\Ro$ and other parameters but stress that the absolute values of $\Ro$ are not directly comparable.

\subsubsection{Magnetic field saturation at fast spin}

Based on the slow rotational evolution of young rapidly rotating stars, there can be no serious doubt that both mass loss rate and the magnetic field saturate at fast rotation \citep[e.g.,][]{Pizzolato2003}. The reason is that, without saturation, the most rapidly rotating stars would spin down much faster than observed. Thus using a critical Rossby number ($\Rom$) we can distinguish between the non-saturated ($\Ro\geq \Rom$) and saturated ($\Ro<\Rom$) regimes as

\begin{equation}
B_0=\begin{cases}
B_\odot\left({\Omega\tauc\over\Omega_\odot\tau_{c\odot}} \right)^a &  \text{if $\Ro\geq \Rom$} \\ 
B_\odot\left({1\over \Rom}{2\pi\over\Omega_\odot\tau_{\rm c\odot}} \right)^a &  \text{if $\Ro < \Rom$}. 
\end{cases}
\label{Bstar}
\end{equation}

An estimate of $\Rom$ can be derived by fitting rotational evolution models to the observational constraints. \cite{Donati2008} and \cite{Reiners2009} find $\Rom \approx 0.1$. This determination of  Romcrit is accurate only within a factor about $2$ and in addition, as discussed in Sect. 2.2.2, it is not defined in exactly the same way. Therefore, we treat $\Rom$ as a free parameter and, by allowing a range $0.05<\Rom<0.2$, we look for a value which best fits observations in section~\ref{sec:Parameters}.

\subsubsection{Suppression of braking at slow spin}

It has recently been shown by \cite{vanSaders2016} that stars more evolved than the Sun rotate faster than previously expected. This indicates that magnetic braking is weaker in intermediate-age and old stars. Therefore, they suggest the existence of another Rossby threshold, which we call $\RoM$, above which magnetic braking is suppressed. A possible explanation for this could be a change in field geometry from a simple dipole to a higher order field which could produce weakened braking \citep{Reville2015}. A realistic treatment should thus consider changes to equation~\eqref{eq:new-rJ} arising from this different field geometry, perhaps through a different $m$. In the absence of any knowledge about the behaviour of these fields, we follow \cite{vanSaders2016} and adopt a simple reduction of the angular momentum loss rate past this critical Rossby number through a change in the $K_{\rm 3}$ coefficient,
\begin{equation}
K_{\rm 3}=\begin{cases}
1 & \text{if $\Ro < \RoM$,}\\
0.1 & \text{otherwise.}
\end{cases}
\label{eq:K3}
\end{equation}
Note that the $K_{\rm 3}=0.1$ above the critical Rossby number is arbitrary. 

Following \cite{vanSaders2016}, we adopt $\RoM=\Ro_\odot$,because of our choice of convective turnover timescale $\rm \tau_{c}$, $\Ro_\odot=1.113$ while this value is equal to 2.16 for \citeauthor{vanSaders2016}.

\subsection{Parameterisation of the mass loss}

The question of how to consider mass loss is often left aside, probably because the change in angular momentum (equation~\ref{new-general-formalism}) does not depend on mass loss when $m=1/2$. This solution was preferred for that reason by \cite{Kawaler1988} and \cite{1997A&A...326.1023B}. For other  $m$ an assumption must be made for $\dot{M}$. One can first note that, for $m>1/2$, magnetic braking increases when mass loss decreases, a slightly problematic result at least asymptotically. For $m<1/2$, braking increases with mass loss, an intuitively more satisfactory result. In the particular case of the numerical simulations by \cite{Matt2012} the spin-down timescale $\tau_\Omega\propto  \dot{M}^{-0.5646}\Omega^{-0.8708}$. Thus the dependence on $\dot{M}$ is not negligible. 

As known since the pioneering work of \cite{1958ApJ...128..664P}, mass loss is strongly linked to heating of the corona, which in itself is linked to the production of energetic photons and particles. Observations clearly indicate that the X and EUV irradiance of young stars (including the Sun) is orders of magnitude higher when young \citep[e.g.][]{2005ApJ...622..680R}. This activity is presumably mostly related to the fast rotation of the stars. Separately, \cite{1992MNRAS.256..269T} show that for fully convective T-Tauri stars, very strong winds of the order of $10^{-7}\,\rm M_\odot\,yr^{-1}$ are expected and are proportional to $\Omega$. Fitting the spin rates of stars on the main-sequence, \cite{Johnstone2015} estimate that the wind mass loss rate scales with stellar parameters as $\dot{M}\propto M^{-3.36}R^{2}\Omega^{1.33}$. 

\cite{See2017} estimate mass loss rates from magnetic field measurements of young stars and a potential field source surface model. Using their results, we consider mass loss to be correlated with the Rossby number, and like the magnetic field, saturate at fast rotation speed. Therefore, mass loss can be parameterised as 

\begin{equation}
\dot{M}=\begin{cases}
\dot{M}_\odot\left({\Omega\tauc\over\Omega_\odot\tau_{c\odot}} \right)^d &  \text{if $\Ro\geq \Rom$} \\ 
\dot{M}_\odot\left({1\over \Rom}{2\pi\over\Omega_\odot\tau_{\rm c\odot}} \right)^d &  \text{if $\Ro < \Rom$} .
\end{cases}
\label{Mdot}
\end{equation}
\cite{See2017} find $d$ ranging between $-1.36$ and $-1.62$ with the bisector ordinary least-squares method. Using linear regression, we were able to fit their data with $d=1.05\pm 0.12$. We treat $d$ as a parameter to be fitted within these extreme values.\footnote{The choice of $a$ and $d$ as names for the exponent is for consistency with a previous version of the paper in which more parameters were considered.}

\subsection{Decoupling of the outer convective envelope}

Up to now, we have implicitly assumed that the convective envelope is tied to the rest of the star which rotates as a solid body. In fact, as proposed by \cite{1991ApJ...376..204M} this is probably not the case and this has to be taken into account: the stellar wind acts on the star's magnetic field lines which originate in the dynamo region, in the convective zone. The rotation state of the stellar interior then depends on the ability and speed at which angular momentum is transferred to the interior. \cite{1991ApJ...376..204M} proposed that a typical angular mixing timescale of the order of $10^7$\,yr was required to explain the small amount of differential rotation observed with seismology between the solar convective zone and radiative interior. \cite{1998A&A...333..629A} applied this prescription to the case of M-dwarfs and showed the effect to be important to explain the efficient spin-up of M-dwarfs in young clusters. We will see that interior angular momentum mixing also plays an essential role in F-dwarfs. 

Following \cite{1991ApJ...376..204M}, we assume that any difference in angular momentum between the outer convective zone (characterised by an angular momentum, moment of inertia and angular rotation linked by $J_{\rm cz}=I_{\rm cz}\Omega_{\rm cz}$)  and the inner radiative layer (characterised by $J_{\rm rad}=I_{\rm rad}\Omega_{\rm rad}$) is mixed with a characteristic timescale $\rm \tau_{mix}$. The equations for the evolution of the rotation frequencies in the outer convective zone (cz) and in the inner radiative zone (rad) are \citep{1998A&A...333..629A}
\begin{eqnarray}
{d\Omega_{\rm cz}\over dt}&=&{I_{\rm rad}\over I}\left(\Omega_{\rm rad}-\Omega_{\rm cz}\right){1\over \tau_{\rm mix}}+{2\over 3}{R_{\rm rad}^2\over I_{\rm cz}}\Omega_{\rm cz}{dM_{\rm cz}\over dt}\nonumber\\
&& -{\Omega_{\rm cz}\over I_{\rm cz}}{dI_{\rm cz}\over dt} -\left.{d\Omega_{\rm cz}\over dt}\right|_{\rm wind}
\end{eqnarray}
and
\begin{eqnarray}
{d\Omega_{\rm rad}\over dt}&=&-{I_{\rm cz}\over I}\left(\Omega_{\rm rad}-\Omega_{\rm cz}\right){1\over \tau_{\rm mix}}-{2\over 3}{R_{\rm rad}^2\over I_{\rm rad}}\Omega_{\rm cz}{dM_{\rm cz}\over dt}\nonumber\\
&& -{\Omega_{\rm rad}\over I_{\rm rad}}{dI_{\rm rad}\over dt},
\label{2shell}
\end{eqnarray}
where $M_{\rm cz}$ is the mass of the convective zone and $R_{\rm rad}$ denotes the radius of the radiative layer. We have thus added another free parameter, $\rm \tau_{mix}$ (the characteristic timescale it takes for angular momentum to travel between radiative and convective zones), which from previous studies is expected to be of the order of $10^7$\,yr \citep{1998A&A...333..629A}. This timescale agrees both with the lack of differential rotation in our Sun and the fast rotation of some M- and G-dwarfs in clusters younger than about $100\,\rm Myr$.

\subsection{Disk locking}
\label{sec:disklock}

The disk-locking process arises from the observational evidence that stars that magnetically interact with their accretion disk during the first few Myr of PMS evolution are prevented from spinning up in spite of contracting towards the ZAMS. This behaviour is believed to result from the magnetic interaction between the young stellar object and its accretion disk, even though the details of the process are still to be understood \citep[see][]{Bouvier2014}.

At the beginning, most stars have disks but lose them in a few Myr \citep[see e.g.][]{2001ApJ...553L.153H}. From observations of the Taurus-Auriga association, \cite{2007A&A...473L..21B} find that the average disk lifetime is a function of stellar mass (from about $0.3$ to $1.6\,\rm M_\odot$) and approximately equal to $4 (\rm M/{\rm M_\odot})^{0.75}\rm Myr$. 

Following \cite{1997A&A...326.1023B} and \cite{Gallet&Bouvier2015}, we assume that fastest rotating stars at any age are those which lost their disk rapidly. Conversely, the stars rotating the slowest should have had their disks the longest. Furthermore, we assume that only the outer convective zone (source of the dynamo) is locked by the disk for a timescale $\rm \tau_{disk}$ which we also treat as a free parameter. 

\section{Observational Constraints and Models}
\label{sec:Observational-constraints}

\subsection{Observational Constraints}
\label{constraints}
The set of free parameters introduced in our rotational evolution model above needs to be constrained by observations in order for us to be able to study the distribution of stellar rotation rates at any age. For this reason, we use rotation periods of stars in a number of open clusters from the literature covering an age range of $1\,\rm Myr$ to $2.5\,\rm Gyr$.  Determining the age of young stellar clusters can be very challenging, and in fact, for very young clusters about $5-10 \rm Myr$, age uncertainties may approximately be equal to the actual age estimate \citep{Bell2013}. Fortunately, the change in rotation rates at these very young ages are relatively mild so that modelling does not suffer much from these uncertainties. The ages of older clusters are better known, with uncertainties close to $10$ percent. Therefore, we neglect the age uncertainties in the rest of this paper.

Considering the fact that many of the stars in our sample are likely to have companions, one might be concerned that the fraction which are interacting would have a significant effect on angular momentum loss. Binaries may impact rotational evolution via gravitational or magnetic interactions. Stars in very close binaries can exert tidal forces on each other, spinning them up or down more rapidly than predicted for a single star. These systems are also close enough for one star to interact with the other's large-scale magnetic field. And at the earliest evolutionary stages, a companion may truncate the protoplanetary disk, minimizing the impact of magnetic braking and allowing the young star to spin faster than its single counterparts  \citep[e.g.,][]{Cieza2009}. However, according to \cite{Duquennoy1991}, only about $4$ percent of binaries have orbital periods short enough to be tidally interacting even on circular orbits. When considering eccentric planets, we estimate that another about $10$ percent with longer periods may be tidally interacting. In addition, the spin properties of spectroscopic binaries appear not to differ from other stars \citep[see][concerning the M34 cluster]{Meibom2011}. This indicates that while duplicity may be important in some cases, it does not play a major role in the process.

\begin{table}
\caption{Statistical properties of the distributions of rotation rates used in this study. $n_\star$ is the number of stars in each cluster, $\overline{P}_{\star}$ is the mean rotation period of stars in the cluster in days and  $\sigma_{P}$(days) shows the standard deviation in the rotation period. The age uncertainties for clusters are generally not provided in the literature and thus have not been included in this table.}              
\label{tab:clusters}      
\begin{threeparttable}
\centering                                      
\hspace*{-1cm}
\begin{tabular}{l c c c c r}          
\hline                     
Cluster & age/Myr & $n_{\star}$ & $\overline{P}_{\star}$/days & $\sigma_{P}$/days & Ref. \\
\hline
ONC & $1.$ & $147$ & $5.424 $ & $3.593 $ & $1$  \\
 NGC2264 & $3.$ & $253$ & $5.252 $ & $3.720 $ & $2$ \\
 CepOB3b & $4.$ & $450$ & $2.389 $ & $2.742 $ & $3$  \\
NGC2362 & $5.$ & $166$ & $5.751 $ & $4.127 $ & $1$ \\
hPer & $13.$ & $323$ & $3.780 $ & $3.327$ & $4$ \\
Pleiades & $125.$ & $50$ &$4.032$ &$3.029$ &$5$\\
M35 & $150.$ & $307$ & $4.136 $ & $3.440 $ & $5$ \\
M34 & $220.$ & $83$ &$6.710$ &$0.887$ &$8$\\
M37 & $550.$ & $705$ & $7.410 $ & $4.677 $ & $1$\\
Praesepe & $600.$ & $51$ & $9.285 $ & $2.999 $ & $6$ \\
Hyades & $650.$ & $53$ & $10.995 $ & $3.865 $ & $6$ \\
NGC6811 & $940.$ & $70$ & $9.259 $ & $3.190 $ & $7$ \\
NGC6819 & $2500.$ & $29$ & $15.572 $ & $6.702 $ &$9$\\
\hline                                             
\end{tabular}
\begin{tablenotes}
 \item[]References: (1)~\cite{Irwin2009}; (2)~\cite{Venuti2016}; (3)~\cite{Littlefair2010}; (4)~\cite{Moraux2013}; (5)~\cite{Hartman2010}; (6)~\cite{Meibom2009}; (7)~\cite{Delorme2011}; (8)~\cite{Meibom2011}; (9)~\cite{Meibom2015}.
\end{tablenotes}
\end{threeparttable}
\end{table}

It is generally common in the literature to report the rotation period of clusters as a function of their magnitudes or colours like $V$, $B-V$, $J-K_s$, etc. This is while the rotational evolution models require the mass of the stars. Our sample includes $13$ clusters for which the statistical properties of the distributions of rotation rates are shown in Table~\ref{tab:clusters}. For four clusters (Praesepe, Hyades, NGC6811 and NGC6819) we have converted the commonly reported colours into masses. For the remaining $9$ clusters, the masses have been inferred directly from the references shown within Table~\ref{tab:clusters}.

In order to calculate the masses, we first extracted the age and metallicity of each cluster from the literature and using the CMD interface\footnote{http://stev.oapd.inaf.it/cgi-bin/cmd} \citep{Girardi2004}, calculated their evolution tracks and isochrones. The isochrones provided stellar masses as a function of different magnitudes and using a number of small adjustments in colour bands we were able to gain the masses in the observed colours. The final step was to interpolate the masses at the same reported colours for each star in the cluster. 

Unlike clusters, masses are usually directly provided for field stars in the literature and that has been used for the objects in our sample. Table \ref{tab:objects} shows the statistical properties of the distributions of rotation rates for the field stars used in our sample.

\subsection{Numerical method}
We solve our ensemble of equations \eqref{new-general-formalism} to \eqref{2shell} using a Runge-Kutta driver with adaptive step size control. All the quantities which depend on stellar evolution, namely radius, moment of inertia, convective timescale are calculated from a stellar evolution grid which extends from $0.5\,\Msun$ to $2.0\,\Msun$ in steps of $0.02\,\Msun$ and metallicities between $\rm [Fe/H]=-0.5$ to $0.1$ in steps of $0.1$. This grid was calculated using CEPAM \citep{ML08} with a mixing length parameter that was fitted to the Sun. Practically this means that for our solar metallicity $1.0\,\Msun$ mass star reaches a radius of $1\,\rm R_\odot$ for an effective temperature of $5777$\,K in 4750\,Myr. The small 200\,Myr offset between the model age and the age of the Solar System is negligible given other sources of uncertainty \citep[see e.g.,][]{Bonanno+Frohlich2015}.

The separation between the two zones defined as the deep radiative zone and the convective zone in equation~\eqref{2shell} was obtained by looking at the first transition from a radiative to a convective zone, neglecting a possible central convective zone.

\subsection{Comparison of existing models}
\label{sec:comparison}
	At this stage, we are ready to compare the predictions of currently existing rotational evolution models to our observational sample. To perform this task, we implement the rotational evolution models of \cite{1997A&A...326.1023B} [B97] and \cite{Johnstone2015} [J15] into our numerical code. Next, we allow each model to evolve for $12$ Gyr and finally look at snapshots from their evolution at the same age as the clusters and field stars in our sample to compare how well models can predict the rotation rates.   For the case of \cite{Gallet&Bouvier2015} [GB15], because this work uses values of $K_1$ which are mass-dependent (see~\ref{GB15 model}), we directly plot their published results. The assumptions of the different models are summarised in Table~\ref{tab:params}.

\begin{table*}
\caption{ Model parameters.}             
\label{tab:params}      
\begin{threeparttable}
\centering                                      
\begin{tabular}{c c c c c c c c c c c}          
\hline                      
Model &Magnetic field & Mass loss & Saturation & Interior mixing & $m$ & $K_1$ & Ref \\
\hline
B97 &  $B\propto \Omega/R^2$ & None & $\Omega > 15\Omega_\odot$ & None & 0.5 & $0.503$ & 1 \\
GB15 & $B\propto \Ro^{-2.6\star}$ & $\dot{M} \propto \Ro^{-1.58\star}$ & $\Omega > 15\Omega_\odot$ & $\tau_{\rm mix}=10$ to 500\,Myr & 0.22 & 1.7 to 8.5  & 2 \\
J15 & $B \propto \Ro^{-1.32}$ & $\dot{M} \propto R^2 \Omega ^{1.33} M^{-3.36}$ &  $\Omega > 15\Omega_\odot$ & None & 0.2177 & 6.76 & 3\\
This work & $B \propto \Ro^{-1.2}$ & $\dot{M} \propto \Ro^{-1.3}$ & $\Ro>0.09$ & $\tau_{\rm mix}=150$\,Myr & 0.2177 & 6.34  & 4\\
\hline                                             
\end{tabular}
\begin{tablenotes}
\item[]References: (1)~\cite{1997A&A...326.1023B}; (2)~\cite{Gallet&Bouvier2015}; (3)~\cite{Johnstone2015}; (4)~This work.
\end{tablenotes}
\begin{tablenotes}
\item[]$^\star$ : This is an approximate relation for $1.5\le \Omega/\Omega_\odot \le 4$.
\end{tablenotes}
\end{threeparttable}
\end{table*}

\subsubsection{The B97 Model}

The model of B97 for the stellar braking of low-mass stars ($0.5 \rm M_\odot< m < 1.1 \rm M_\odot$ ) with an outer convective zone stems from \cite{Kawaler1988}'s formulation. They use a dynamo relation which saturates for fast rotators beyond the spin rate of $\omega_{\rm sat}$ and is described by the following equations:
\begin{equation}
\left({dJ\over dt}\right)_w=\begin{cases}
\displaystyle -K\Omega^3\left({R\over R_\odot}\right)^{1/2} \left({M\over M_\odot}\right)^{-1/2} & {(\Omega < \omega_{\rm sat})},\\
\displaystyle -K\Omega \omega_{\rm sat}^2 \left({R\over R_\odot}\right)^{1/2} \left({M\over M_\odot}\right)^{-1/2} & {(\Omega > \omega_{\rm sat})}
\end{cases}
\label{eq:B97}
\end{equation}
where $K$ is a calibration constant. The critical rotation rate in equation~\eqref{eq:B97} is the only free parameter in B97.  The values of the parameters used in this model are listed in table~\ref{tab:params}.

B97 can be made equivalent to our model by neglecting the $\Omega^2 R^2$ term in equation~\eqref{eq:new-rJ}, assuming $m=1/2$ and assuming that $B\propto \Omega/R^2$ rather than $B\propto \Ro^{-1}$. We thus obtain that ${dJ/ dt}={(2/3)}{(K_{\rm 1}^2/ {K_{\rm 2})}v_{\rm {esc}}}\Omega^3(B_\odot^2R_\odot^4/\Omega_\odot^2)$ and as a result $K={(2/3)}{(K_{\rm 1}^2/ {K_{\rm 2}})}(B_\odot R_\odot^4/\Omega_\odot^2 v_{\rm{esc\odot}})$. Assuming $\Omega_\odot=2.67\times 10^{-6}$ and $B_\odot=1\,$G, and using $K=2.7\times 10^{47}$\,cgs, B97 implies $K_1^2/K_2\approx 5$ and $K_1\approx 0.503$. 

\subsubsection{The GB15 Model}
\label{GB15 model}

\citet{Gallet&Bouvier2013, Gallet&Bouvier2015} develop an angular momentum evolution model for 0.5, 0.8 and 1.0 $M_\odot$ stars. They adopt \cite{Matt2012}'s prescriptions for the wind braking law and following their work choose $K_2 = 0.0506$ and $m = 0.22$ as the input parameters in equation~\ref{eq:new-rJ}. The mean magnetic field and mass loss relations used in their model are adopted from \cite{Cranmer2011} with slight modifications. They are a function of stellar density, effective temperature, and angular velocity. As shown in Table~\ref{tab:params}, for angular velocities between 1.5 and 4 times that of the Sun, they find relatively steep dependences on Rossby number, i.e., $\Ro^{-2.6}$ for the magnetic field strength  and $\Ro^{-1.58}$ for the mass loss rate. They also use a prescription that smoothly joins the non-saturated and saturated magnetic regimes.

In addition, GB15 fit several other parameters. The value of $K_1$ is allowed to change as a function of mass. The value of the core-envelope mixing timescale (our $\rm \tau_{mix}$) changes both as a function of stellar mass and initial rotation rate. Thus, for 3 mass bins and 2 rotation rates, 9 additional parameters must be found  in addition to $K_1$, $K_2$, $a$, $d$, critical spin rate and disk lifetime $\rm \tau_{disk}$. As a result, $K_1$ varies between $1.7$ for $1\,\Msun$ stars to $8.5$ for $0.5\,\Msun$ stars. The value of $\rm \tau_{mix}$ range for 10\,Myr for fast-rotating $1\,\Msun$ stars to 500\,Myr for slow-rotating $0.5\,\Msun$ stars.

\subsubsection{The J15 Model}

 J15 construct a model to describe rotation and wind properties of low-mass main-sequence stars. Like GB15 they adopt \cite{Matt2012}'s prescriptions for the wind braking law and choose $K_2 = 0.0506$ and $m = 0.22$. The mean magnetic field used in their study is adopted from \cite{Vidotto2014}. They derive the mass loss rates by fitting their rotational evolution model to the observations by assuming that, in the unsaturated regime, the mass loss rate per unit surface area of a star has power-law dependences on its mass and rotation rate. Their model works with six parameters summarised in table~\ref{tab:params}. An important point to note about the J15 model is that it does not assume a decoupling of internal layers of stars but considers that they rotate as solid bodies. 
 Since the model is limited to stars older than $100\,\rm Myr$, the consequence of this assumption is that they do not consider the pre-main-sequence phase.

\subsubsection{Comparison of Models}

\begin{figure*}
\begin{subfigure}{.32\textwidth}
  \centering
  \includegraphics[width=1.\linewidth]{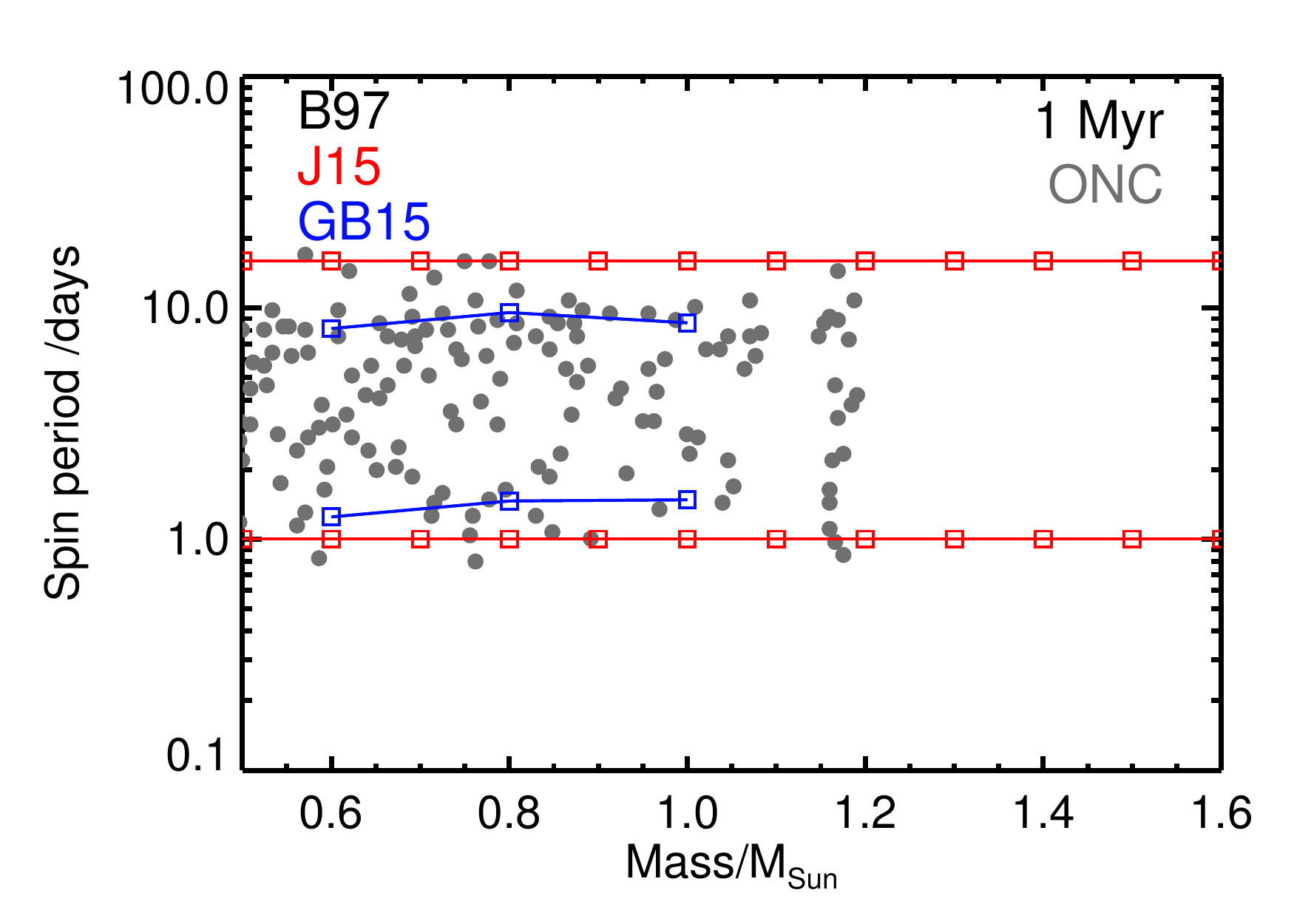}
  \label{fig:sfig1}
\end{subfigure}%
\begin{subfigure}{.32\textwidth}
  \centering
  \includegraphics[width=1.\linewidth]{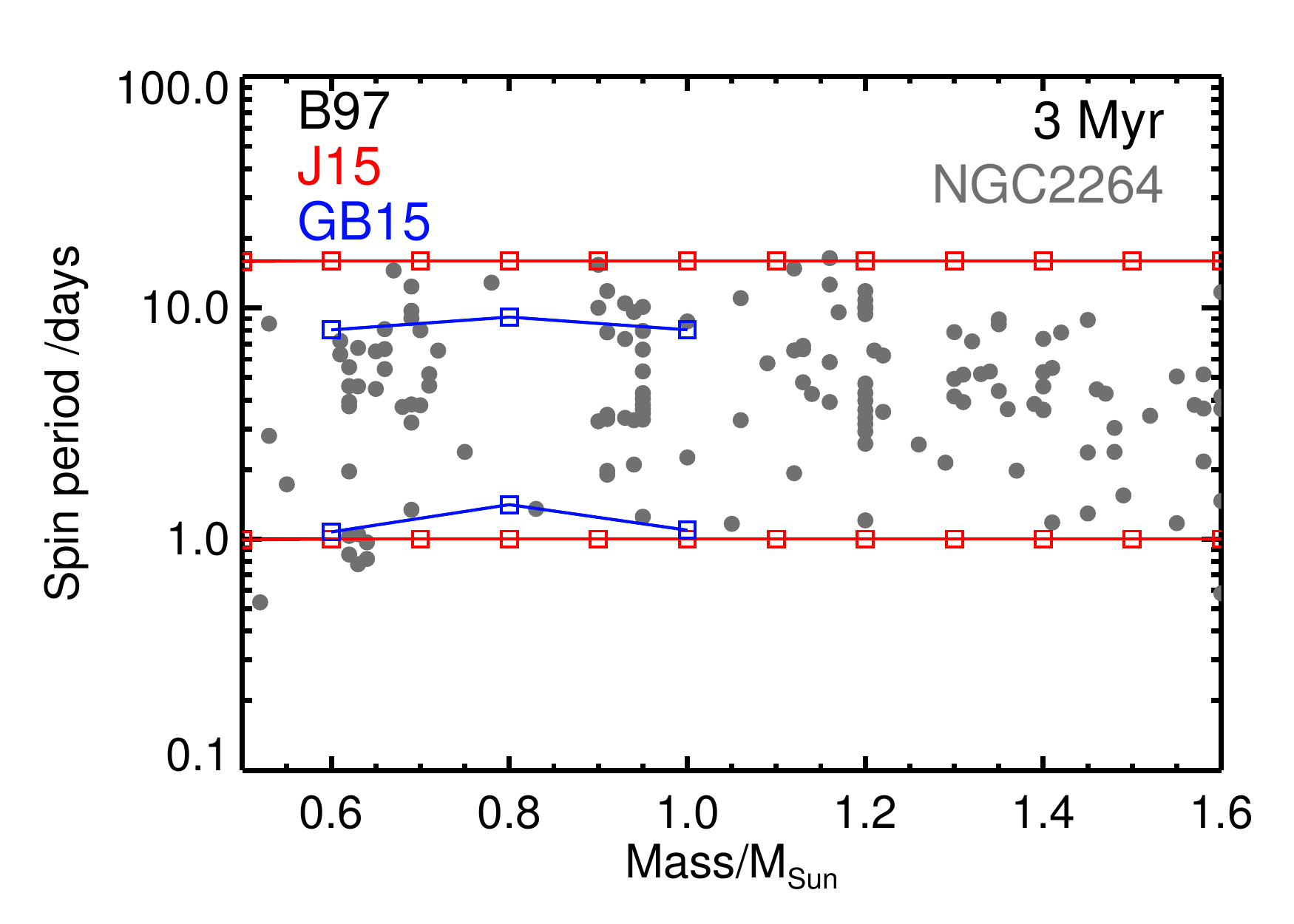}
  \label{fig:sfig111}
\end{subfigure}%
\begin{subfigure}{.32\textwidth}
  \centering
  \includegraphics[width=1.\linewidth]{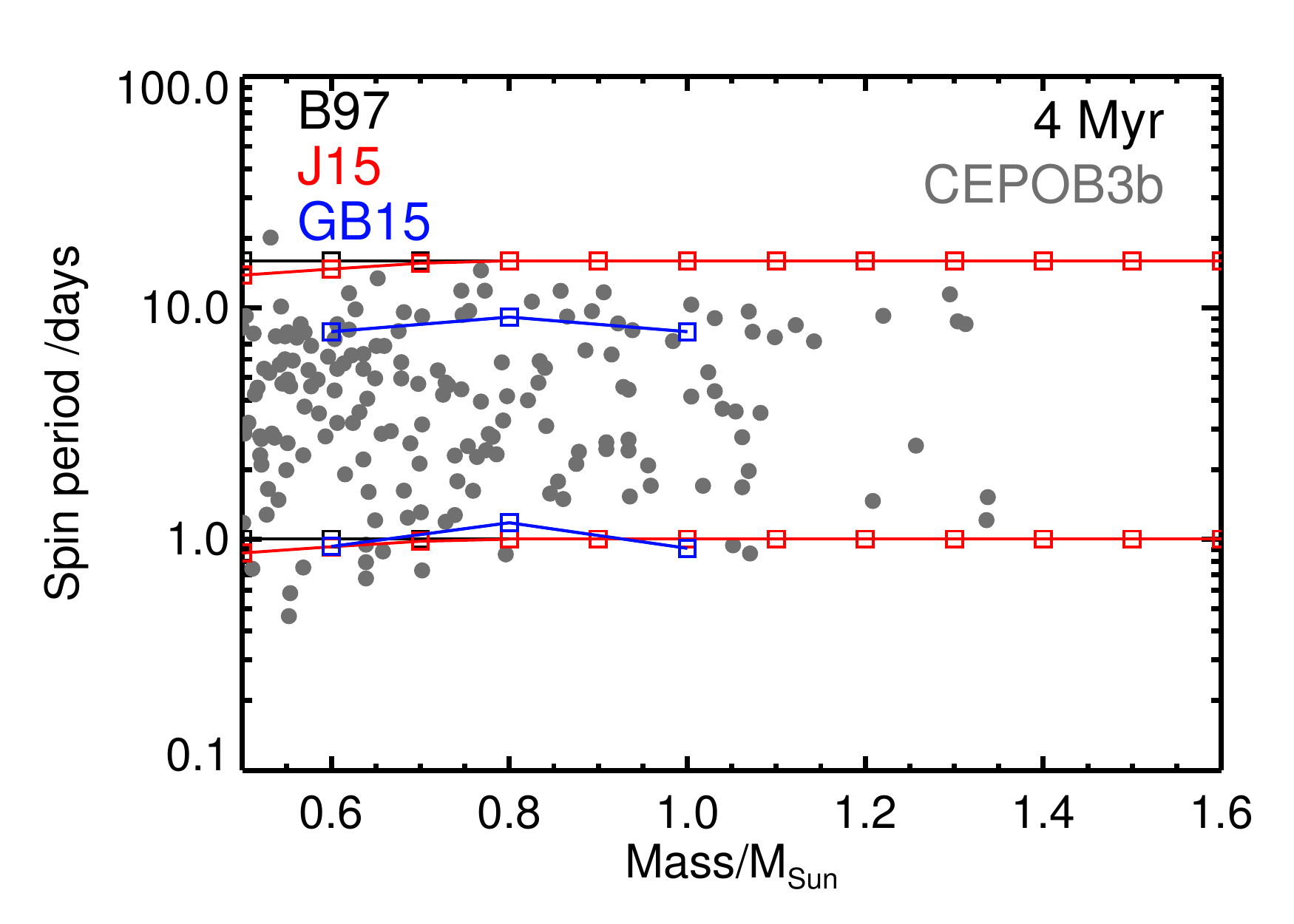}
  \label{fig:sfig2}
\end{subfigure}
\begin{subfigure}{.32\textwidth}
  \centering
  \includegraphics[width=1.\linewidth]{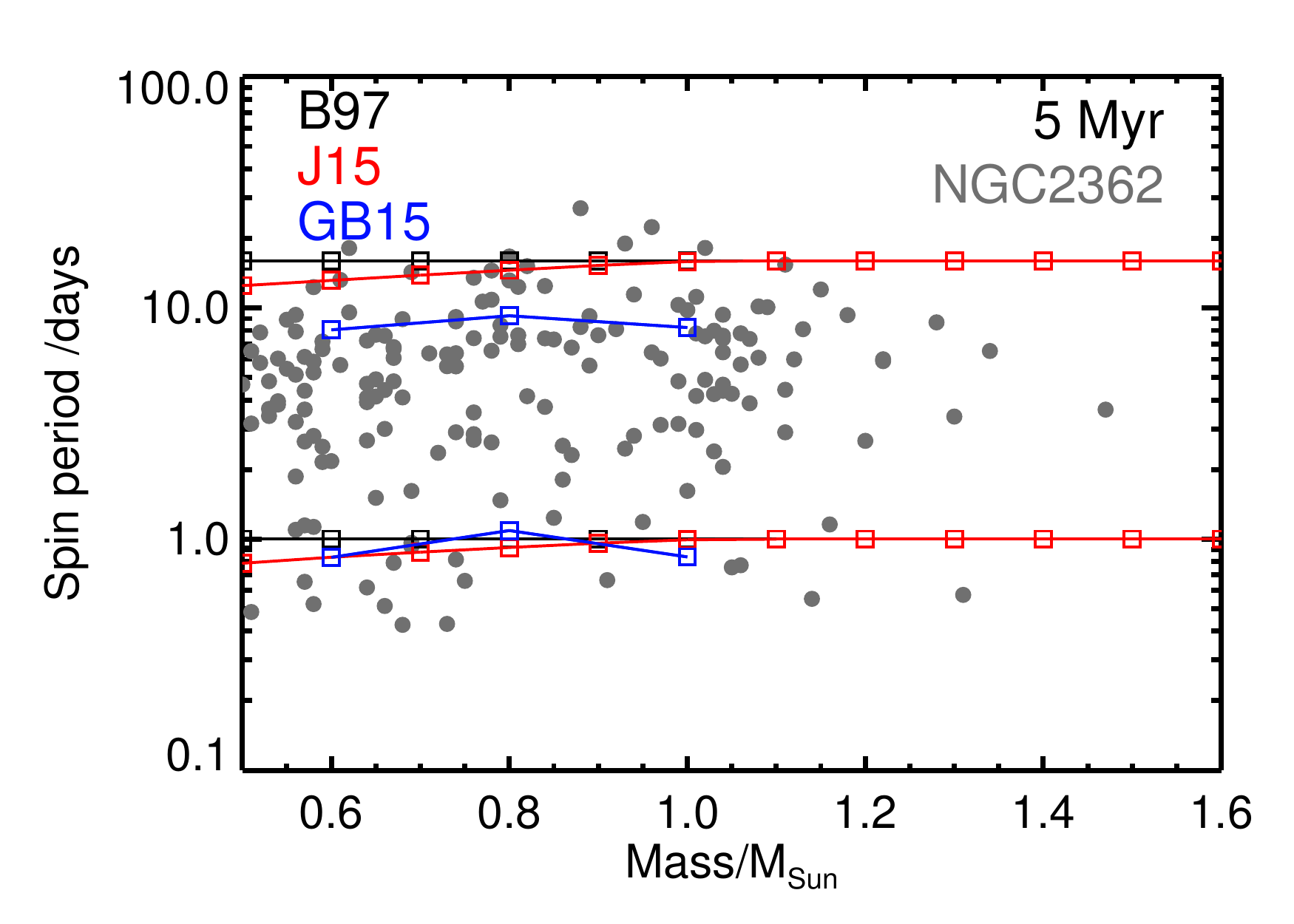}
  \label{fig:sfig3}
\end{subfigure}
\begin{subfigure}{.32\textwidth}
  \centering
  \includegraphics[width=1.\linewidth]{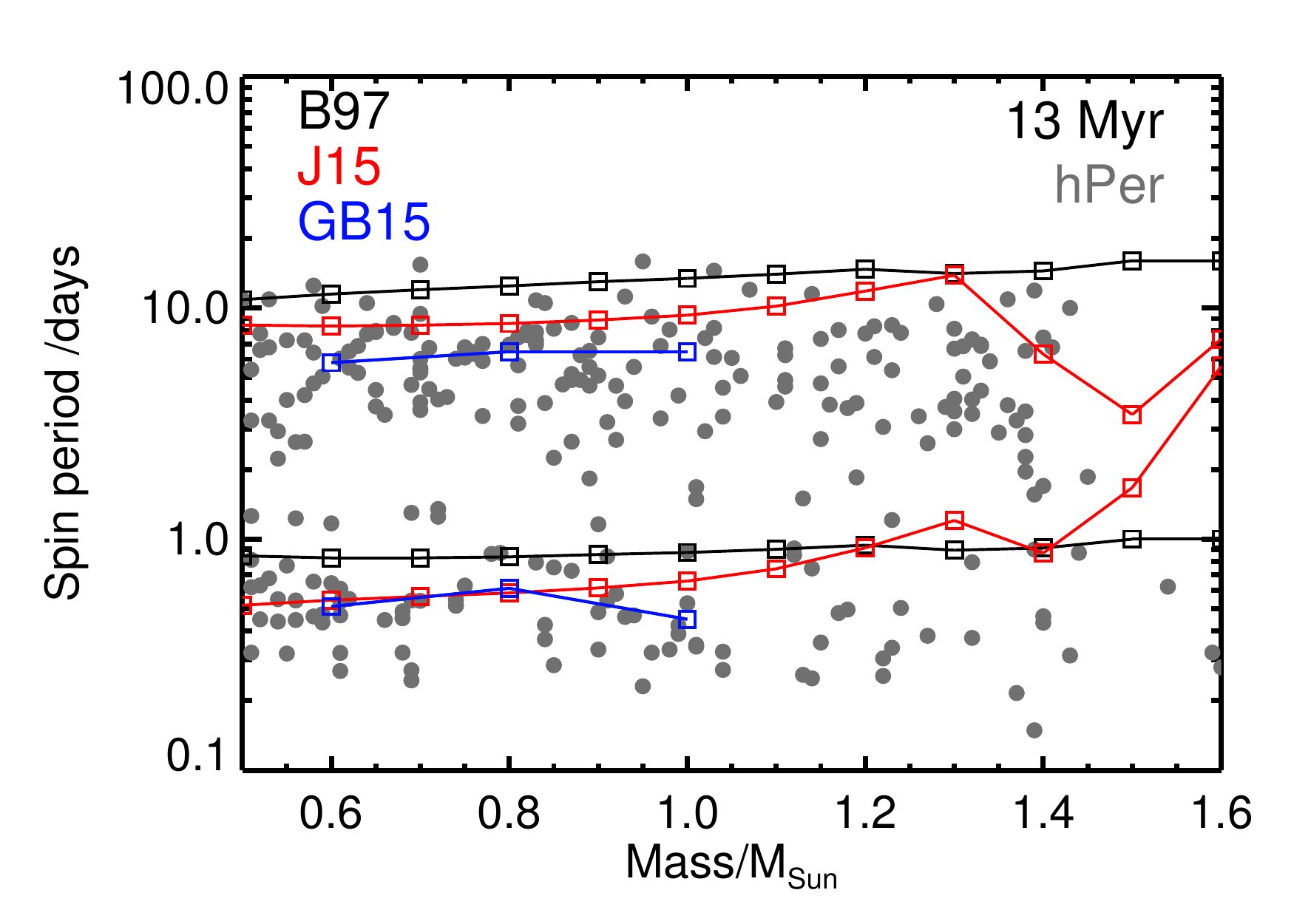}
  \label{fig:sfig4}
\end{subfigure}
\begin{subfigure}{.32\textwidth}
  \centering
  \includegraphics[width=1.\linewidth]{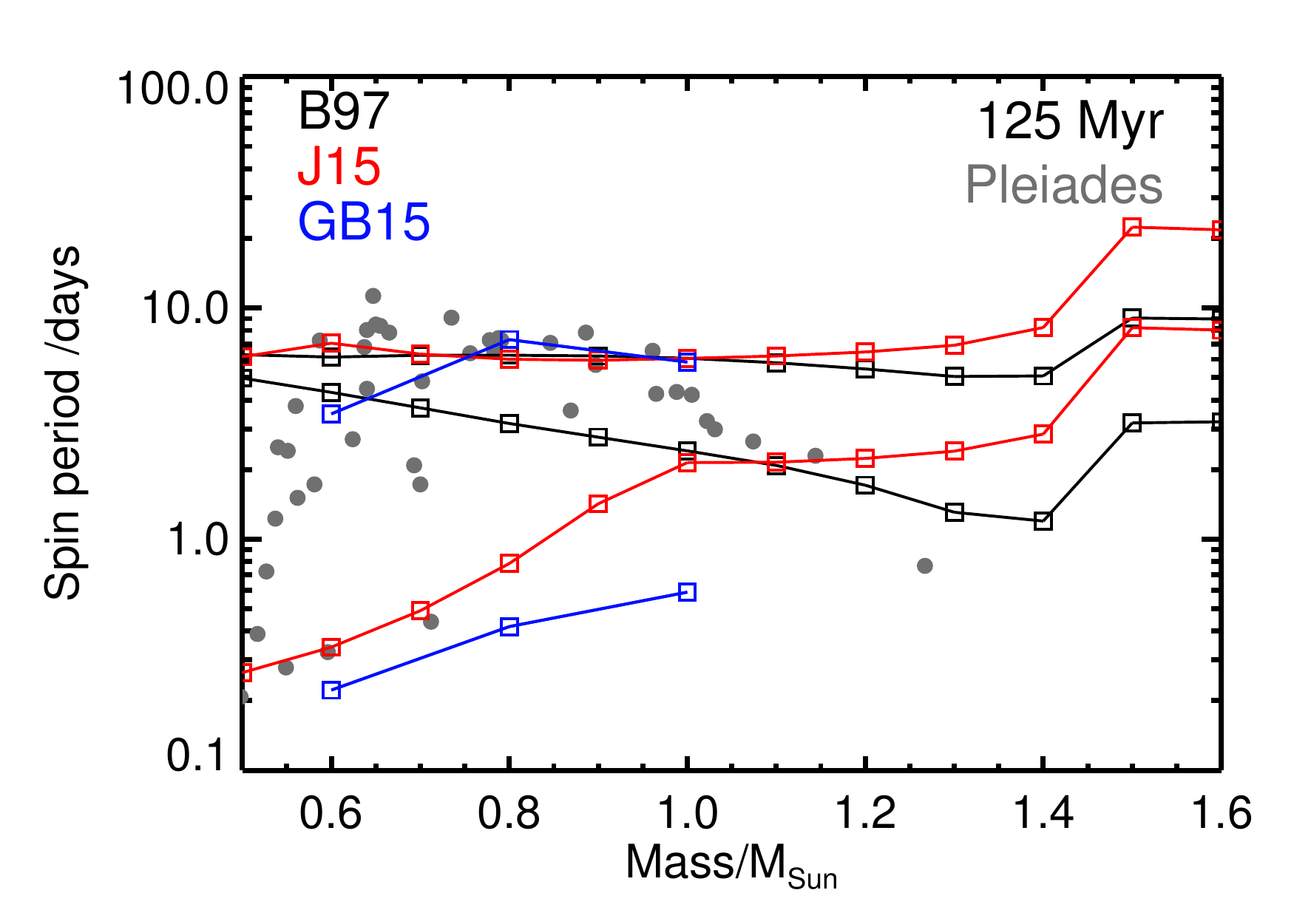}
  \label{fig:sfig5}
\end{subfigure}
\begin{subfigure}{.32\textwidth}
  \centering
  \includegraphics[width=1.\linewidth]{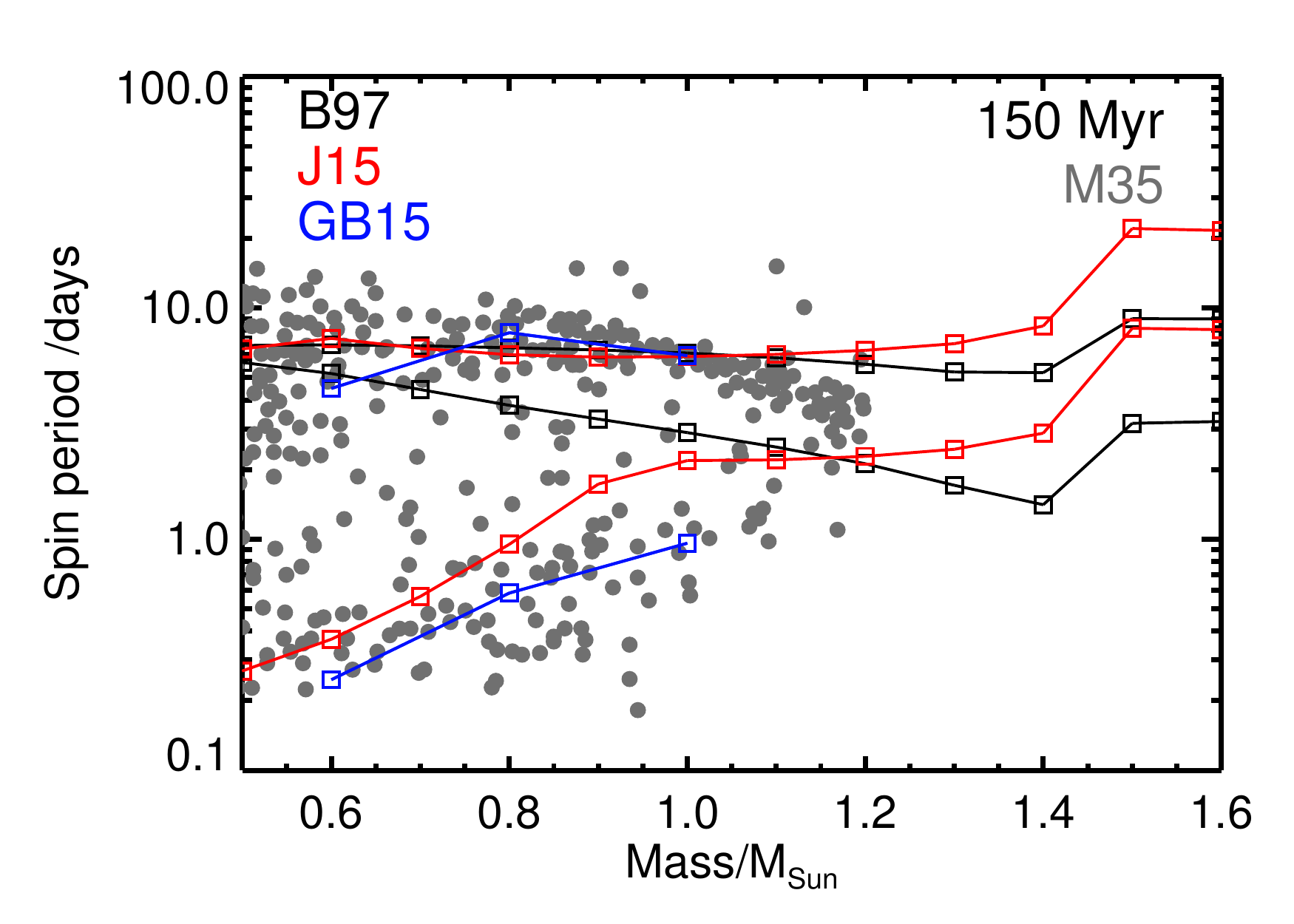}
  \label{fig:sfig6}
\end{subfigure}
\begin{subfigure}{.32\textwidth}
  \centering
  \includegraphics[width=1.\linewidth]{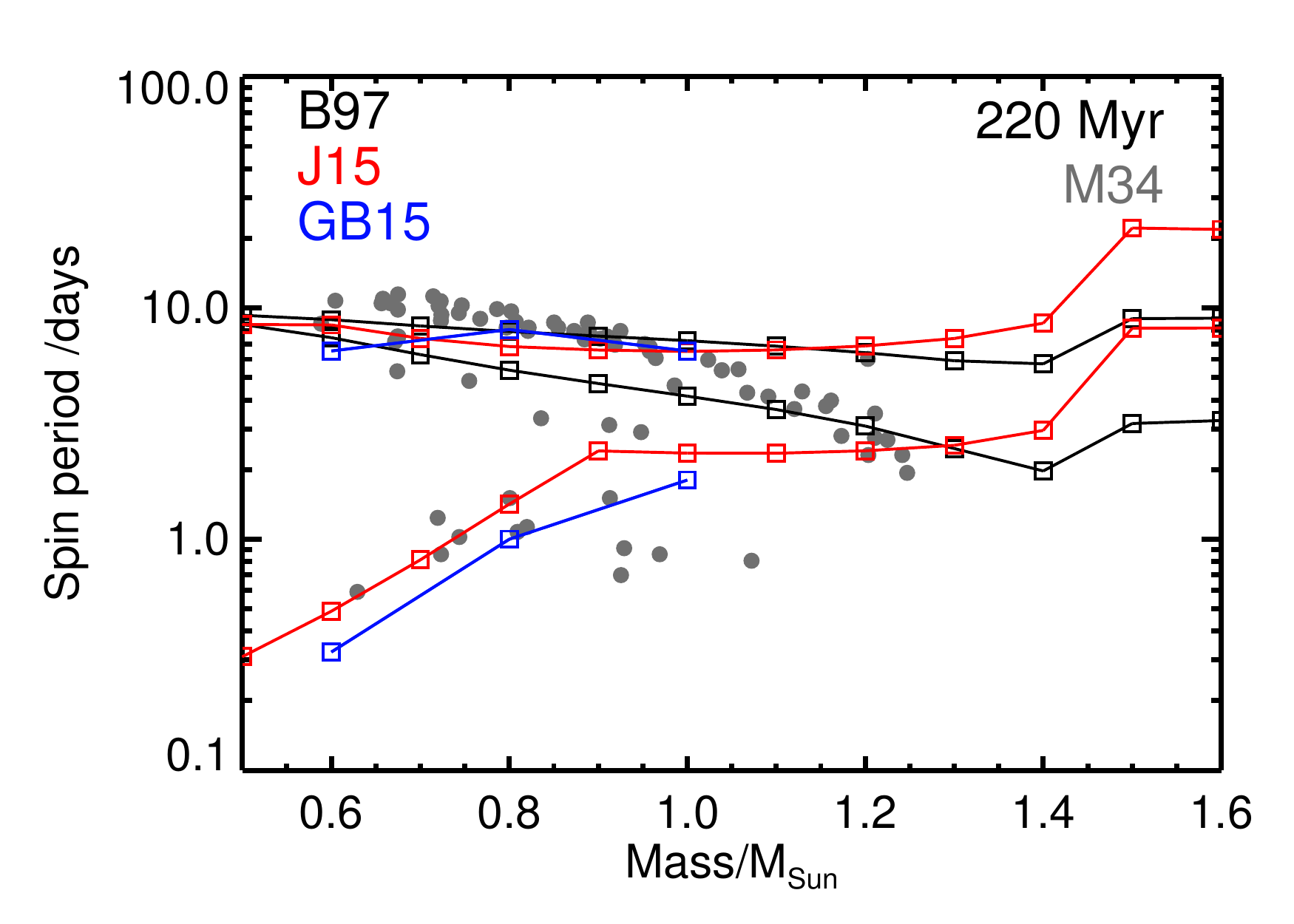}
  \label{fig:sfig7}
\end{subfigure}
\begin{subfigure}{.32\textwidth}
  \centering
  \includegraphics[width=1.\linewidth]{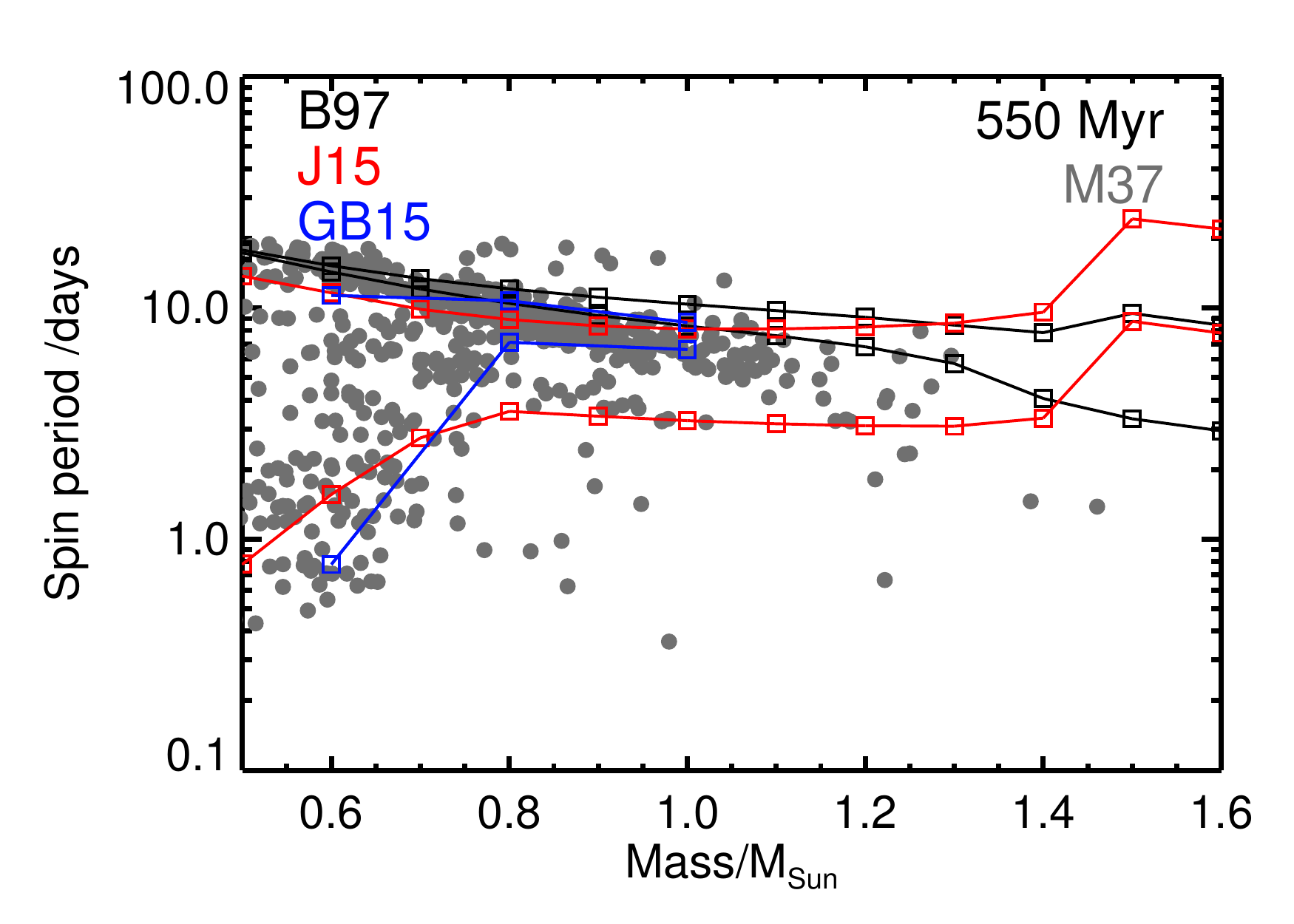}
  \label{fig:sfig8}
\end{subfigure}
\begin{subfigure}{.32\textwidth}
  \centering
  \includegraphics[width=1.\linewidth]{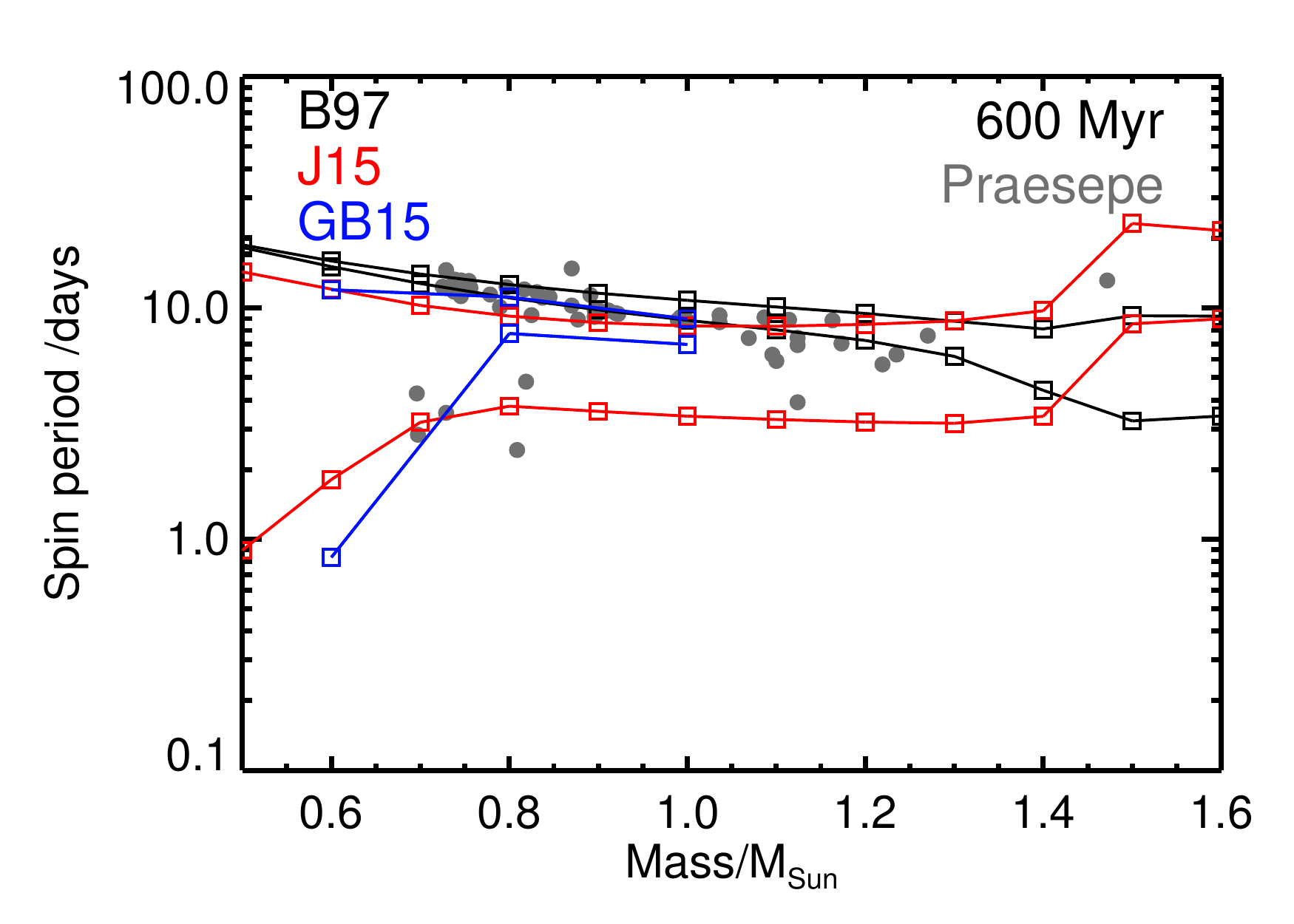}
  \label{fig:sfig9}
\end{subfigure}
\begin{subfigure}{.32\textwidth}
  \centering
  \includegraphics[width=1.\linewidth]{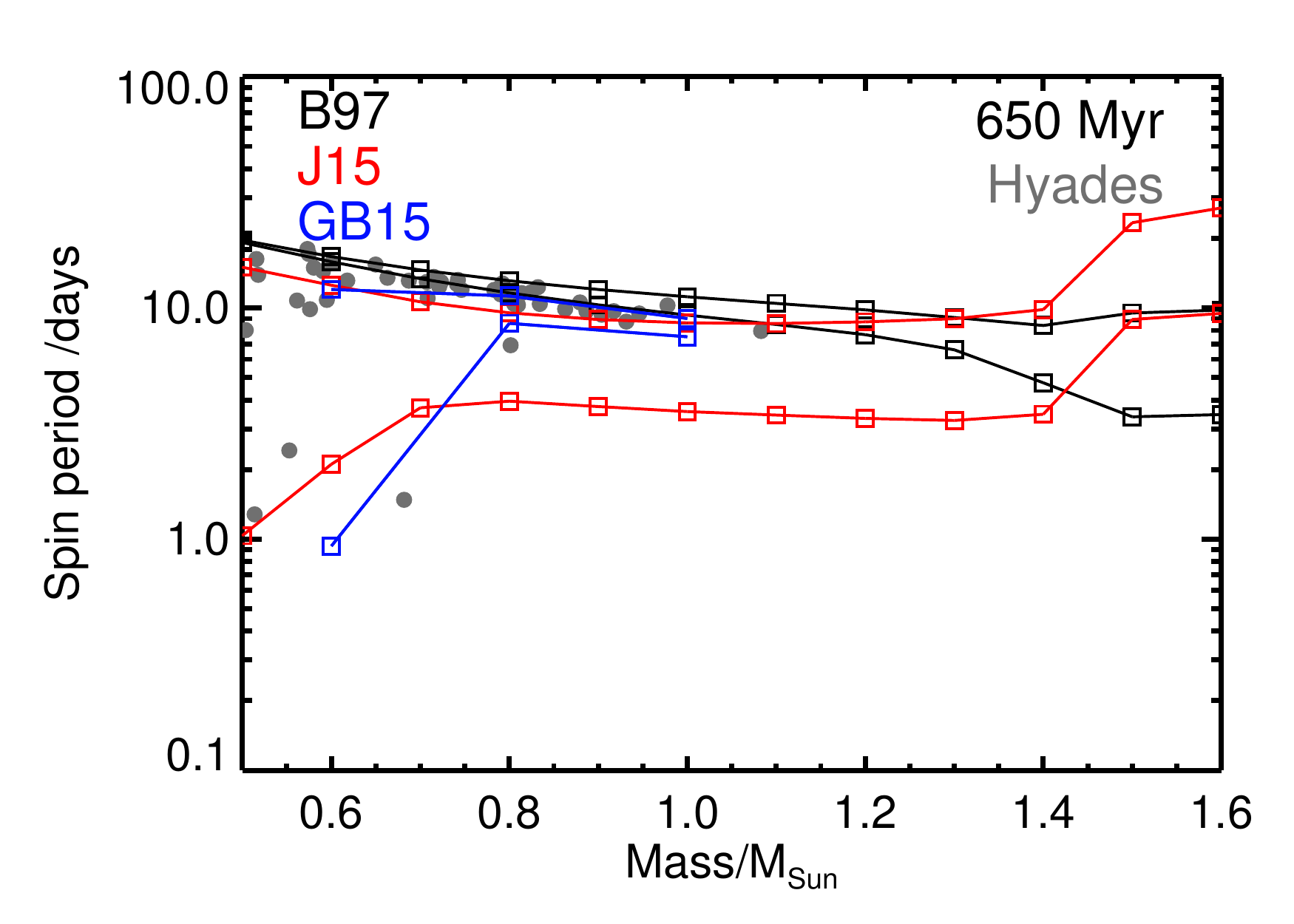}
  \label{fig:sfig10}
\end{subfigure}
\begin{subfigure}{.32\textwidth}
  \centering
  \includegraphics[width=1.\linewidth]{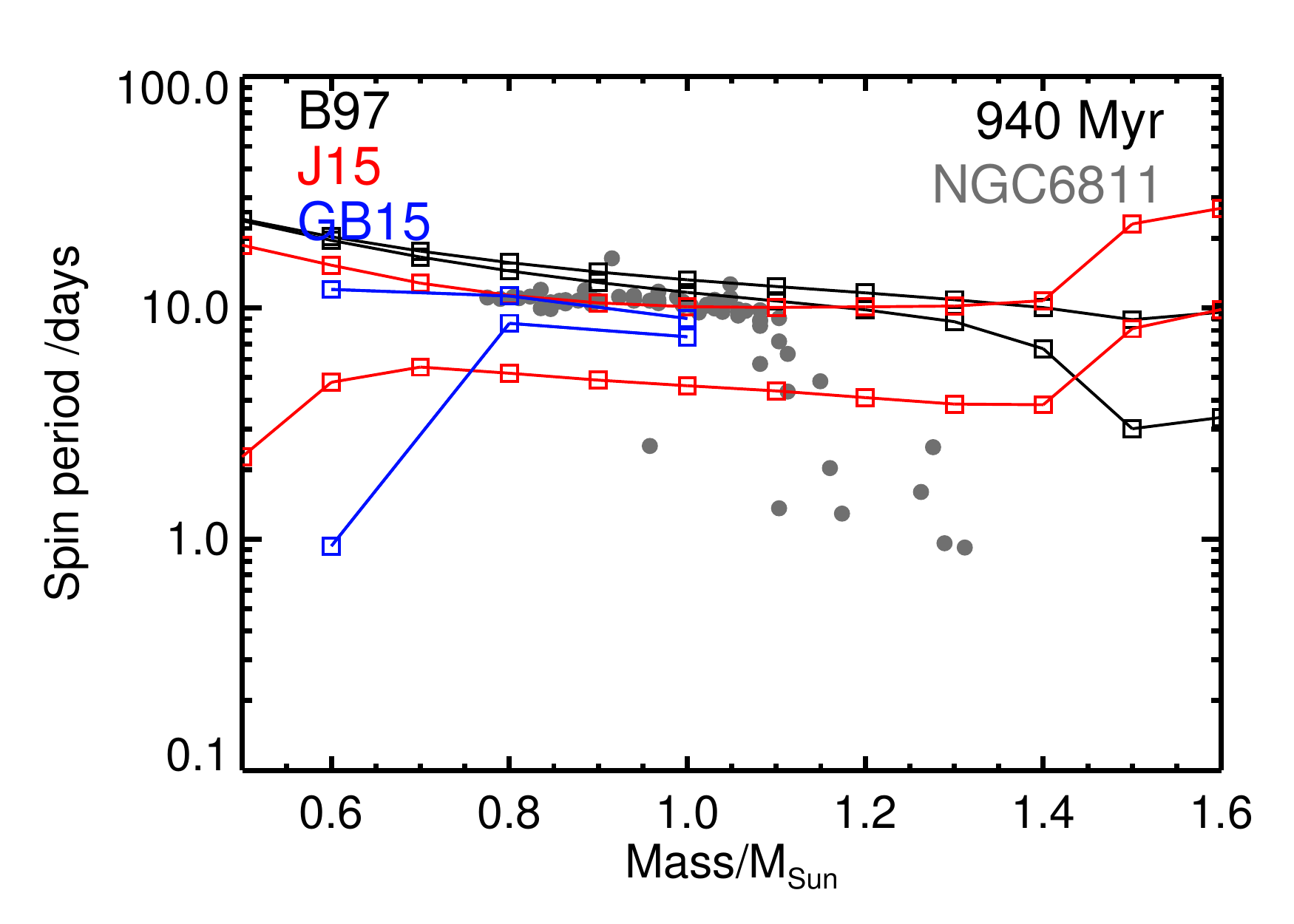}
  \label{fig:sfig11}
\end{subfigure}
\begin{subfigure}{.32\textwidth}
  \centering
  \includegraphics[width=1.\linewidth]{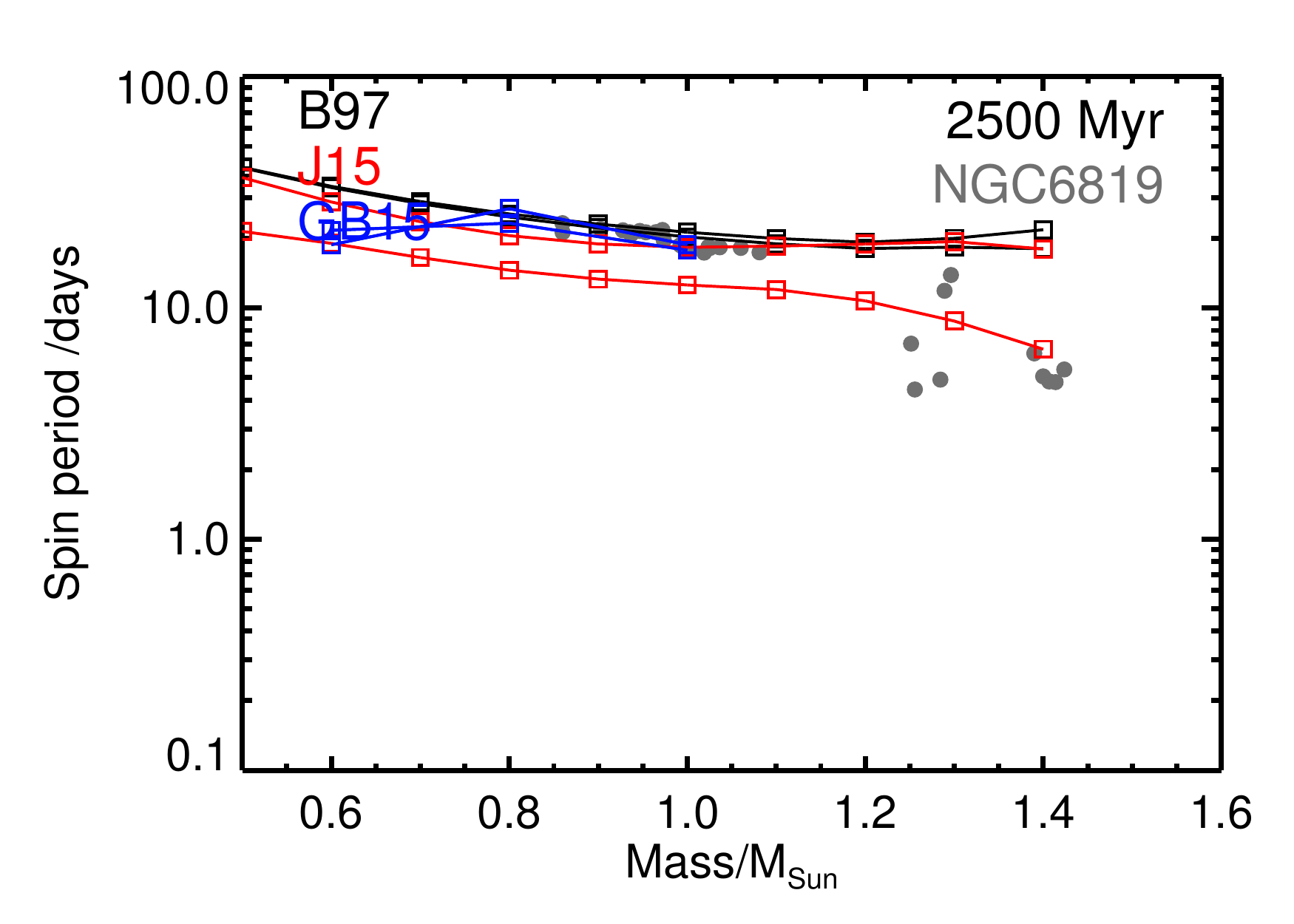}
  \label{fig:sfig11}
\end{subfigure}
\begin{subfigure}{.31\textwidth}
  \centering
  \includegraphics[width=1.\linewidth]{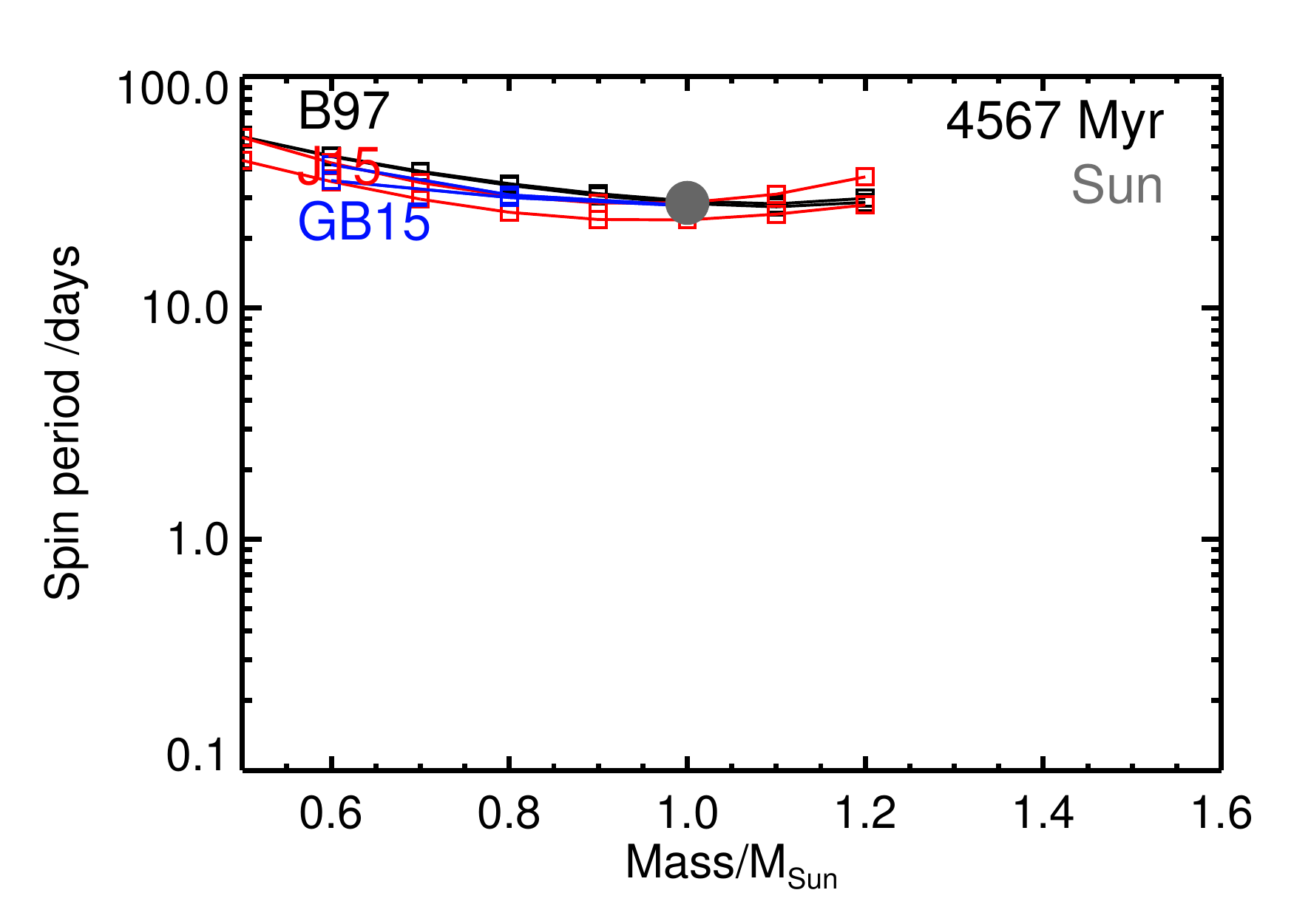}
  \label{fig:sfig11}
\end{subfigure}
\begin{subfigure}{.31\textwidth}
  \centering
  \includegraphics[width=1.\linewidth]{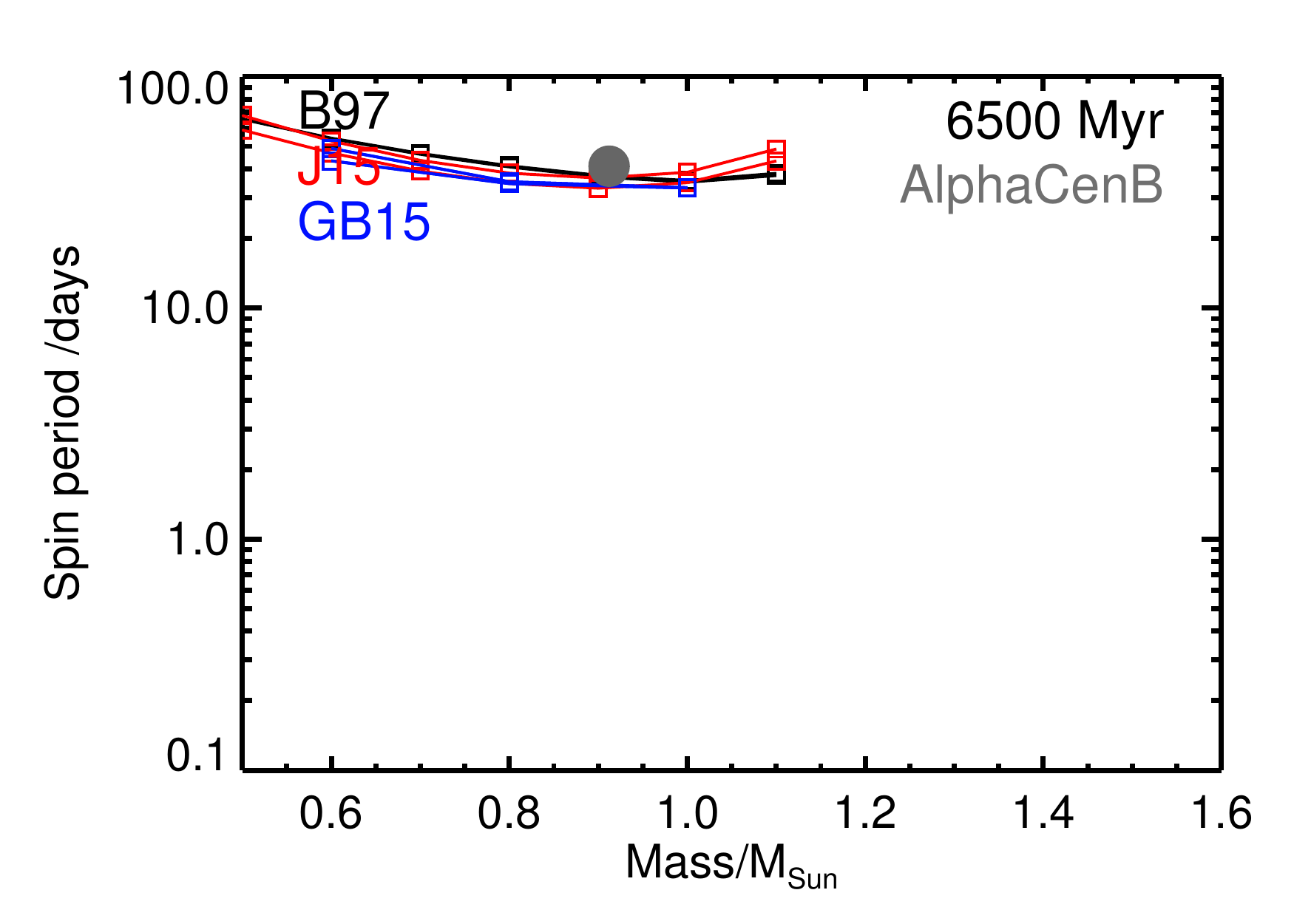}
  \label{fig:sfig11}
\end{subfigure}

\caption{Predictions of the evolution of stellar spin period made by three previous models B97, GB15 and J15 plotted over observed spin periods of our sample. The B97 model is shown in solid black lines, J15's in red and that of GB15's is in blue. It should be noted that B97 and GB15 assume the mixing timescale to be different for slow and fast rotators, an assumption that we do not make here.  The position of the Sun and AlphaCenB have been marked with  larger circles.}
\label{fig:model_comparison}
\end{figure*}

Fig.~\ref{fig:model_comparison} shows predictions of the evolution of stellar spin period made by the above models versus stellar mass. These have been plotted over observations of the rotation periods of the clusters and field stars of our sample at different ages (Tables~\ref{tab:clusters} \& \ref{tab:objects}). The two lines for each model represent the trend of the slow and fast rotators. All three models fitted $90$ percent of the stars to estimate their parameters and thus  $10$ percent of the stars rotating the fastest are not included in their fit. The stars first rapidly spin up due to the contraction during the PMS stage. But after entering the main-sequence at about $30\,$ to $40\,\rm Myr$, magnetic braking starts to slow the stars down.

Looking at fig.~\ref{fig:model_comparison}, the three models provide a fair fit to the ensemble of clusters that have been observed. Globally, the models do reproduce the observed spin down of stars with age and some of the observed trends in terms of mass.  However, for the case of B97 we can see that, starting from two initial spin periods that cleanly encompass the observations at 1\,Myr for the ONC, the situation deteriorates with more stars being outside this envelope, particularly for M35 at 150\,Myr and for M37 at 550\,Myr. B97 tends to explain the slow rotators in M37 and older clusters but not the less numerous fast rotators. Because of the early slow down of the fast rotators, this parameterisation creates poor fits to many of the stars in Pleiades at $125$ \rm{Myr} as well as M35 at $150$ \rm{Myr}.

 GB15 obtains a quite good fit for the evolution of low-mass stars at the expense of using many free parameters. It is also important to stress that these fair fits are obtained by assuming a dependence of mixing timescale on rotation rate. However, this is a hypothesis that we view as unnecessary at this point and that we choose not to adopt. The J15 parameterisation (which assumes solid-body rotation) is more successful, including for the Pleiades. It generally explains the spin rates for about 90 percent of the observed stars. However, it also leaves out at least 10 percent of the fast rotators in M35 and M37. 

Furthermore, these parameterisations are valid for stars with an outer convective zone but generally fail to reproduce the fast-spinning F-dwarfs beyond the Kraft break \citep{Kraft1967}, as seen for M37, NGC6811 and NGC6819. This is not surprising because they were not specifically designed for that purpose, but it calls for a model that would work for a wider range of stars.

\section{Parameter Calibrations}
\label{sec:Parameters}

In this section we describe our method to calibrate our model parameters. We adopt the approach of trying to fit nearly all stars seen in clusters rather than focusing on a limited fraction which has usually been done in previous studies. In this way, we hope that the parameterisation can be used to identify more reliably outliers or stars which have been spun up by their exoplanet \citep[e.g.,][]{Poppenhaeger+Wolk2014,Guillot+2014}. Also, using this approach, we are looking at the evolution of single clusters instead of individual mass bins.

 Considering the collection of equations introduced in the previous sections and the assumptions made so far, we are left with $8$ free parameters which are listed in Table~\ref{tab:modelparameters}. Two of the parameters we do not vary. First, we do not calibrate $\RoM$ because we are mainly focusing on clusters and stars at younger ages whereas $\RoM$ mostly affects older stars.  We adopt  $\RoM=\Ro_\odot$. Secondly, the $K_{\rm 2}$ parameter is constant because it has generally a small effect. Therefore, we are left with $6$ parameters to vary.
\begin{table}
\caption{ Description of all the parameters introduced throughout this study.}
\label{tab:modelparameters}      
\begin{threeparttable}
\centering                                      
\hspace*{-1cm}
\begin{tabular}{l l}          
\hline                      
Parameter & Usage \\
\hline
\textit{$K_1$} & Calibration to the Sun \\
\textit{$K_2$} & Accounting for rapid rotators\\
\textit{a} & $B_0\propto (\Omega\tau_c)^a $\\
\textit{d} & $\dot{M}\propto (\Omega\tau_c)^d$\\
\textit{$\rm \tau_{mix}$} & Timescale of angular momentum transfer\\ & between radiative and convective zones\\
\textit{$\rm \tau_{disk}$} & Timescale of keeping stellar rotation \\ & constant due to the presence of the disk\\
$\Rom$ & Discerning between critical and \\ &sub-critical regimes\\
$\RoM$ & Suppression of magnetic field at old ages\\
\hline                                             
\end{tabular}
\end{threeparttable}
\end{table}

 Because of the large number of parameters and the inhomogeneous data to be used, we proceed as follows. We start from the parameters that, from initial tests, are shown to have higher effects on the evolution of angular momentum.  We choose to fit individual clusters by eye and find one optimal solution per parameter. 

In the next subsections, we present the results of each parameter search around our preferred values for the other parameters. For simplicity we present these results only for the cluster(s) which led to our choice.

\subsection{Calibration to the Sun: $K_{\rm 1}$ and $K_{\rm 2}$}

We use the ensemble of relations defined by equations~\eqref{new-general-formalism}, ~\eqref{Bstar},~\eqref{Mdot} \& ~\eqref{2shell} to calculate the evolution of our Sun and constrain $K_{\rm 1}$ as a function of the other parameters of the model. $K_{\rm 1}$ is adjusted so that the model correctly reproduces the Sun's rotation period at its current age. 

We adopt an initial rotation period of 8 d as typical of the rotation of T-Tauri stars \citep[e.g.][]{2009IAUS..258..363I}. The corresponding initial CESAM model has been evolved from an initially extended state for $1$\,Myr. At that point, it is a fully convective pre-main-sequence star with a radius of $2.2\rm\,R_\odot$. 

\begin{figure}
\hspace*{-0.25in}
  \resizebox{1.12\hsize}{!}{\includegraphics{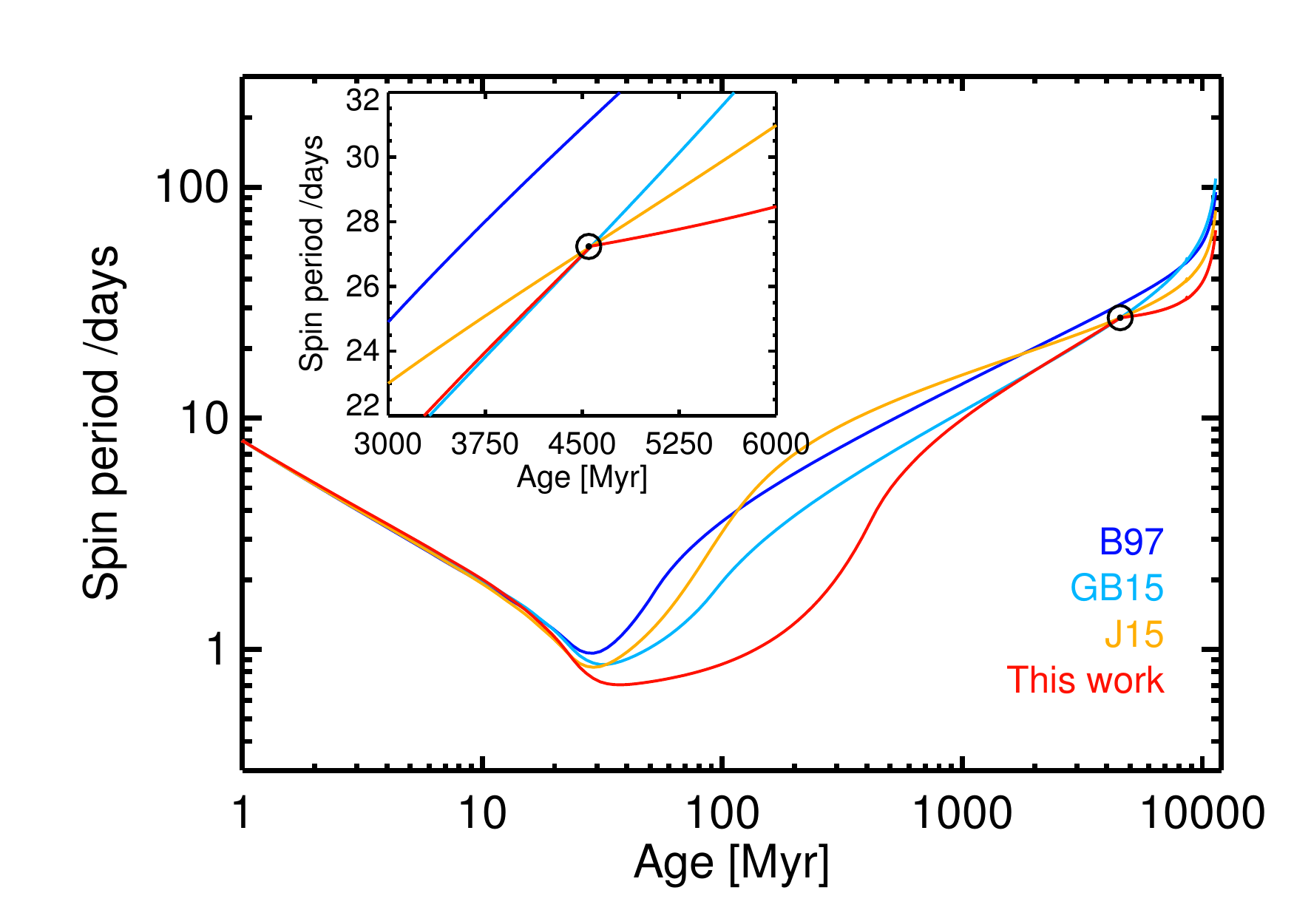}}
  \caption{Evolution of the spin period of the Sun with age in million years for different models, as labelled (see Table~\ref{tab:params} for details). The inset shows the plot in linear scale. The circle indicates the present state of the Sun.}
  \label{fig:sun_calib}
\end{figure}

Fig.~\ref{fig:sun_calib} shows the resulting spin evolution of the Sun from various sources in the literature and for our preferred model. The details of the parameters of each model are presented in Table~\ref{tab:params}. For $K_{\rm 2}$, following the results obtained by \cite{Matt2012}, we adopt a value of $0.0506$ and, as described earlier because of the small effect it has on the overall evolution of angular momentum, we choose not to vary this parameter.

By construction (through an optimization of the parameter $K_1$), all parameterisations reproduce the solar rotation but the evolution histories differ. In the pre-main-sequence evolution phase, the spin period is independent of the parameterisation: the spin-up is controlled by the stellar contraction. The evolutionary tracks begin to deviate from each other after 30\,Myr around the beginning of the main-sequence. The J15 parameterisation produces the strongest braking for solar-mass stars younger than the Sun. In this phase, the GB15 parameterisation yields spin periods which can be shorter by up to a factor $2$. The parameterisation that we present here results in a spin period which can be almost one order of magnitude smaller than J15 around an age of 200\,Myr. For older stars, the trends are inverted, with GB15 producing the slowest spinning stars and J15 predicting slightly faster rotation speeds. However, our parameterisation yields still faster rotation speeds because of the assumption of the suppression of magnetic braking when the Rossby number exceeds that of the Sun. This yields a non-smooth curve for our model in fig.\ref{fig:sun_calib}.

\subsection{Disk lock parameter: $\rm \tau_{\rm disk}$}
\label{sec:disklock_param}

It was shown in section \ref{sec:disklock} that the disk lock effect and the value of the parameter controlling it in the model $\rm \tau_{disk}$ is very important in terms of specifying the initial content of angular momentum in stars.  Because magnetic braking has a timescale of at least 10\,Myr, $\rm \tau_{\rm disk}$ is fully decoupled from the other parameters and may be determined independently.

In addition, it was also highlighted in \ref{sec:disklock} that the fastest rotating stars at, say, $5\,\rm Myr$ are those which lose their disks rapidly. As a result we should expect the fastest rotators to have a $\rm \tau_{\rm disk}\approx 1$ to $2\,\rm Myr$. Therefore, following \cite{2001ApJ...553L.153H}, we started by adopting a disk lifetime of $6\,\rm Myr$, corresponding to the mean overall disk lifetime observed in young clusters and then tested different values in its proximity to the slow rotator branch. In addition, for the fast rotators we tested values close to $1\,\rm Myr$. \cite{Bertout2007} suggested that there is a dependence between $\rm \tau_{\rm disk}$ and stellar mass. However, our calculations of spin rates in young clusters do not enable us to distinguish between the Bertout et al. relation and one with a value of $\rm \tau_{\rm disk}$ that is independent of stellar mass.  As a result we simply assume a constant timescale for all masses but acknowledge that for small mass fast rotators, there may be a slight tendency for an even smaller $\rm \tau_{\rm disk}$.

\begin{figure}
 \hspace*{-0.25in}
 \resizebox{1.12\hsize}{!}{\includegraphics{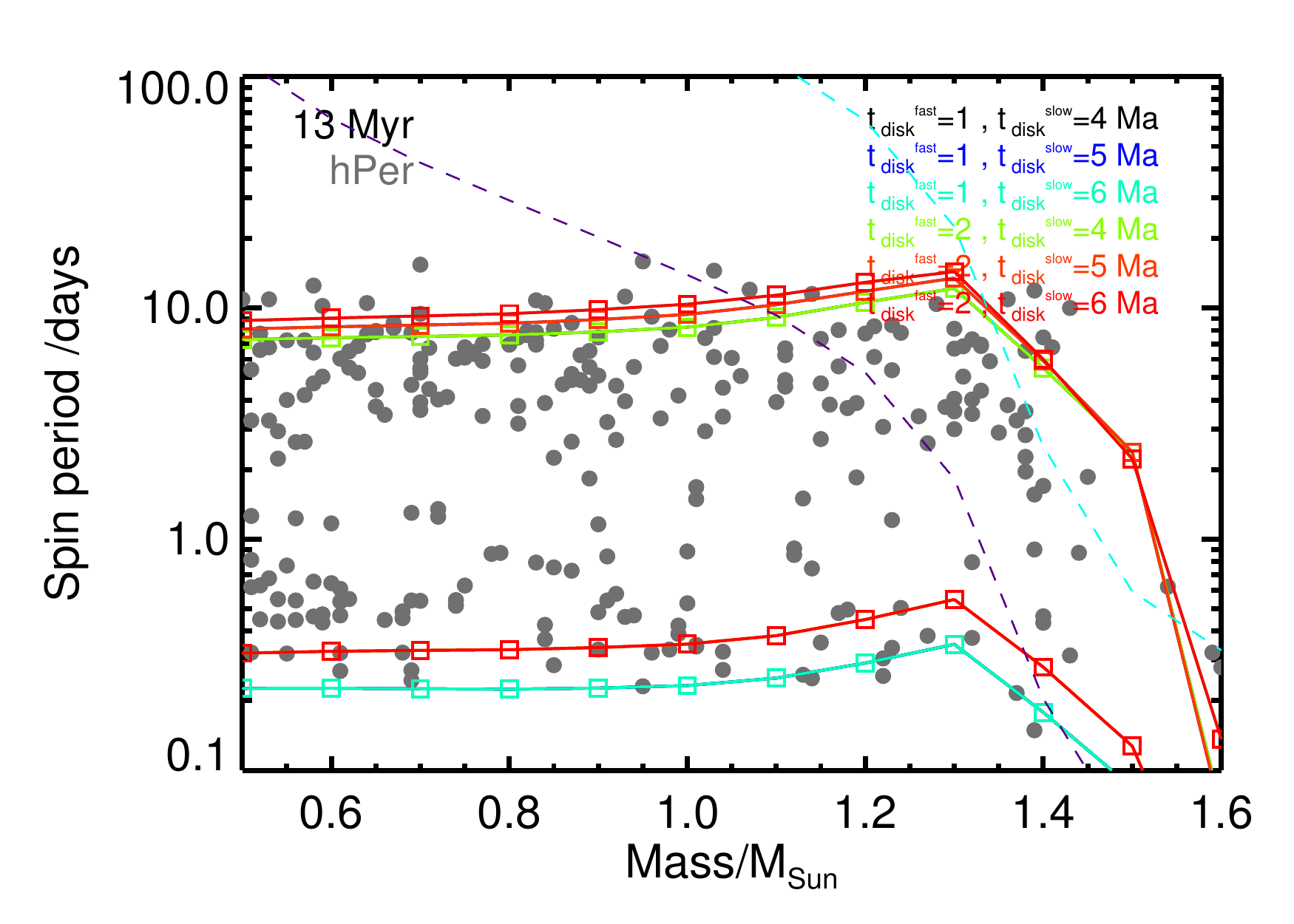}}
	\caption{Models of different $\tau_{\rm disk}$ on top of the spin period profile of the hPer cluster as a function of stellar mass. The numbers in legends represent the value of $\tau_{\rm disk}$ in fast and slow rotators branch respectively. }
\label{fig:param_timelock}
\end{figure}

Based on fig.~\ref{fig:param_timelock} we have adopted a value of $\rm \tau_{\rm disk}=1\,\rm Myr$ and $6\,\rm Myr$ for best fit to the fast and slow rotators, respectively.

\subsection{Magnetic field and mass loss parameters: $\Rom$, $a$ and $d$}
\label{B_Mdot_degeneracy}
 According to equations~\eqref{Bstar} and \eqref{Mdot}, the magnetic field of the star and its mass loss are controlled by three parameters in our model, $a$ which describes the dependence of the magnetic field to the star's Rossby number, $d$ which plays the same role for mass loss and $\Rom$ which determines when the magnetic field saturates and magnetic braking is weakened. 
 
In the non-saturated regime our model predicts that the stars have an angular momentum decay rate of $dJ/dt\propto {\rm Ro}^{-d(1-2m)-4am}$. This shows $a$ and $d$ are degenerate.  In addition, because of the calibration of our models to the solar spin rate, constant~$K_1$ depends on $\Rom$ which therefore also makes $\Rom$ degenerate with $a$ and $d$.

In order to explore the extent of this degeneracy we used the results of \cite{See2017} and \cite{Folsom2016} who provide predictions on the dependency of mass loss and magnetic field strength on Rossby number using Zeeman Doppler Imaging. It should be stressed that the estimates obtained for $a$ \& $d$ parameters by \cite{See2017} and \cite{Folsom2016} have been found by adopting a value of $\Rom =0.1$. This shows that the $\Rom$ parameter is also degenerate. Hence, we vary this parameter alongside $a$ \& $d$.
 
On this basis, we choose to test values of $0.8<a<1.8$, $0.7<d<1.7$ and $0.05<\Rom<0.2$ to explore the degeneracy between these parameters and look for cases that could provide fair fits to our clusters. In appendix~\ref{app:a and d} we show how $\Rom$, $a$ \& $d$ affect the models against the stars in the M35 and M37 clusters.

\begin{figure*}
    \centering  
    \includegraphics[width=1.\linewidth]{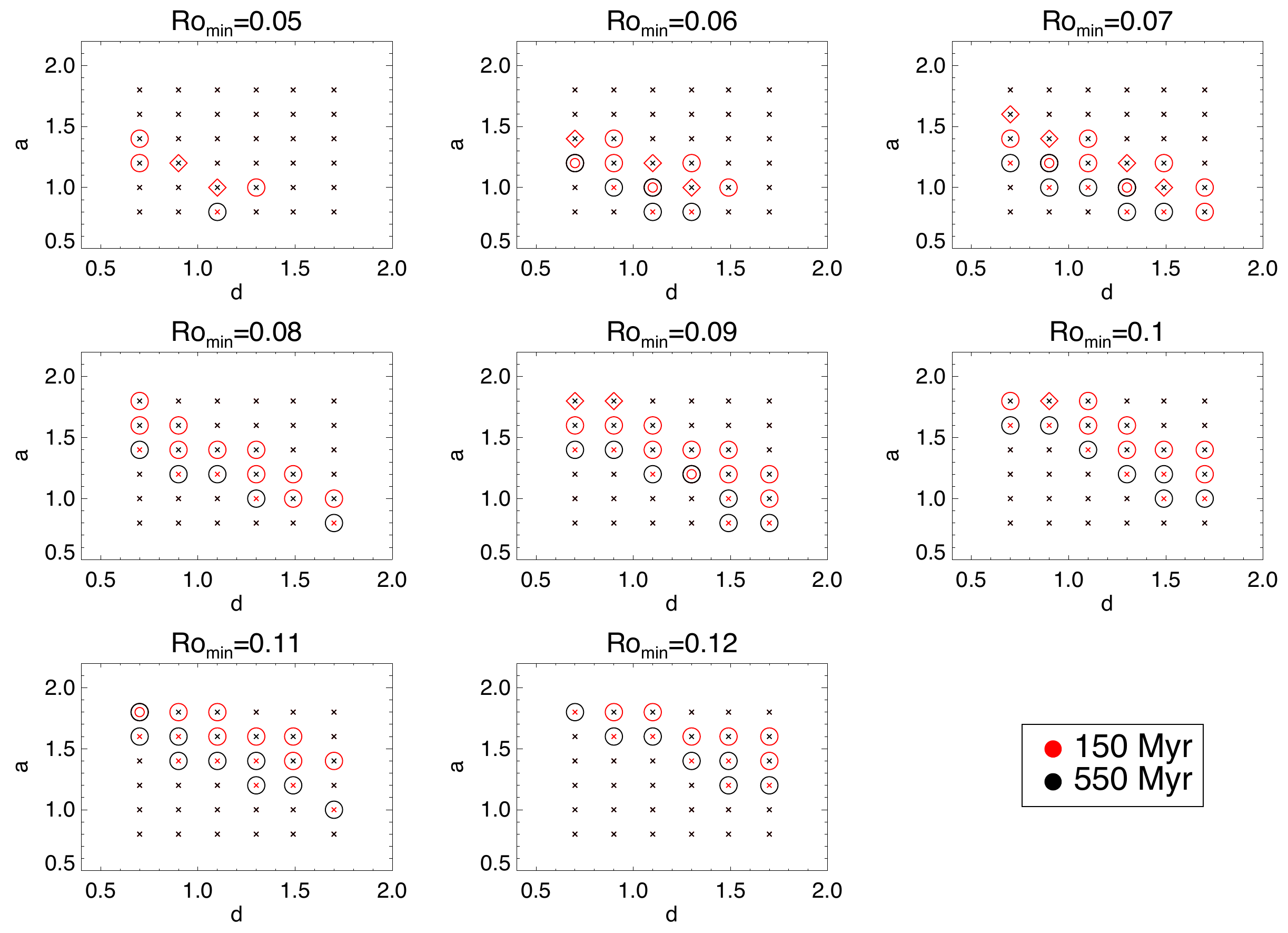}
    \caption{ Exploring the degeneracy between parameters controlling magnetic field and mass loss $a$, $\Rom$ and $d$. We test values of $0.8<a<1.8$, $0.7<d<1.7$ and $0.05<\Rom<0.2$. Since none of the cases with $\Rom>0.12$ create a good fit, we do not include them in this figure. The red and black colours show the results of the tests at 150\,Myr and 550\,Myr respectively. Circles represent fits that were fair while diamonds show a good fit to a cluster and crosses are for tests which were ruled out. The $6$ concentric black and red circles show cases which were successful at both ages.}
    \label{fig:summary}
\end{figure*}

In fig.~\ref{fig:summary} we explore the degeneracy between parameters  $\Rom$, $a$ \& $d$. It should be noted that since none of the cases in the $\Rom>0.12$ limit were successful, we do not include them in fig.~\ref{fig:summary}. In each panel we have tested different values of the $a$ parameter versus $d$ and for different $\Rom$ values. The plots show whether or not a certain set of parameters fits our sample at two different ages; 150\,Myr and 550\,Myr. These two ages have been chosen because they contained the most stars among our clusters and in all cases if a model provided a good fit for these two, it was also successful at the other ages.The red and black colours show the results of the tests at 150\,Myr and 550\,Myr respectively. 

The three symbols represent how well a model could explain the rotation profile of a cluster. Circles represent fits that were fair, i.e. covered most of the stars in a cluster. For example, in fig.~\ref{fig:param_rossby} the $\Ro=0.06$ curve represents a fair fit for the 150\,Myr cluster. Diamonds show a good fit, i.e. The majority of the stars in a cluster are situated between the two fastest and slowest rotating envelopes while the curves are not far away from the stars either. For instance, in fig.~\ref{fig:param_rossby} the $\Ro=0.07$ is considered as a good fit while although the $\Ro=0.12$ curve covers more stars, it is located far away from the envelope and thus not regarded as a good fit. Finally, crosses show tests which were ruled out. 

Fig.~\ref{fig:param_rossby} to \ref{fig:param_d} show that the slope at 550\,Myr is not steep enough to explain both low-mass fast rotators and the lack of massive fast rotators. Hence, none of the models were accepted as good fits at 550\,Myr. The models which created fair fits at both (and thus all) ages are shown with two concentric black and red circles. We find that only in 6 out of 576 cases we can find a successful fit. This also gives an acceptable limit for the range of each parameter.

\begin{figure*}
\begin{subfigure}{.5\textwidth}
  \centering
  \includegraphics[width=1.\linewidth]{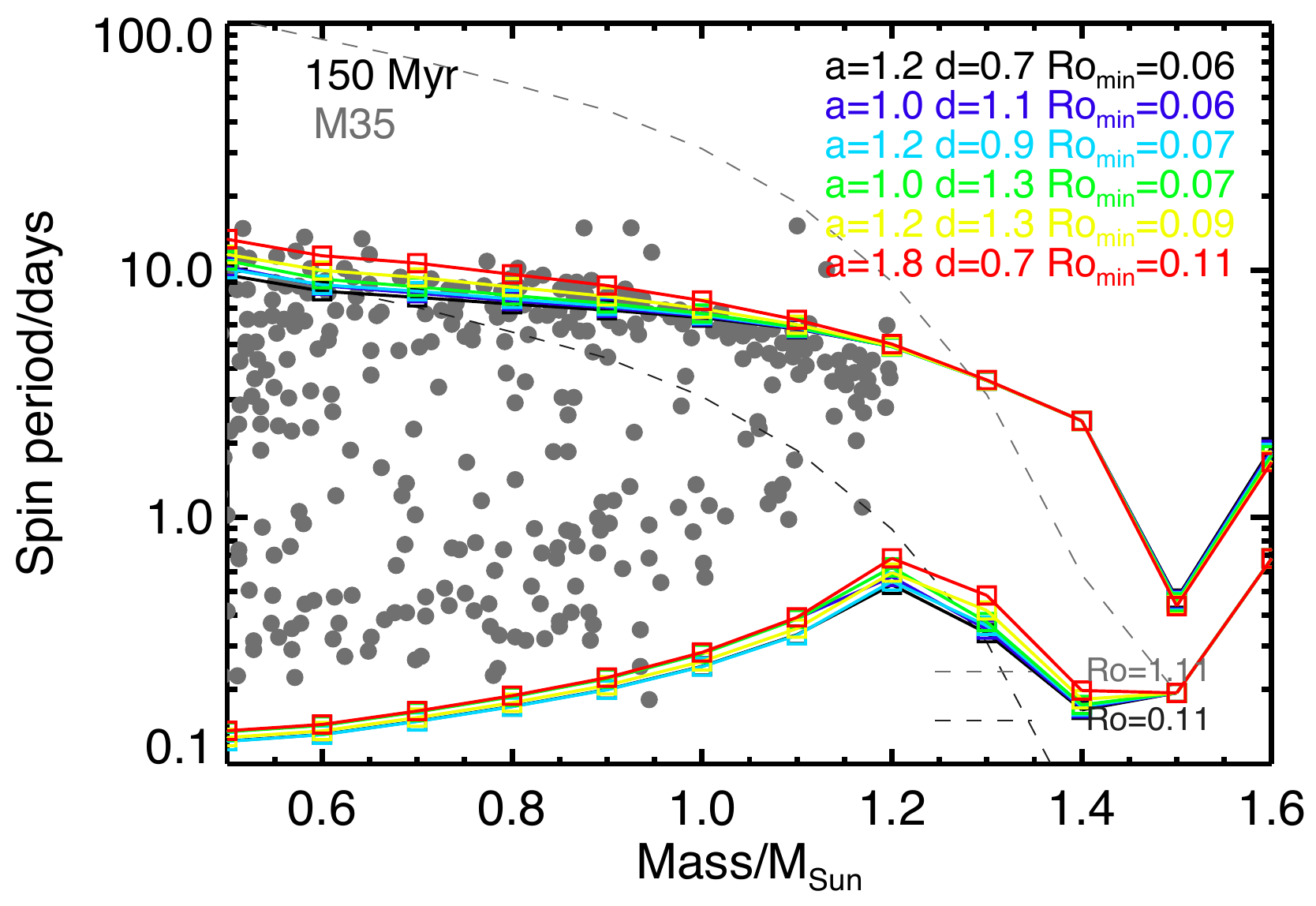}
\end{subfigure}%
\begin{subfigure}{.5\textwidth}
  \centering
  \includegraphics[width=1.\linewidth]{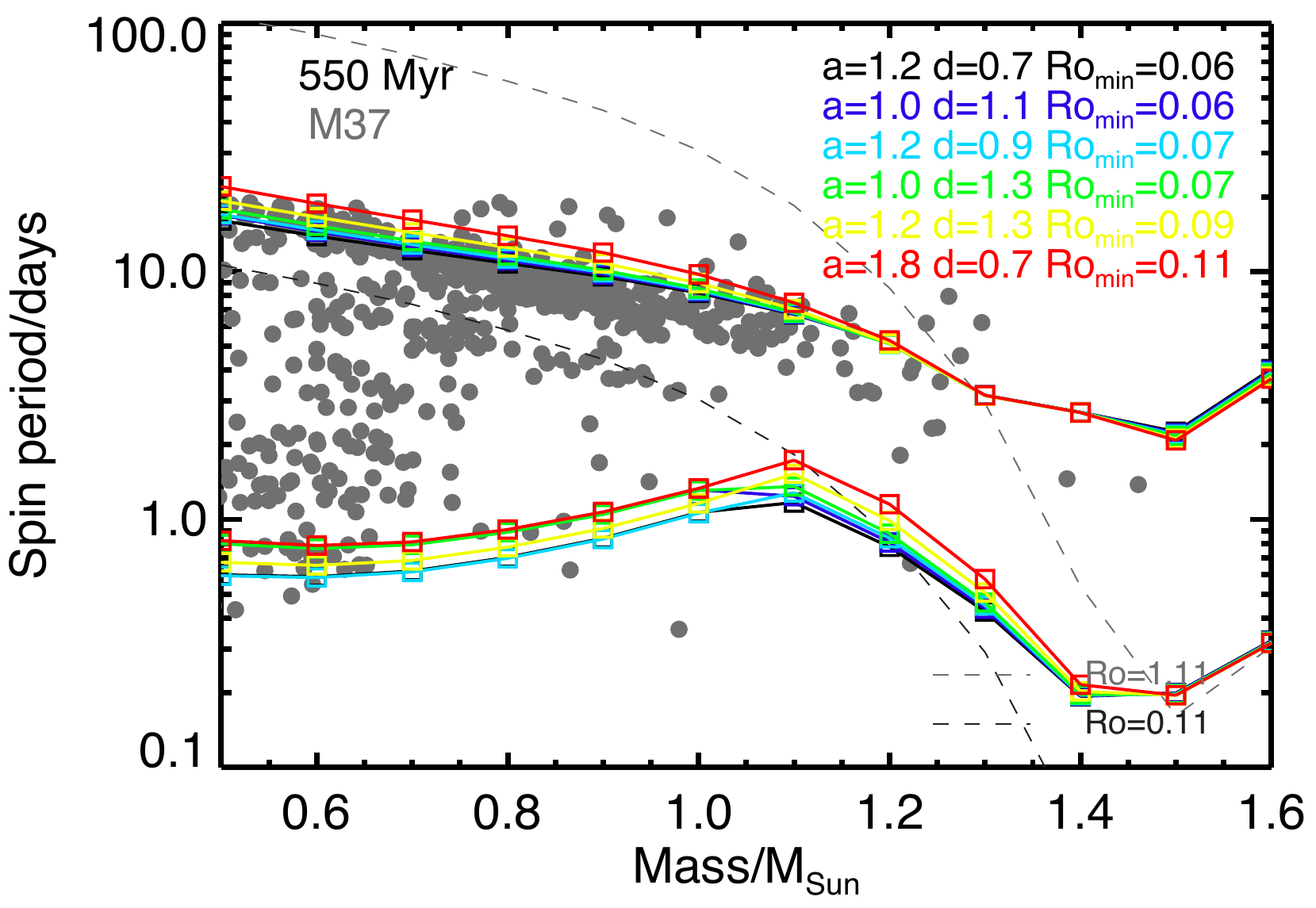}
\end{subfigure}%
\caption{ An overlap of the 6 successful models shown by two concentric circles in figure~\ref{fig:summary} at two different ages; 150\,Myr and 550\,Myr.}
\label{fig:overlap}
\end{figure*}

 Fig.~\ref{fig:overlap} compares these successful models at 150\,Myr and 550\,Myr. The degeneracy between the three parameters $a$, $d$ and $\Rom$ is apparent. We point out that, for the 150\,Myr-old cluster, our envelope for fast rotating stars is slightly below the cloud of points. On the other hand, at 550\,Myr, we cannot explain a few fast rotating stars below 0.6 $\rm M_\odot$ and our braking model seems to be slightly too weak for fast rotating stars above 0.8 $\rm M_\odot$. On the other hand, our model is quite successful at explaining most of the slow rotating stars at both ages (and, as we will see afterwards, for older clusters). Overall, while our solutions appear to be capturing the dominant features of the spin--mass--age relation, we still have problems for fast rotating stars. This indicates that either the observations are incomplete or more probably that magnetic braking is more complex than assumed here. 

Since all the successful models are extremely close to one another, we cannot select one over the others based on these spin--mass--age relations. We therefore choose our preferred model based on the proximity of the parameters to what has been found in the literature. These parameters are $a = 1.2$, $d = 1.3$ \& $\Rom = 0.09$. We adopt them for the rest of the paper. 

\subsection{Decoupling of internal layers: $\tau_{\rm mix}$}
\label{sec:decoupling}

The timescale over which angular momentum is transferred between the inner core and the outer convective zone, the mixing timescale $\tau_{\rm mix}$, is a free parameter that we adjust to fit observations at this stage. Several previous studies have included a decoupling timescale in their models and found it to be of order of a few $10\,\rm Myr$. J15 have focused on ages above $100\, \rm Myr$ and found that they were able to fit their sample without assuming core-envelope decoupling.  We also performed a series of tests to check whether we could indeed describe observations without $\tau_{\rm mix}$. In order to do this, we repeated our tests in section~\ref{B_Mdot_degeneracy}.  However, our results showed that a major defect of such models is that they are generally unsuccessful at reproducing a good fit to both young and old clusters with the same parameters. We could not find a set of parameters that described all ages successfully and thus infer that core-envelope decoupling is a necessary component of our model.

 GB15 on the other hand, assume $\tau_{\rm mix}$ to depend on mass and initial rotation simultaneously. They argue that the reason for adopting a different mixing timescale for the slow and fast rotators, is that angular momentum mixing may depend on the width of the tachocline (between the radiative core and the convective zone) which is influenced by shear. We acknowledge that this is a possibility, but it introduces yet another source of freedom to the model.

\begin{figure*}
\begin{subfigure}{.5\textwidth}
  \centering
  \includegraphics[width=1.\linewidth]{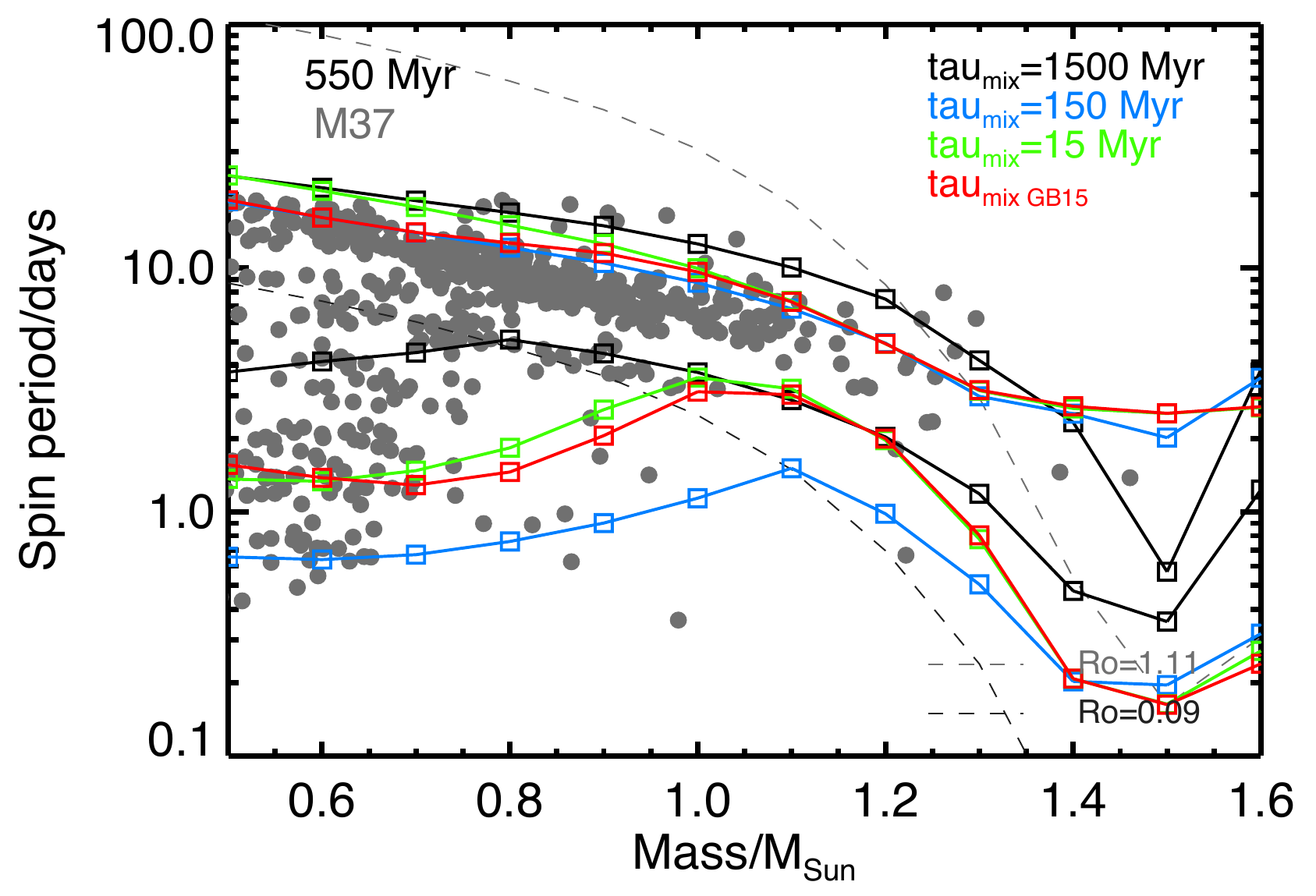}
\end{subfigure}%
\begin{subfigure}{.5\textwidth}
  \centering
  \includegraphics[width=1.\linewidth]{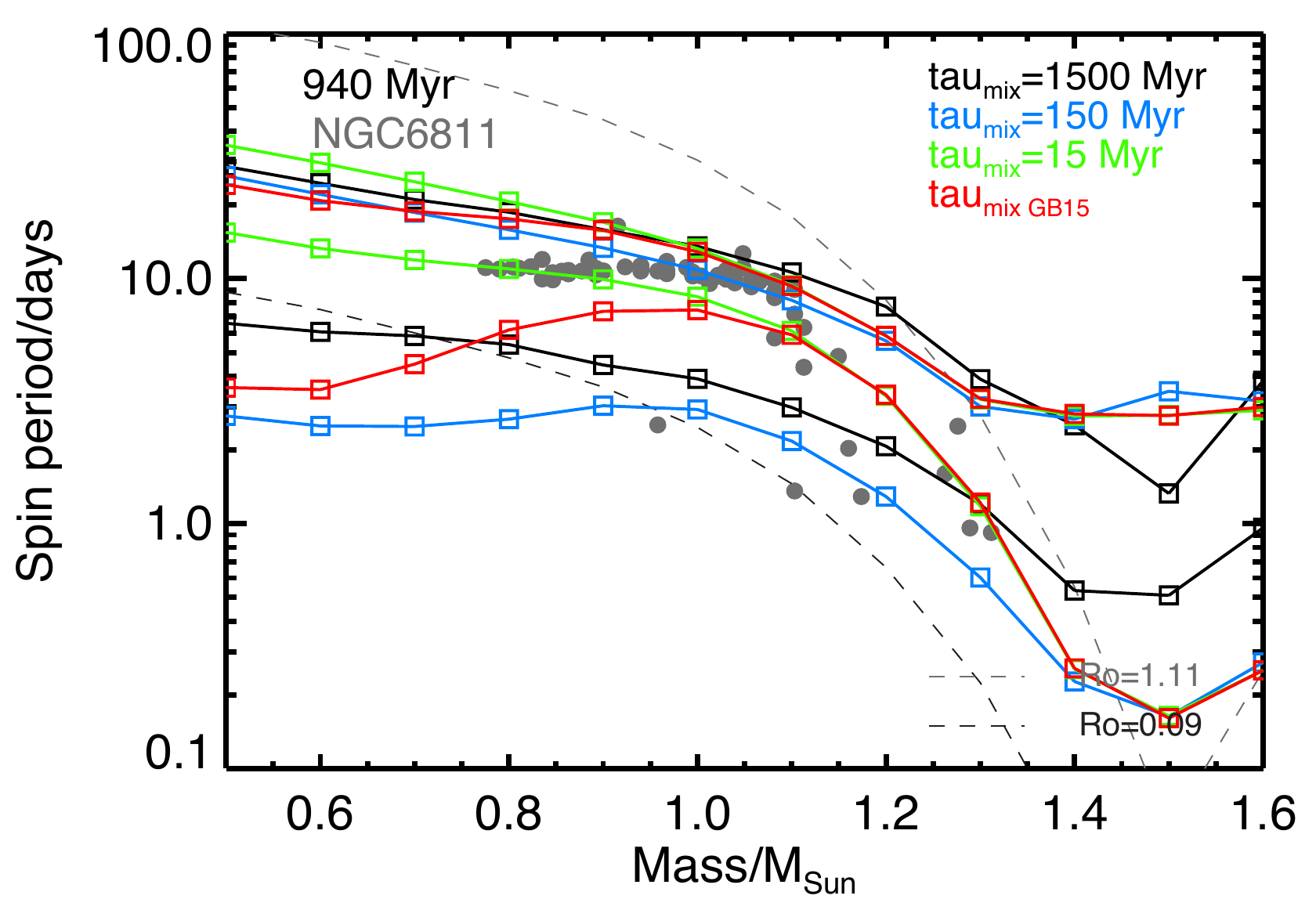}
\end{subfigure}%
\caption{ Test of different  $\tau_{\rm mix}$ in comparison with   $\tau_{\rm mix}$ in GB15 at two different ages 550\,Myr and 940\,Myr}
\label{fig:mixing_orders}
\end{figure*}

Fig.~\ref{fig:mixing_orders} shows the result that we get after adopting GB15's $\tau_{\rm mix}$. It also shows that we were able to fit our sample using a constant $\tau_{\rm mix}=150\, \rm Myr$ for all masses and rotation regimes. We did not attempt to fit all possible combinations of $\tau_{\rm mix}(M)$ but instead tried fixed  $\tau_{\rm mix}$ and found that $\tau_{\rm mix}$150\,Myr improved the fit in the sense that it could explain more stars in the low-mass fast rotators branch at young and intermediate ages. However, we are also getting a larger range of spin periods at old ages. This indicates that the solution is more complex and beyond our model.

Finally, we tested how different orders of magnitude of our adopted $\tau_{\rm mix}$ would fit observations (fig.~\ref{fig:mixing_orders}). An interesting point in the behaviour of the mixing timescale parameter is how either lower or higher values generally predict slower rotation rates. 

\section{Results: Confronting The Model With Observations}
\label{sec:results}

Based on the investigations performed in the previous sections we are now at a stage where we can provide a model with our preferred set of parameters (see Table~\ref{tab:params}) for predicting the rotational evolution of stars vs. their mass.

\subsection{Comparison to Cluster Observations}

\begin{figure*}
\begin{subfigure}{.32\textwidth}
  \centering
  \includegraphics[width=1.\linewidth]{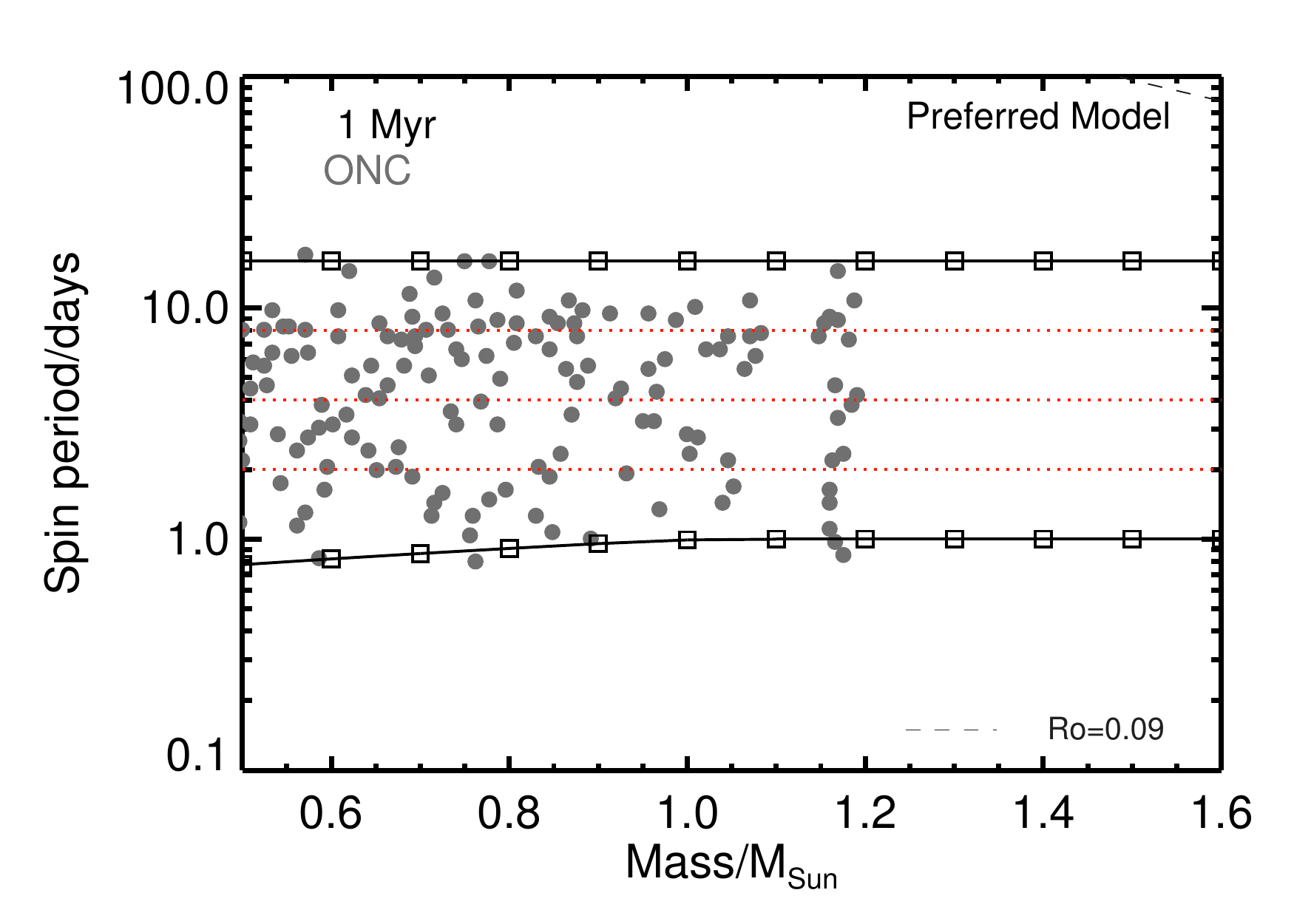}
\end{subfigure}%
\begin{subfigure}{.32\textwidth}
  \centering
  \includegraphics[width=1.\linewidth]{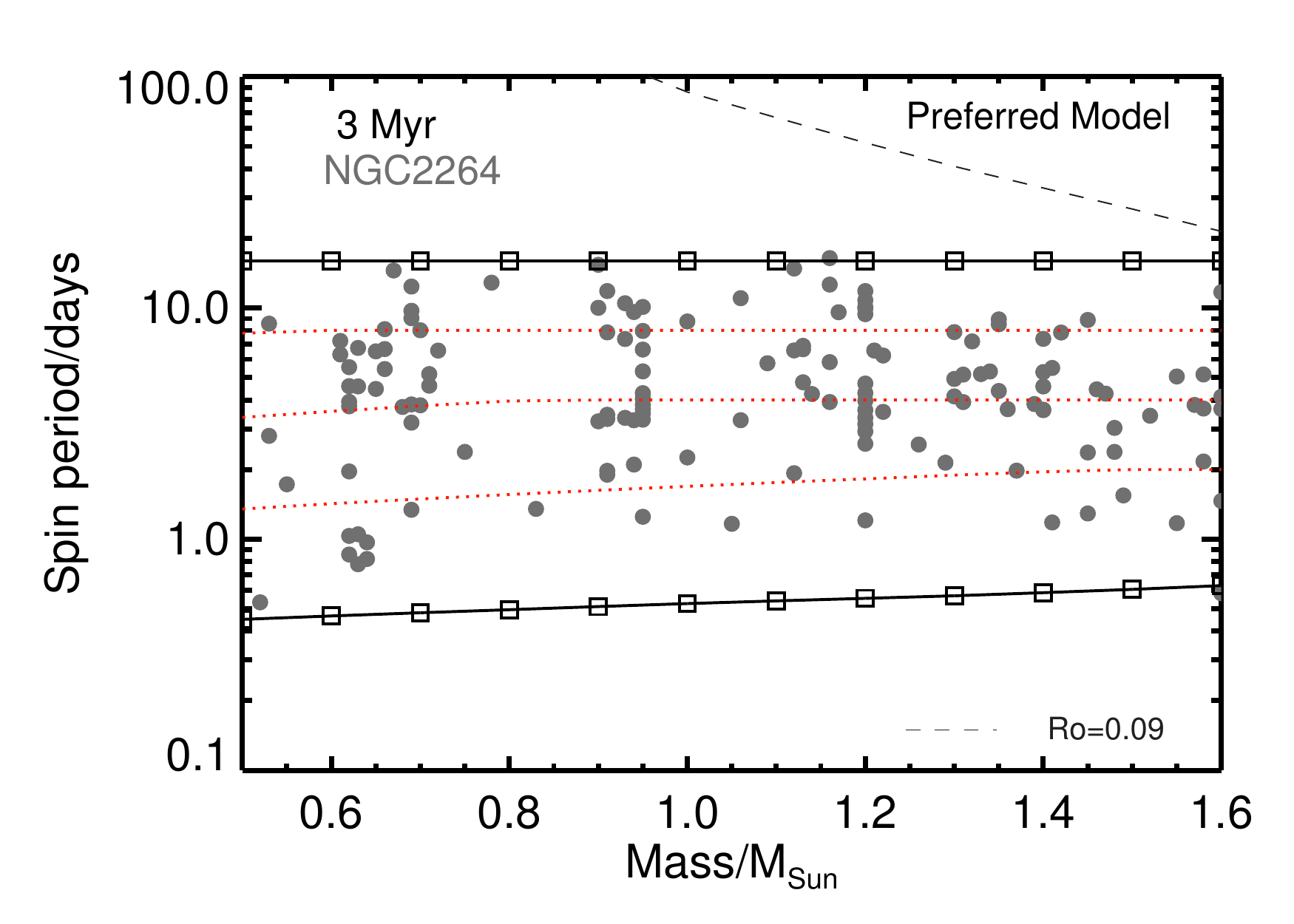}
\end{subfigure}%
\begin{subfigure}{.32\textwidth}
  \centering
  \includegraphics[width=1.\linewidth]{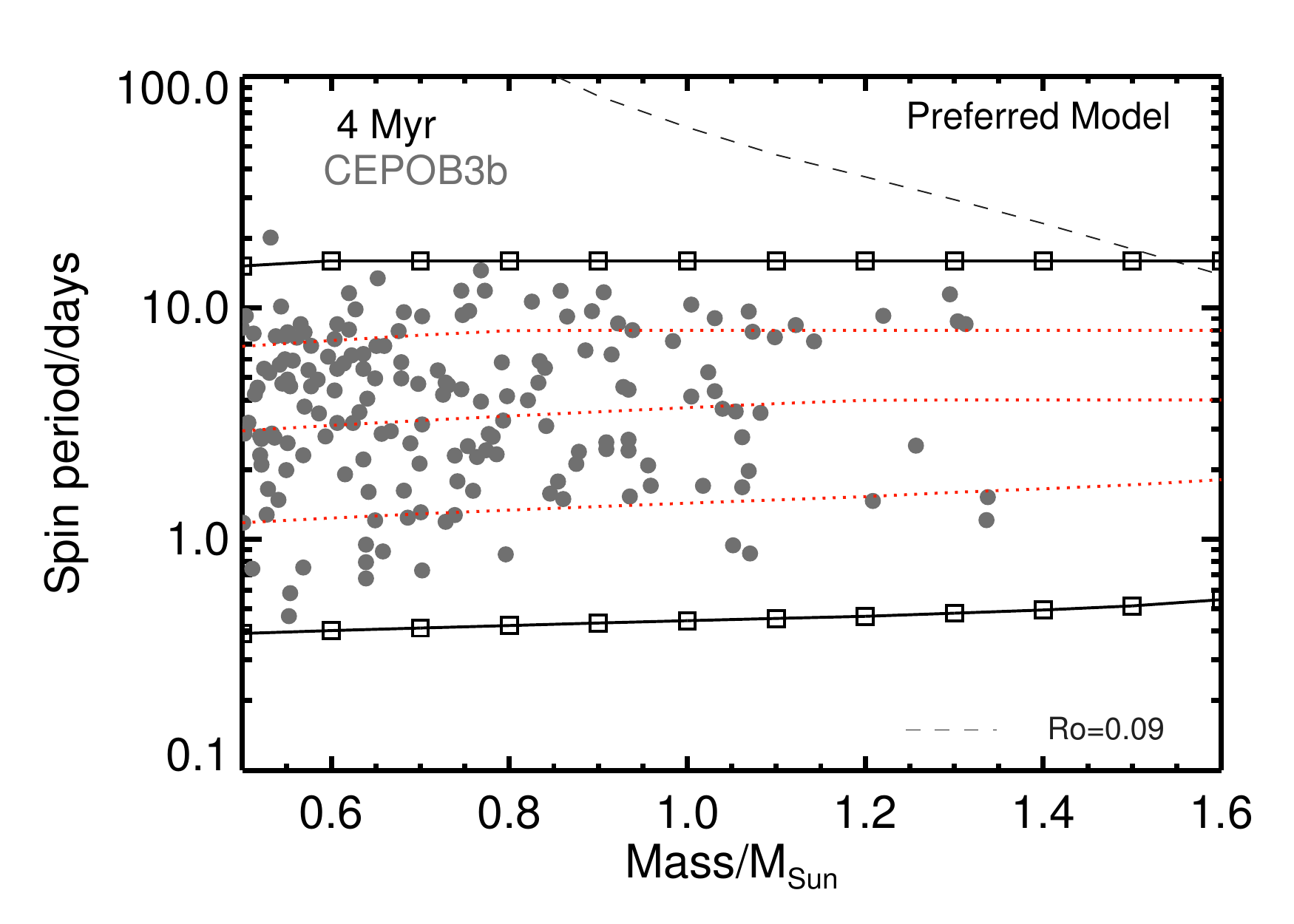}
\end{subfigure}
\begin{subfigure}{.32\textwidth}
  \centering
  \includegraphics[width=1.\linewidth]{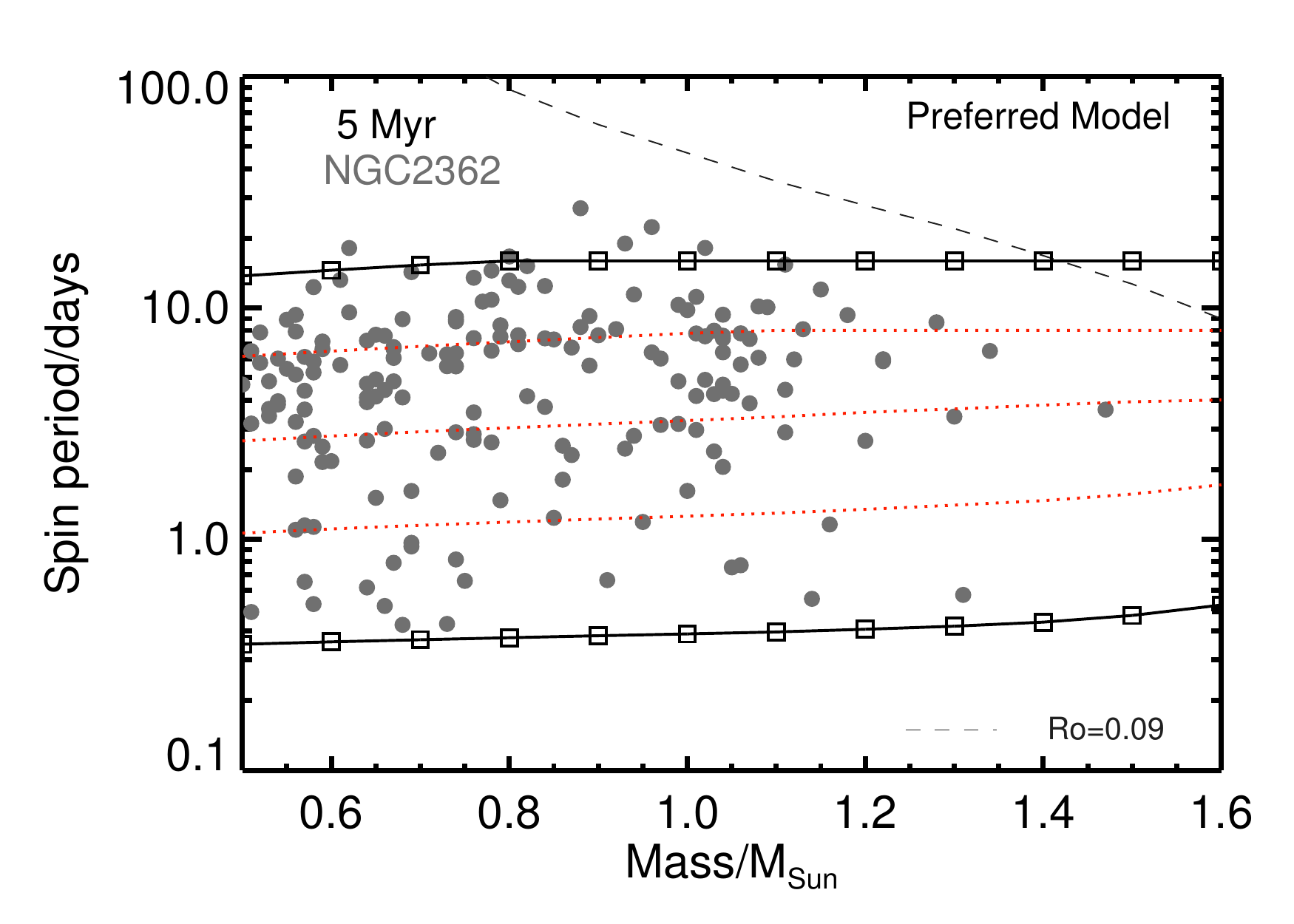}
\end{subfigure}
\begin{subfigure}{.32\textwidth}
  \centering
  \includegraphics[width=1.\linewidth]{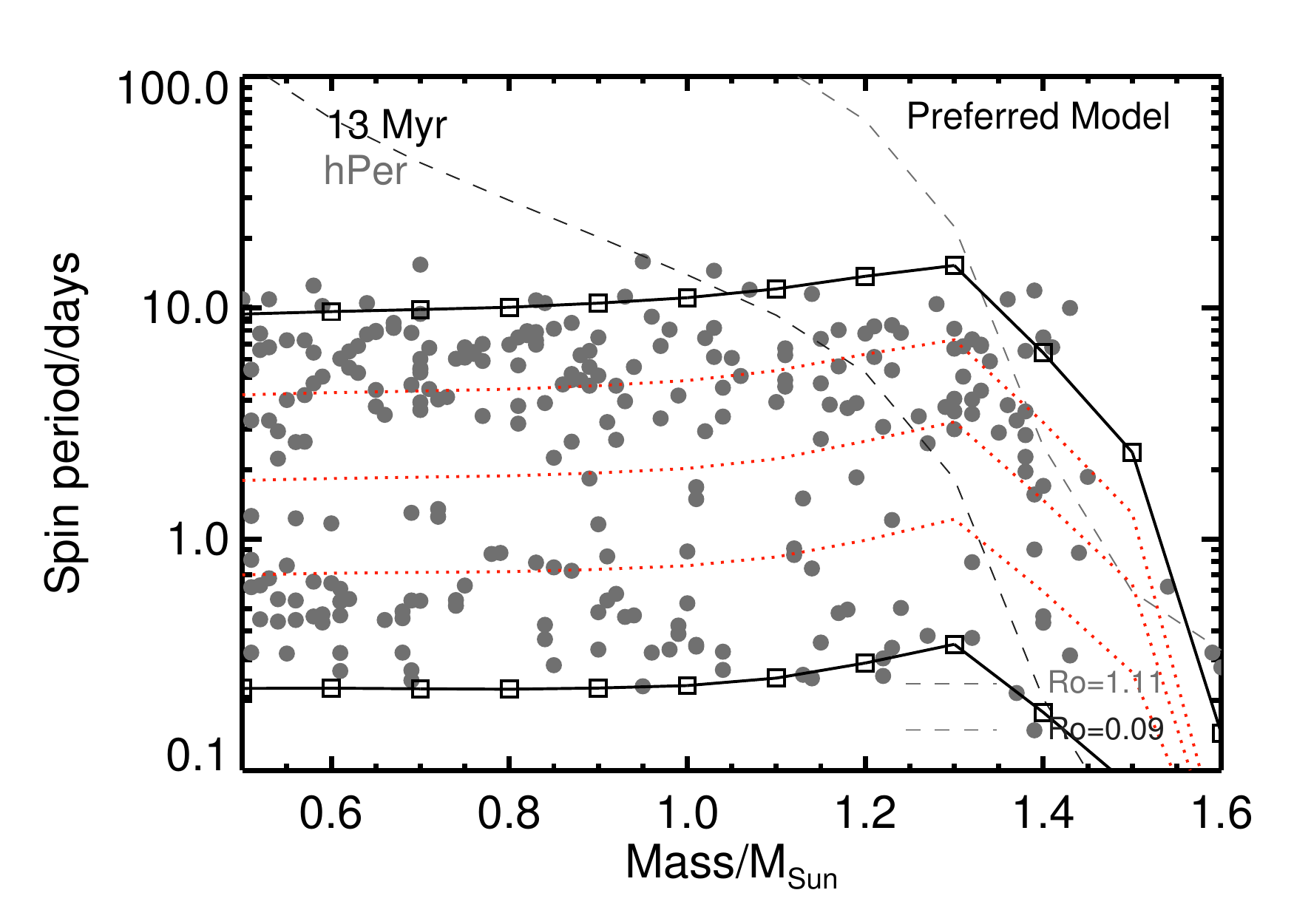}
\end{subfigure}
\begin{subfigure}{.32\textwidth}
  \centering
  \includegraphics[width=1.\linewidth]{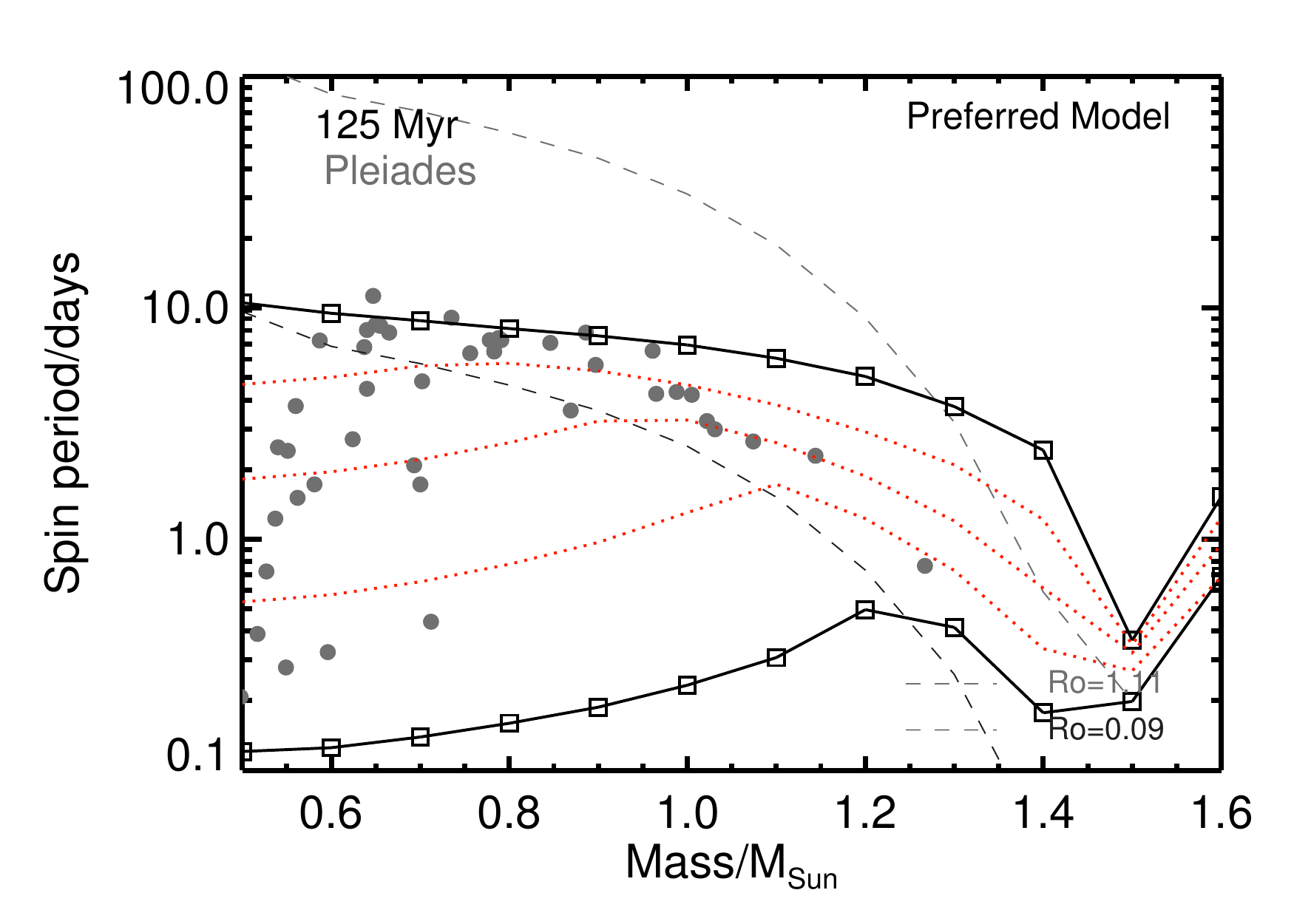}
\end{subfigure}
\begin{subfigure}{.32\textwidth}
  \centering
  \includegraphics[width=1.\linewidth]{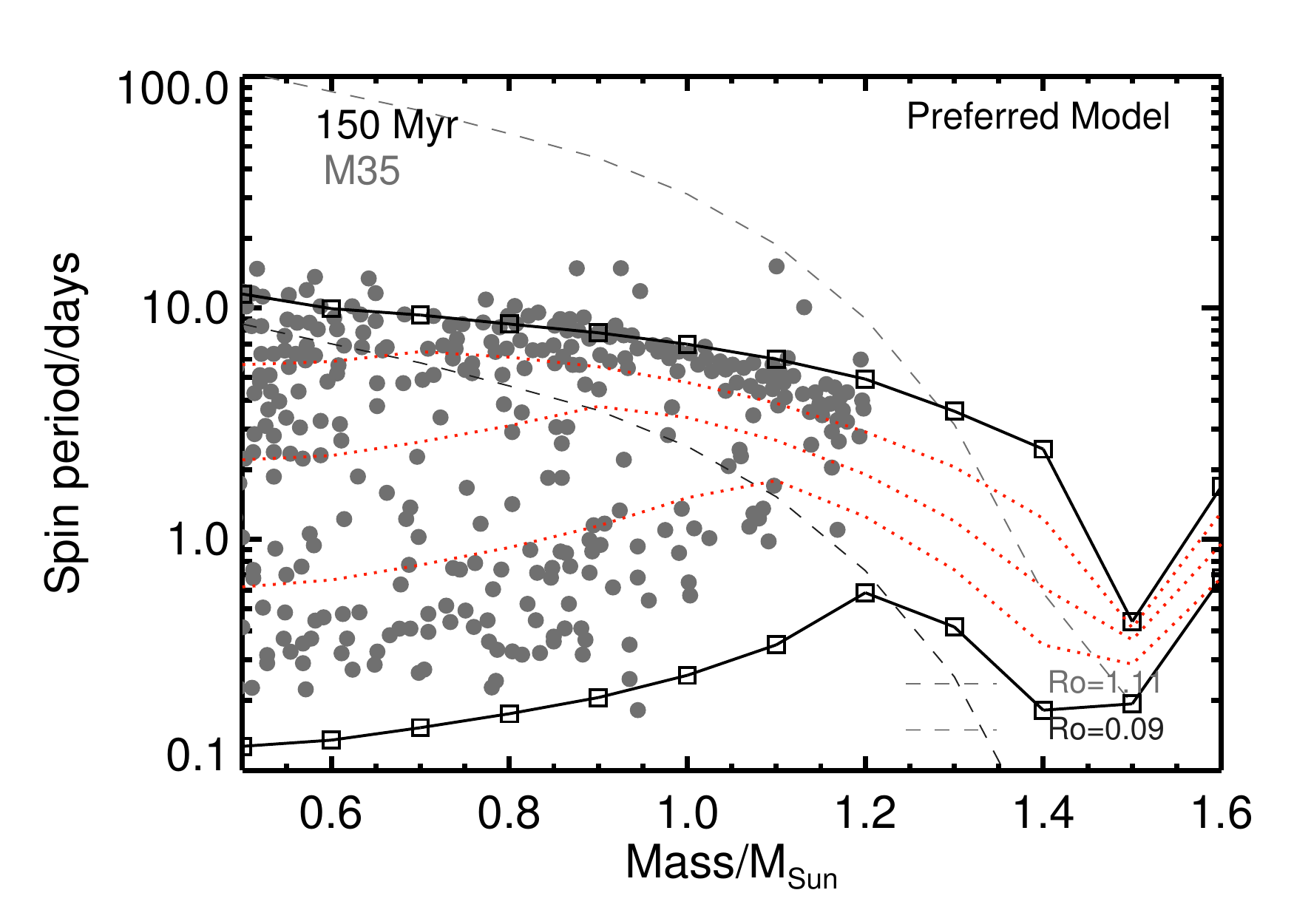}
\end{subfigure}
\begin{subfigure}{.32\textwidth}
  \centering
  \includegraphics[width=1.\linewidth]{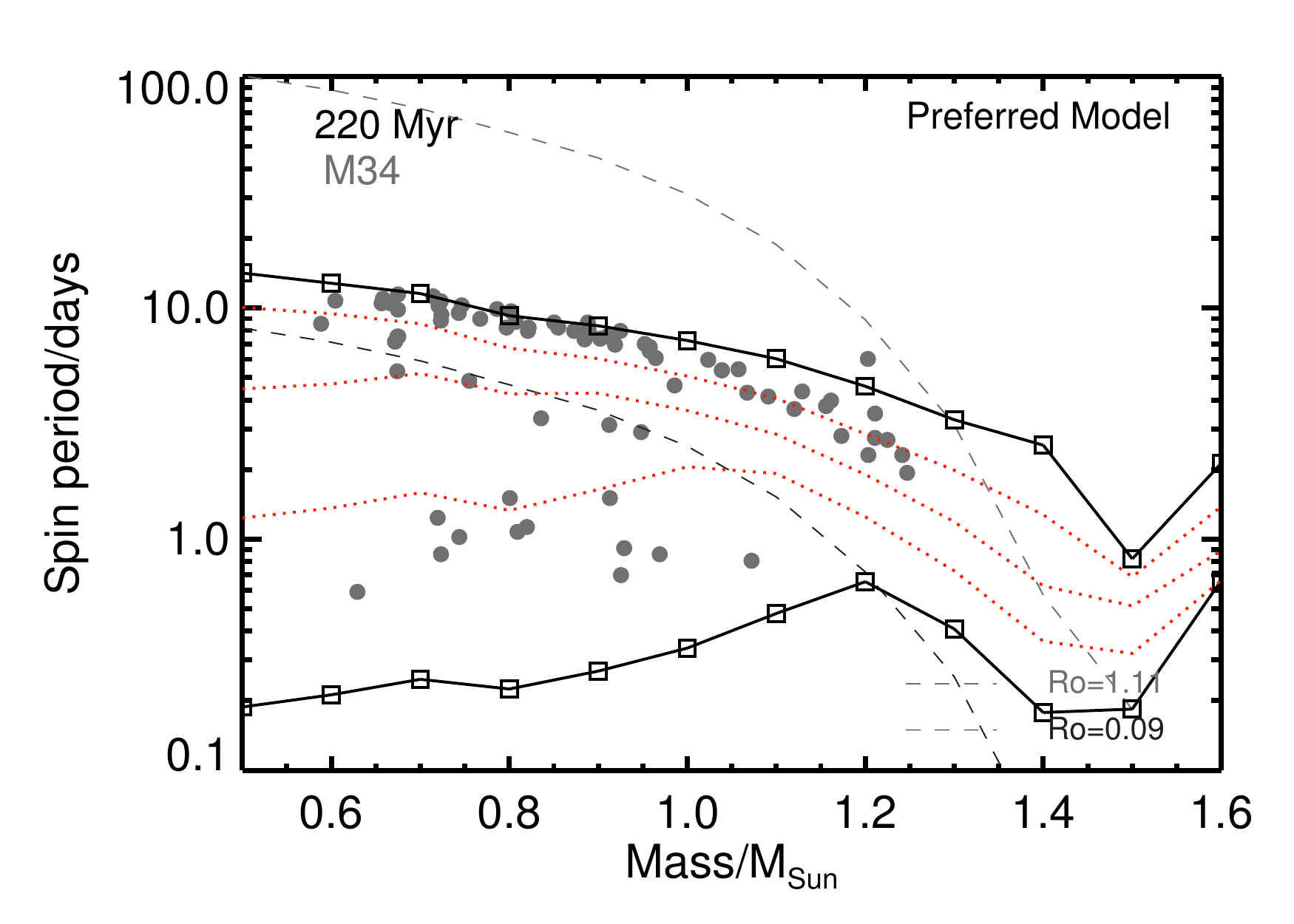}
\end{subfigure}
\begin{subfigure}{.32\textwidth}
  \centering
  \includegraphics[width=1.\linewidth]{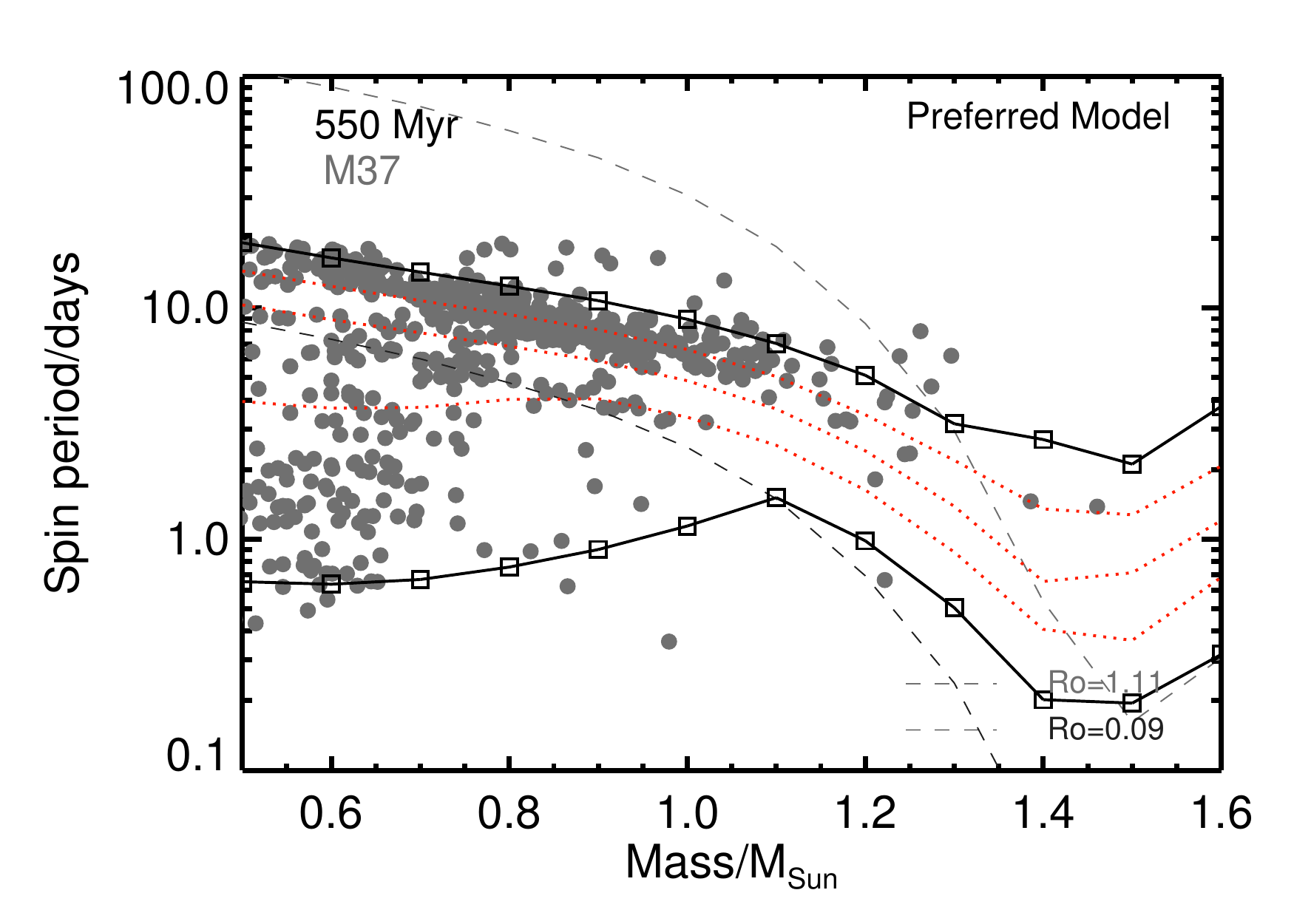}
\end{subfigure}
\begin{subfigure}{.32\textwidth}
  \includegraphics[width=1.\linewidth]{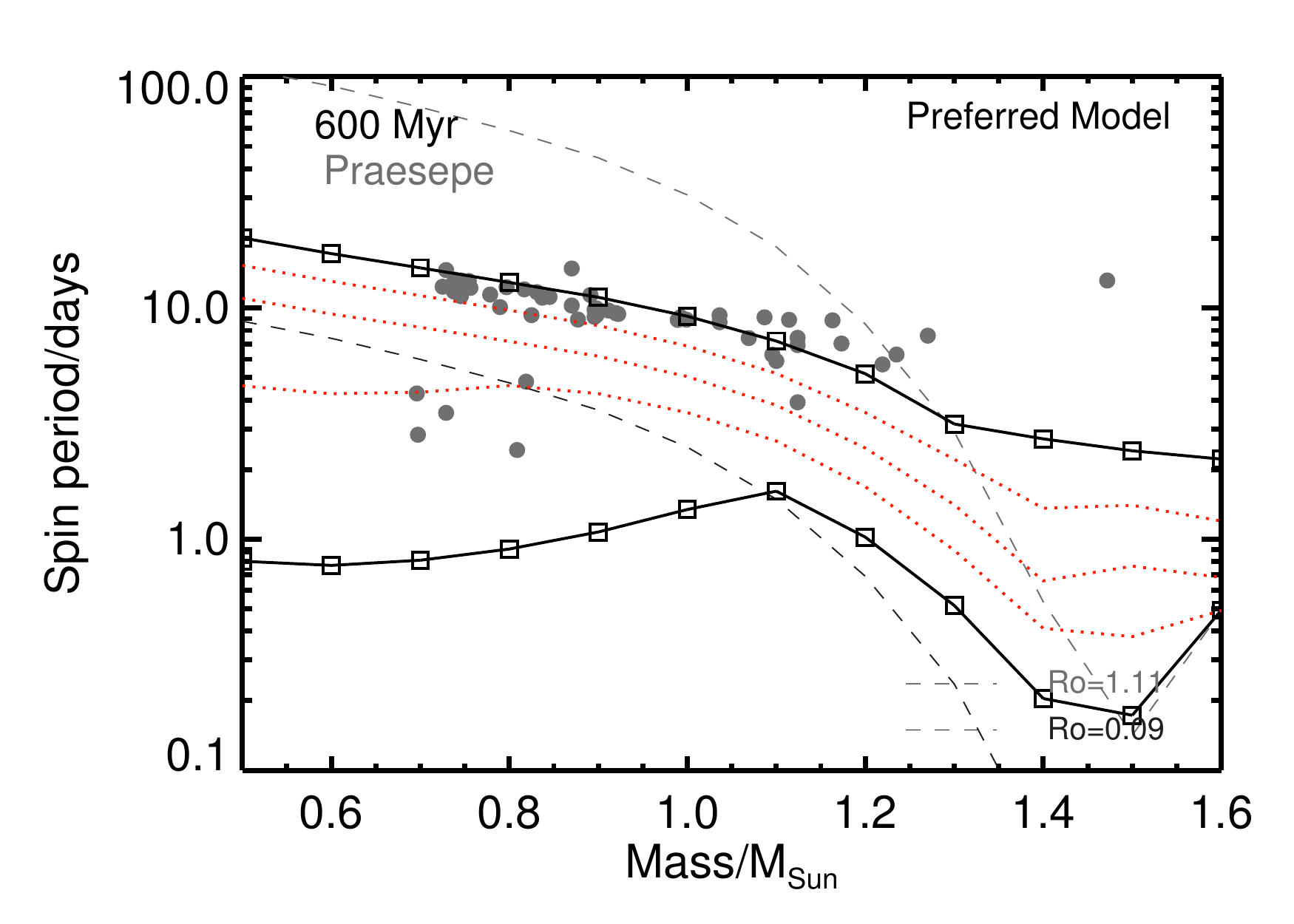}
\end{subfigure}
\begin{subfigure}{.32\textwidth}
  \centering
  \includegraphics[width=1.\linewidth]{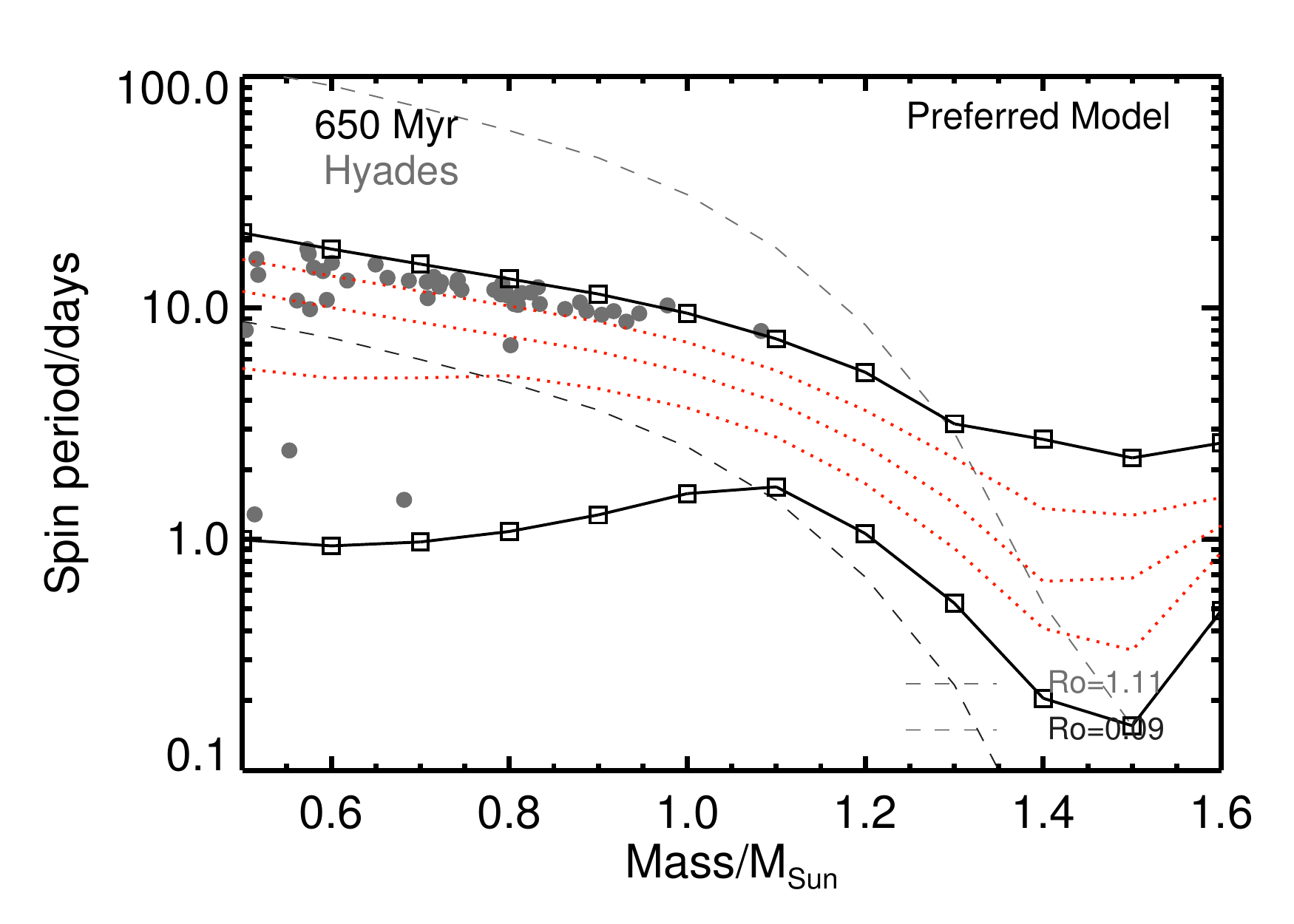}
\end{subfigure}
\begin{subfigure}{.32\textwidth}
  \centering
  \includegraphics[width=1.\linewidth]{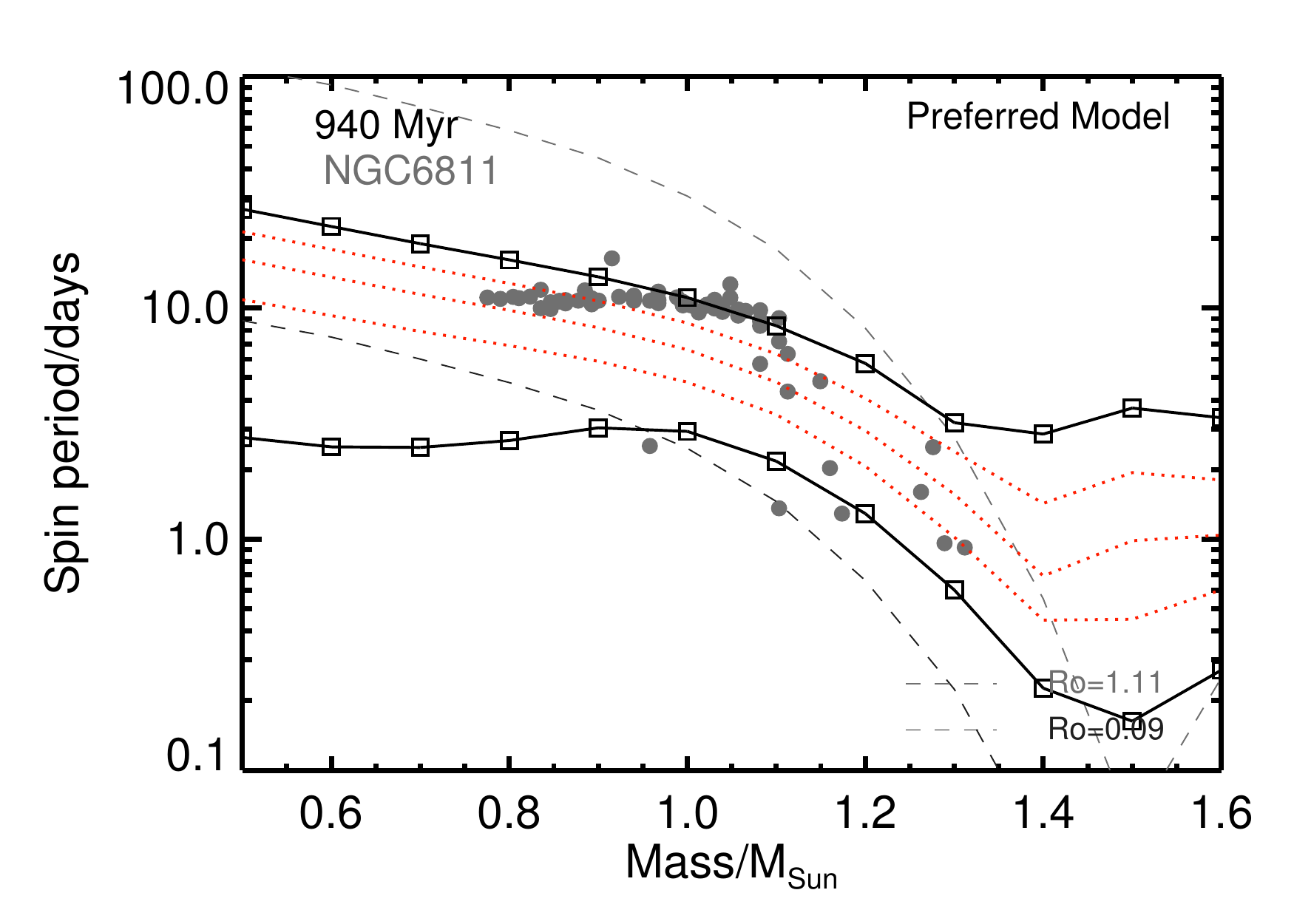}
\end{subfigure}
\begin{subfigure}{.32\textwidth}
  \centering
  \includegraphics[width=1.\linewidth]{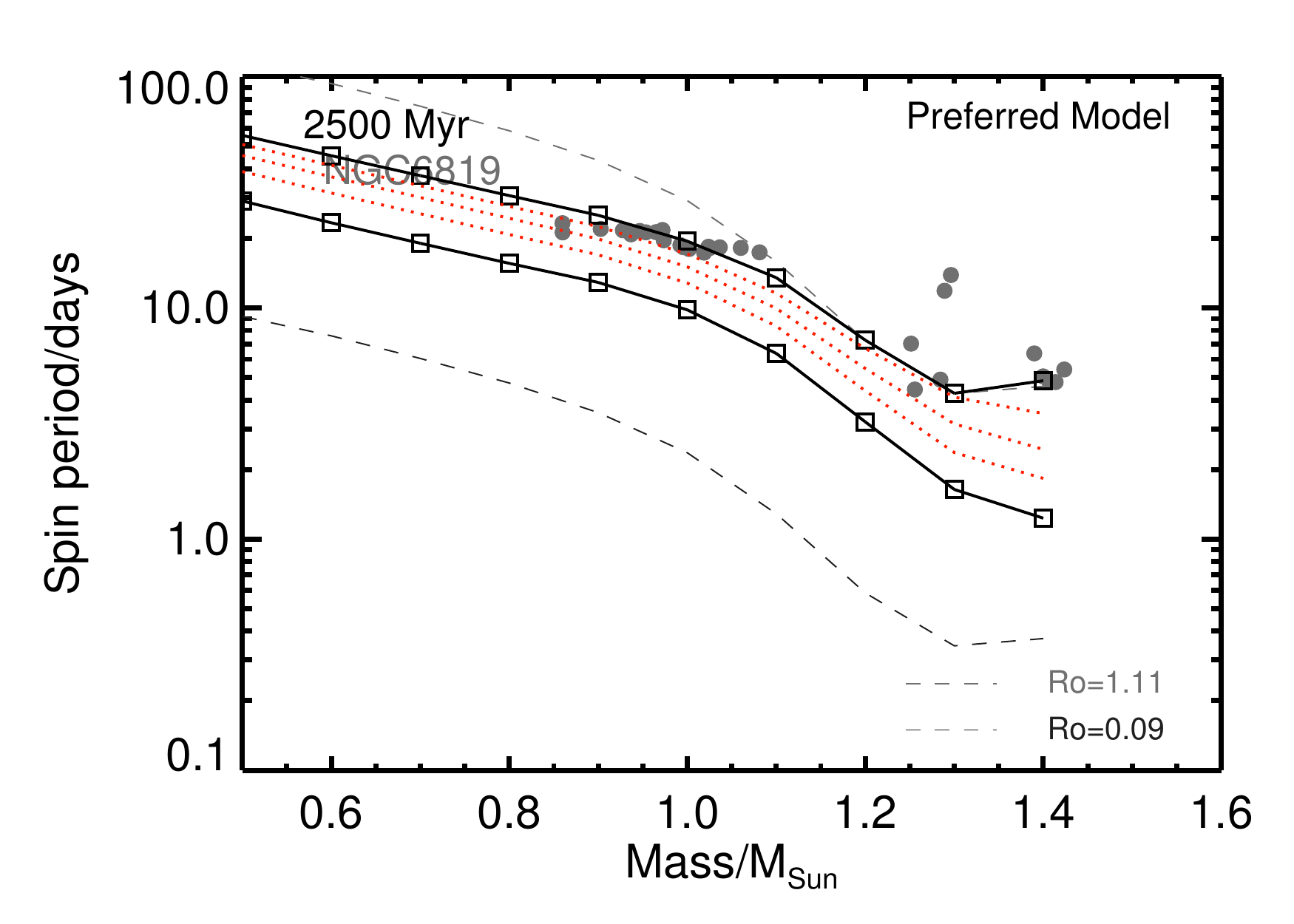}
\end{subfigure}
\begin{subfigure}{.31\textwidth}
  \centering
  \includegraphics[width=1.\linewidth]{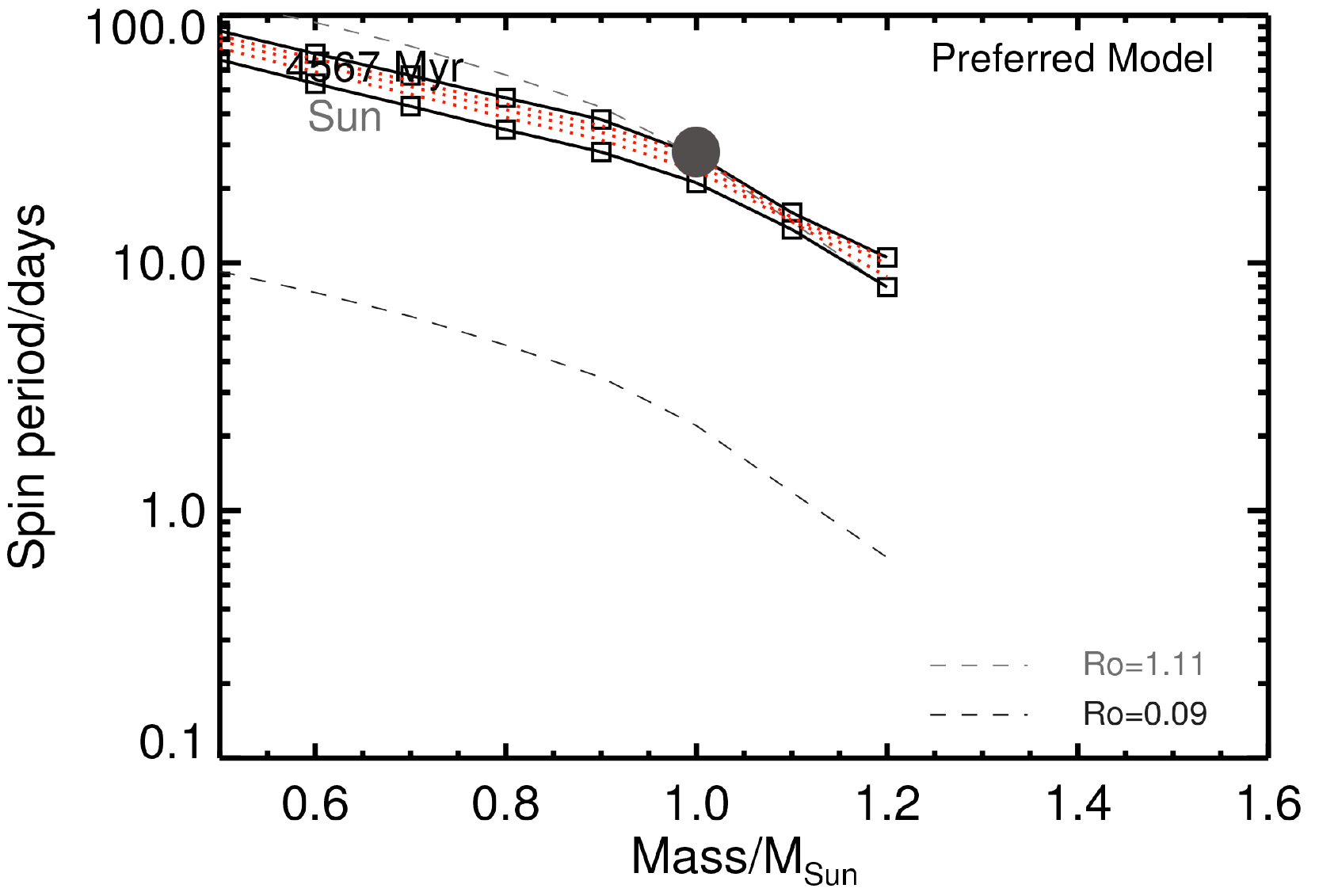}
\end{subfigure}
\begin{subfigure}{.3\textwidth}
  \centering
  \includegraphics[width=1.\linewidth]{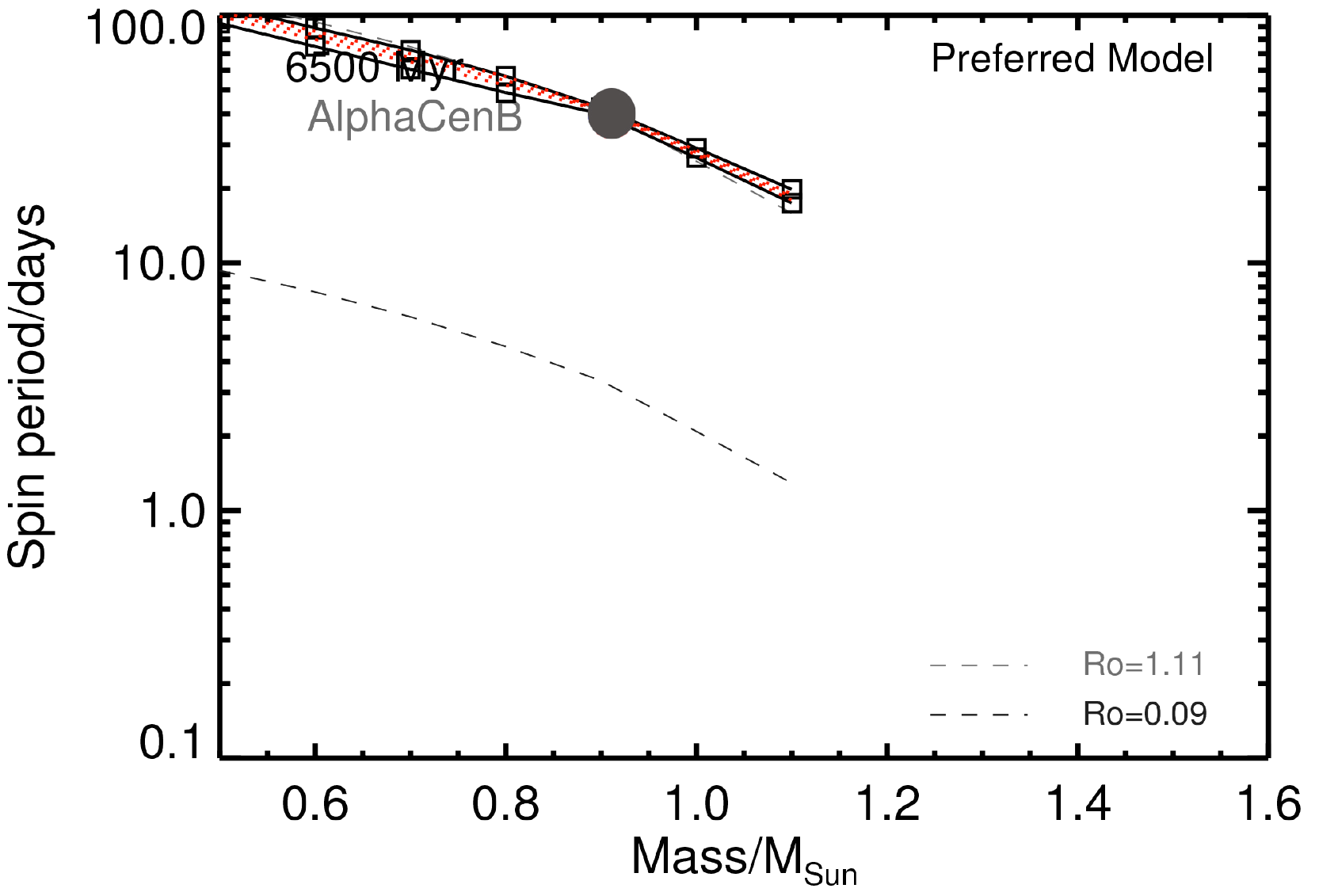}
\end{subfigure}
\caption{Evolution of spin period vs. stellar mass predicted by our preferred model on top of observations of individual clusters at different ages. The two lines represent the slow and fast rotators trend with initial spin periods of 16.0 and 1.0\,d respectively. The set of parameters used for this model are listed in Table~\ref{tab:params}. The two dashed lines show the critical Rossby numbers first for weakening the  magnetic braking beyond $\Rom=0.09$ and then practically stopping breaking at $\RoM=\Ro_\odot$ according to \protect\cite{vanSaders2016} prescription. The red dotted lines correspond to models with intermediate initial periods compared to those of the fast and slow rotator branches. From top down, they have initial spin periods of 8.0, 4.0 and 2.0\,d respectively.  The position of the Sun and AlphaCenB have been marked with a larger circle.} 
\label{fig:preferred model}
\end{figure*}

Fig.~\ref{fig:preferred model} summarises the spin period evolution of fast and slow rotators as a function of mass and at different ages. The two solid black lines represent the slow (upper curve) and fast (lower curve) rotator population envelopes. The dotted red lines represent models with intermediate initial spin periods of $8.0$, $4.0$ \& $2.0$ days. Our models provide a good overall fit to the observational constraints.


As previously discussed in \ref{sec:disklock_param}, the spin rates of the outer convective zones of the stars are assumed to be held constant during the first $1\,\rm Myr$ for solar-mass fast rotators and $6\,\rm Myr$ for solar-mass slow rotators. We have also assumed a linear variation of $\tau_{\rm disk}$ between these two  for the intermediate models in fig. \ref{fig:preferred model}.

After the disk disappears, stars spin up due to their PMS contraction until they approach the main-sequence and magnetic braking begins to be important. As a result, rotation periods which are uniformly distributed between about 1 and 16 days in the ONC at 1\,Myr start decreasing at 5\,Myr in NGC2362, in particular for fast rotators which supposedly lost their disk early and began their spin-up.  At 13\,Myr the high-mass stars have spun up significantly due to pre-main-sequence contraction. There are a few outliers with $a \approx 10$ days spin period and a mass of $1.4 \rm M_\odot$, but these could easily be explained by a $0.1 \rm M_\odot$ or so error on the mass determination. After entering the main-sequence, at the age of about 30\,Myr, stars spin down through magnetic braking. Fig.~\ref{fig:preferred model} shows that after about 100\,Myr slow rotators appear to converge towards a relatively well-defined sequence whereas fast rotators remain scattered relatively significantly. The envelopes for the fast and slow rotators appear to globally capture the evolution of these populations. We see that for the M37 cluster, we slightly underestimate the spread in rotation rates. The difference, however, is small. \cite{Johnstone2015} also report a similar issue in their predictions for this cluster.

 We acknowledge that between 600 \,Myr and 940\,Myr, our model seems to predict a slightly too fast envelope for fast rotators. At older ages, up to 2.5\,Gyr, the slow rotators branch of our model reproduces the observations well. However, there are very few fast rotators. This raises the question of whether our envelope for fast rotators is correct. In fact, as shown in fig.~\ref{fig:model_comparison}, the B97 and GB15 parameterisations tend to have a much faster convergence so that, after about 500\,Myr, the initial conditions have been lost. It is not clear, however, if the observations require that. The clusters which have been observed at these relatively old ages contain somewhat fewer stars and it seems plausible that the absence of these fast rotating stars is due to the fact that they are rarer than the slow-rotating ones. In addition, we can see that the Hyades does possess a number of low-mass, fast rotating stars which are not explained by the B97 and G15 parameterisations. We hence believe that our parameterisation provides a plausible conservative estimate for rapidly rotating stars. 

Furthermore, our model also captures the essence of the Kraft break the fact that F-type stars, above about $1.2\,\rm \rm M_\odot$, rotate much more rapidly than G-type stars. This was not the case of the B97, G15 and J15 parameterisations (see fig.~\ref{fig:model_comparison}). The agreement between model and observations is not as good for these massive stars as for the lower-mass stars however. 

Finally, we point out that the increase in spin periods seen in the NGC6819 cluster is due to stars leaving the main-sequence. Because they expand, conservation of angular momentum requires that their spin rate decreases. 

\subsection{Bimodality in Rotation Rates}

It has long been known that in most of the intermediate-age clusters such as M35 \& M37, many of the stars lie on one of two principal sequences, identified as C and I by \cite{Barnes2003}. These sequences are not necessarily clearly defined with a significant number of stars found in between them. 

The I-sequence consists of relatively slow rotators. With increasing age, the I-sequence becomes more tightly defined, and moves to longer periods. Stars on the C-sequence are rapid rotators. In young clusters ,$t\leqslant 150\,\rm Myr$, the C-sequence has members that span the mass range of solar-like stars. In older clusters the C-sequence fades away, starting with the higher-mass stars. By the age of the Hyades ($600\,\rm Myr$) it seems as if the only remaining fast rotators are low-mass stars. However, as discussed previously, this may be an artefact from the relatively limited number of stars associated with these old clusters. 

Our model reproduces this feature qualitatively. This is because slow rotators with $\Ro>\Rom$ have an angular momentum decay rate $dJ/dt\propto\Omega^{(1-2m)d+4ma+1}=\Omega^{2.60}$ whereas for fast rotators with $\Ro\le\Rom$, $dJ/dt\propto\Omega$ (assuming $K_2v_{\rm esc}>\Omega R$). Extremely fast rotators, characterised by $K_2v_{\rm esc}< \Omega R$, are found on the PMS and for spin periods shorter than about a day. For these, the magnetic braking is even weaker, $dJ/dt\propto\Omega^{1-2m}=\Omega^{0.56}$.  

Thus, the stronger dependence of magnetic braking on $\Omega$ for slow rotators leads to a faster convergence towards a relatively well-defined relation. On the other hand, the weaker dependence on $\Omega$ for fast rotators with already maximal field strength implies that the scatter in spin rates is conserved for longer.  We can see, however, by focusing on the intermediate initial spin rates (red curves) in fig.~\ref{fig:preferred model} that although our model captures the broad behaviour of these populations, it fails to reproduce the fact that the slow rotators appear to rapidly outnumber the fast rotators. This probably requires a different physics from that assumed here \citep[see e.g.,][]{Brown2014}.

\subsection{Application to individual field stars}

Our model parameters have been calculated based on cluster observations and a calibration to the Sun. We now turn to the comparison of our model predictions to those observed on individual field stars, listed in Table~\ref{tab:objects}. Most of those were obtained by an analysis of Kepler observations combining photometrically determined spin periods to precise determination of age, mass and metallicity from astroseismic modelling \citep{vanSaders2016}. The ensemble spans a range of age between about 1 and more than 10\,Gyr, masses between $0.8$ and $1.3\,\Msun$ and [Fe/H] between $-0.25$ and $+0.4$. It is thus well suited for our purposes. 

\begin{table*}
  \centering
  \caption{Statistical properties of the distributions of rotation rates in individual field stars used in this study. The observed stellar spin period $\overline{P}_{\star}$ is compared to that calculated assuming $\RoM=\infty$ or assuming $\RoM=\Ro_\odot$ (our preferred model).}              
  \label{tab:objects} 
  \begin{threeparttable}
    \centering
\begin{tabular}{l c c c c c c r}
\hline
Star & age [Ma] & $M_{\star}\ \rm [M_\odot]$ & $\rm [Fe/H]$ & \multicolumn{3}{c}{$\overline{P}_{\star}$\,[days]} & Ref. \\
\cline{5-7}
  &  &  &  & observed & $\RoM=\infty$ & $\RoM=\Ro_\odot$ & \\
\hline
 KIC10644253 & $ 1070 \pm  250$ & $  1.13 \pm   0.05$ & $ 0.12 \pm 0.09$ & $10.91 \pm 0.87$ & $   7.1_{-   3.0}^{+   3.3}$\ \cmark & $   7.1_{-   3.0}^{+   3.3}$\ \cmark &  1\\[1ex]
  KIC9139151 & $ 1710 \pm  190$ & $  1.14 \pm   0.03$ & $ 0.11 \pm 0.09$ & $11.00 \pm 2.20$ & $   9.3_{-   3.2}^{+   3.6}$\ \cmark & $   9.3_{-   3.2}^{+   3.6}$\ \cmark &  1\\[1ex]
 KIC10454113 & $ 2030 \pm  290$ & $  1.19 \pm   0.04$ & $-0.06 \pm 0.09$ & $14.60 \pm 1.10$ & $   5.2_{-   2.5}^{+   3.1}$\ \xmark & $   5.2_{-   3.1}^{+   2.4}$\ \xmark &  1\\[1ex]
  KIC3427720 & $ 2230 \pm  170$ & $  1.13 \pm   0.04$ & $-0.03 \pm 0.09$ & $13.90 \pm 2.10$ & $  10.1_{-   3.6}^{+   3.9}$\ \cmark & $  10.1_{-   3.6}^{+   3.9}$\ \cmark &  1\\[1ex]
  KIC5184732 & $ 4170 \pm  400$ & $  1.27 \pm   0.04$ & $ 0.38 \pm 0.09$ & $19.80 \pm 2.40$ & $  22.0_{-   3.7}^{+   4.2}$\ \cmark & $  20.9_{-   3.4}^{+   4.1}$\ \cmark &  1\\[1ex]
 KIC10963065 & $ 4360 \pm  460$ & $  1.09 \pm   0.02$ & $-0.20 \pm 0.10$ & $12.40 \pm 1.20$ & $  13.5_{-   4.4}^{+   4.4}$\ \cmark & $   9.3_{-   3.4}^{+   4.1}$\ \cmark &  1\\[1ex]
   AlphaCenA & $ 4400 \pm  300$ & $  1.10 \pm   0.02$ & $ 0.20 \pm 0.02$ & $22.00 \pm 3.00$ & $  26.0_{-   2.9}^{+   2.8}$\ \cmark & $  25.8_{-   3.2}^{+   1.5}$\ \cmark &  3\\[1ex]
         Sun & $ 4567 \pm    1$ & $  1.00 \pm   0.00$ & $ 0.00 \pm 0.00$ & $27.20 \pm 1.00$ & $  27.2_{-   2.2}^{+   2.1}$\ \cmark & $  27.0_{-   2.0}^{+   0.5}$\ \cmark &  2\\[1ex]
  KIC8006161 & $ 5040 \pm  170$ & $  1.04 \pm   0.02$ & $ 0.34 \pm 0.09$ & $29.80 \pm 3.10$ & $  35.2_{-   3.8}^{+   3.5}$\ \cmark & $  35.2_{-   3.8}^{+   3.5}$\ \cmark &  1\\[1ex]
  KIC6196457 & $ 5510 \pm  710$ & $  1.23 \pm   0.04$ & $ 0.17 \pm 0.11$ & $16.40 \pm 1.20$ & $  25.8_{-  25.8}^{+   9.7}$\ \cmark & $  18.9_{-  18.9}^{+   5.1}$\ \cmark &  1\\[1ex]
 KIC11401755 & $ 5850 \pm  930$ & $  1.03 \pm   0.05$ & $-0.20 \pm 0.06$ & $17.20 \pm 1.40$ & $  23.2_{-   6.2}^{+   7.1}$\ \cmark & $  16.2_{-   5.7}^{+   6.4}$\ \cmark &  1\\[1ex]
      70OphA & $ 6200 \pm 1000$ & $  0.89 \pm   0.02$ & $ 0.04 \pm 0.02$ & $39.80 \pm 1.00$ & $  42.7_{-   5.5}^{+   5.7}$\ \cmark & $  42.7_{-   6.3}^{+   4.4}$\ \cmark &  3\\[1ex]
  KIC6116048 & $ 6230 \pm  370$ & $  1.01 \pm   0.03$ & $-0.24 \pm 0.09$ & $17.30 \pm 2.00$ & $  26.2_{-   5.3}^{+   3.0}$\ \cmark & $  18.6_{-   5.3}^{+   2.5}$\ \cmark &  1\\[1ex]
  KIC6521045 & $ 6240 \pm  370$ & $  1.04 \pm   0.02$ & $ 0.02 \pm 0.10$ & $25.30 \pm 2.80$ & $  31.4_{-   4.2}^{+   4.4}$\ \cmark & $  24.0_{-   4.7}^{+   5.4}$\ \cmark &  1\\[1ex]
 KIC10586004 & $ 6350 \pm 1370$ & $  1.16 \pm   0.05$ & $ 0.29 \pm 0.10$ & $29.80 \pm 1.00$ & $  35.2_{-   7.8}^{+  10.5}$\ \cmark & $  27.5_{-   5.4}^{+   7.9}$\ \cmark &  1\\[1ex]
  KIC9955598 & $ 6430 \pm  470$ & $  0.96 \pm   0.01$ & $ 0.08 \pm 0.10$ & $34.70 \pm 6.30$ & $  40.9_{-   4.2}^{+   4.2}$\ \cmark & $  38.6_{-   5.4}^{+   6.2}$\ \cmark &  1\\[1ex]
 KIC11244118 & $ 6430 \pm  580$ & $  1.10 \pm   0.05$ & $ 0.35 \pm 0.09$ & $23.20 \pm 3.90$ & $  38.0_{-   3.8}^{+   5.2}$\ \xmark & $  32.7_{-   3.7}^{+   8.1}$\ \xmark &  1\\[1ex]
   AlphaCenB & $ 6500 \pm  300$ & $  0.91 \pm   0.02$ & $ 0.23 \pm 0.03$ & $41.00 \pm 1.80$ & $  47.7_{-   3.0}^{+   4.2}$\ \cmark & $  47.7_{-   3.0}^{+   4.2}$\ \cmark &  3\\[1ex]
      16CygB & $ 6820 \pm  280$ & $  1.06 \pm   0.03$ & $ 0.05 \pm 0.02$ & $23.20 \pm 7.40$ & $  33.3_{-   2.1}^{+   4.4}$\ \xmark & $  23.5_{-   1.6}^{+   5.8}$\ \cmark &  1\\[1ex]
      16CygA & $ 7070 \pm  460$ & $  1.10 \pm   0.03$ & $ 0.09 \pm 0.02$ & $23.80 \pm 1.70$ & $  36.7_{-   3.5}^{+   4.2}$\ \xmark & $  25.6_{-   1.8}^{+   3.5}$\ \cmark &  1\\[1ex]
  KIC9098294 & $ 7280 \pm  510$ & $  1.00 \pm   0.03$ & $-0.13 \pm 0.09$ & $19.80 \pm 1.30$ & $  35.2_{-   4.8}^{+   5.2}$\ \xmark & $  25.1_{-   4.8}^{+   6.1}$\ \cmark &  1\\[1ex]
  KIC8349582 & $ 7930 \pm  940$ & $  1.19 \pm   0.04$ & $ 0.30 \pm 0.10$ & $51.00 \pm 1.50$ & $  49.9_{-  49.9}^{+  60.6}$\ \cmark & $  33.8_{-  33.8}^{+  43.8}$\ \cmark &  1\\[1ex]
  KIC3656476 & $ 8130 \pm  590$ & $  1.17 \pm   0.03$ & $ 0.25 \pm 0.09$ & $31.70 \pm 3.50$ & $  62.2_{-  24.1}^{+ 292.5}$\ \xmark & $  39.5_{-  12.4}^{+ 215.8}$\ \cmark &  1\\[1ex]
  KIC7871531 & $ 9150 \pm  470$ & $  0.84 \pm   0.02$ & $-0.24 \pm 0.09$ & $33.70 \pm 2.60$ & $  50.7_{-   4.6}^{+   2.9}$\ \xmark & $  40.8_{-   5.5}^{+   2.8}$\ \cmark &  1\\[1ex]
       55Cnc & $10910 \pm 1620$ & $  0.91 \pm   0.03$ & $ 0.31 \pm 0.04$ & $39.00 \pm 9.00$ & $  68.5_{-   6.7}^{+   6.7}$\ \xmark & $  59.1_{-   3.0}^{+   5.5}$\ \xmark &  4\\[1ex]
\hline
\end{tabular}
    \begin{tablenotes}
    \item[1]\cite{vanSaders2016}, 
    \item[2]Based on Carrington's spin period and the meteoritic datations with approximate uncertainties to account for differential rotation and a possible zero-age offset, respectively \citep[e.g.,][]{Thompson+2003, Bonanno+Frohlich2015},
    \item[3]\cite{Delorme2011}, except age of AlphaCen A+B from \cite{Bazot+2012},
    \item[4]\cite{Maxted2015}
    \end{tablenotes}
  \end{threeparttable}
\end{table*}

In order to compare observations and models, we proceed as follows. For each star, we calculate a nominal model for the stellar age, mass and metallicity and an initial spin period of $4$\,d (taken as the geometrical mean of the values observed in the ONC). We also calculate models for ages, masses and metallicities which differ by $\pm 1\sigma$ around the observational values, and models for the nominal stellar parameters but initial spin periods of $1$ and $16$\,d, respectively. We note the difference between the spin periods thus obtained and those of the nominal model $\delta P_{\rm age}$, $\delta P_{\rm mass}$, $\delta P_{\rm [Fe/H]}$ and $\delta P_{P \rm ini}$ respectively. We then calculate our model uncertainties from a quadratic mean, $\delta P_{\rm model}=(\delta P_{\rm age}^2+\delta P_{\rm mass}^2+\delta P_{\rm [Fe/H]}^2+\delta P_{P \rm ini}^2)^{1/2}$. The model generates asymmetries so negative and positive uncertainties are calculated separately. We then compare the observed and modelled values within uncertainties. A model is successful for one star if the observed and modelled values overlap. It is unsuccessful otherwise. 

Following the approach taken by \cite{vanSaders2016}, two kinds of model are calculated. A standard one with a magnetic braking that continues unhindered at large Rossby numbers (slow spin rates) and one that is stalled for $\Ro>\Ro_\odot$. As shown in Table~\ref{tab:objects}, the former model is successful for $13/24$ stars (excluding the Sun, which is used for the calibration). It fails mostly for old, slowly spinning stars. The latter model that we adopted is successful for $21/24$ stars, which thus points to a tapering of magnetic braking for slowly spinning stars, in line with the results of \cite{vanSaders2016}. 

The three cases that still present a mismatch with our preferred model are:

\begin{itemize}
\item{KIC10454113}: This midlife, solar-metallicity, F star (with an estimated $T_{\rm eff}\approx 6400$\,K based on our model calculations) is rotating much more slowly ($P_{spin}\approx15$\,days) than expected from our models which predict spin periods between about 2 and 8\,d. The difference is significant and to be compared to our results for the NGC6819 cluster which also showed that two stars were rotating more slowly than our predictions. At this point, we do not know how to explain, on one hand, the fast rotators in NGC6811 (0.94\,Gyr) and, on the other hand, KIC10454113 and the slow rotators in NGC6819 (2.5\,Gyr). 

\item{KIC11244118}:  Another star that our model has difficulty explaining is this 6.4\,Gyr F star, with a metallicity of (+0.35) which is rotating faster ($P_{spin}\sim 23$\,days) than our model predictions (30 to 41\,days). We notice that this star is characterised by $\Ro>\RoM$. 

\item{55Cnc}: This is a binary system consisting of a G-type star (55 Cancri A), known to host at least five planets, and a smaller red dwarf (55 Cancri B). The main star spins in 30 to 48 days, faster than the model prediction of 56 to 65 days, even when accounting for the suppression of magnetic braking after $\Ro>\RoM$. 

\end{itemize}

The engulfment of a hot Jupiter \citep[e.g.][]{Poppenhaeger+Wolk2014, Guillot+2014}  could be invoked to explain the anomalously fast rotation of KIC11244118 and 55Cnc. However, given that the list is limited to only 24 stars and given that hot Jupiters are quite rare, this is highly unlikely. For KIC10454113, the fact that the inferred spin period is two to three times longer than the model predictions is even more puzzling. We are therefore not able to explain the spin rates of these stars and recommend further studies, in particular to confirm their characteristics.  

The special case of $\alpha$\,Cen A \& B illustrates the importance of a precise characterization of the targets. These two stars are known for their proximity to the Sun and in the case of $\alpha$\,Cen B for the possible presence of an Earth-mass planet in a short orbit that has recently been invalidated \citep{Rajpaul2016}. \citet{Delorme2011} indicate, from astroseismic constraints, an age of $6.5\pm 3$\,Gyr which yields a spin period clearly in excess of the observations for both stars. However, the detailed study of \citet{Bazot+2012} on $\alpha$\,Cen A yields a younger age of $5.0\pm 5$\,Gyr. With this age, our preferred model reproduces the observed spin periods of both components. 

\section{Conclusion}

We have presented a parametric semi-empirical model for the magnetic braking of stars with masses between 0.5\, and 1.6\ $\rm M_\odot$ and from pre-main-sequence to the end of their main-sequence lifetime. The parameters prescribe the stellar mass loss rate and magnetic field amplitude as a function of the physical characteristics of the star obtained from a stellar evolution model and of its spin rate. In the spirit of previous studies (B97, GB15), we also assume that the spin of the outer convective layer of stars is maintained constant by their circumstellar disks and that the slowest (fastest) spinning stars are those which had a disk for the longest (shortest). Two critical Rossby numbers define when the magnetic field saturates and stops increasing with increasing spin rates (for $\Ro<\Rom$) and when the magnetic field changes form, leading to a decrease in angular momentum loss efficiency (for $\Ro>\RoM$). Contrary to previous models which parameterised the magnetic field intensity and mass loss as a function of the rotation rate, we tie them to the star's Rossby number calculated at the base of the outer convective zone. This feature enables applying the same model to a wide variety of stellar types. All models are calibrated to the Sun, in order to reproduce a $27$\,d spin period after $4.5$\,Gyr of evolution of a solar metallicity $1\,\Msun$ star. 

 The model parameters were determined from an analysis of the observation of young clusters with ages between 1\,Myr and 2.5\,Gyr.  For non-saturating magnetic fields $\Ro\ge \Rom=0.09$, we determined that $B\propto \Ro^{-1.2}$ and $\dot{M}\propto\Ro^{-1.3}$ led to a fair fit of the observations for a wide range of stellar types (M to F). We found that the suppression of magnetic braking in F stars is naturally explained by the shortening of the convective turnover timescale at the base of their (small) outer convective zone. We also investigated the degeneracy between the parameters controlling magnetic field strength and mass loss rate (parameters~$\Rom$, $a$ \& $d$). We found that in only 6 out of 576 cases we could find a successful fit. For instance, the following values could all provide fair fits: $\Rom=0.06$, $a=1.0$, $d=1.1$; $\Rom=0.09$, $a=1.2$, $d=1.3$; \& $\Rom=0.11$, $a=1.8$, $d=0.7$. Therefore, our parametric model provides a way to analyse the spin rate of stars and to identify outliers which may have been spun up or down by external processes.

One of the consequences of our parameterisation is that the convergence in spin rates with age is slower than generally assumed. At the age of the Sun, we predict a spread of rotation rates of about 10 percent that depends on the angular momentum acquired by stars at the end of the circumstellar disk phase. This spread increases at lower masses. 

Following \citet{vanSaders2016}, we could confirm that stars of old age and slow rotation rates see their braking suppressed. This is plausibly a consequence of a change in the magnetic field geometry but should be investigated further. 

 Although our model provides a fair fit to the spin evolution of stars between $0.5\,\Msun$ and $1.6\,\Msun$, we acknowledge that it does not provide the final answer to the problem. For instance a remaining issue that needs to be addressed is that our model predicts too many fast rotators at older ages which is not in line with observations.

We are providing our model both for fast- and slow-rotating stars as a function of mass and age. Our model is thus a useful basis to analyse and predict stellar spin periods. A more exhaustive exploration of the ensemble of solutions would be desirable in order to understand and constrain the physical mechanisms at play. Beyond that, progress on the evolution of the spin rates of stars will require a combination of dedicated studies and observations of spin rates and magnetic field properties of stars as a function of age, mass and metallicity.  

Finally, although many of the stars in our sample are likely to have companions, as we showed in section~\ref{constraints} binarity does not play a major role for the purposes of this study.

\section*{Acknowledgements}
We thank the anonymous referee for providing useful comments on the paper that have helped us to improve the original manuscript. We also thank Colin Johnstone, Habib G. Khosroshahi and Rapha\"el Raynaud for their comments and help on the paper. We acknowledge support from the Centre for International Scientific Studies and Collaboration (CISSC) and from the Cultural Section of the French Embassy through a Gundishapur fellowship, as well as the OCA BQR and the IPM School of Astronomy.



\bibliographystyle{mnras}
\bibliography{tevolstarspin.bib}



\appendix

\section{On the convective turnover timescale}\label{sec:tconv}

\subsection{Calculation method}

In the present work, we used $\tauc=H_P/v_{\rm conv}$, the ratio of the pressure scale height $H_P$ to the convective velocity $v_{\rm conv}$ calculated with the mixing length approach. We used the CESAM stellar evolution code \citep{2008Ap&SS.316...61M}, and these quantities were calculated at half a pressure scale height $H_P$ over the base of the convective zone. 

\begin{figure}
\centerline{\includegraphics[width=95mm]{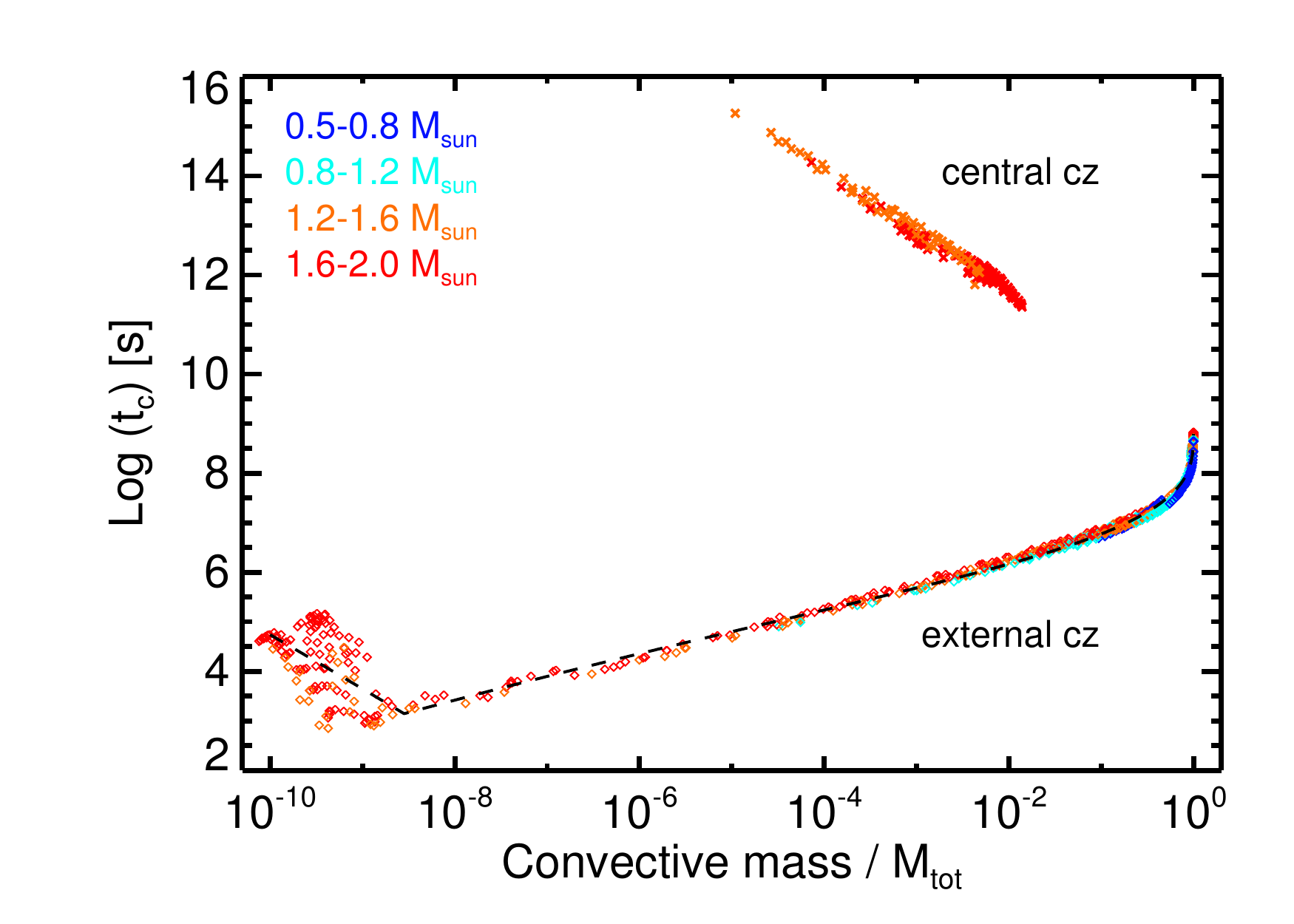}}
\caption{Convective turnover timescale $\tauc$ as a function of the mass of the convective zone divided by the total stellar mass for stars of all ages (from the pre-main-sequence to the end of the main-sequence), metallicities between $\rm [Fe/H]=-0.5$ and $0.5$ and masses from 0.5 to 2.0\,M$_\odot$ (as labelled). The top points correspond to cases with a central convective zone, in which case the turnover timescale is evaluated at half the mass of convective zone (this case is not used in our calculations). The bottom points correspond to the more standard situation of an external convective zone, in which case $\tauc$ is evaluation half a pressure scale height over the base of the convective zone. The dashed curve shows the fit to the external convective zone case (equation~\eqref{eq:tauc}).}
\label{fig:tauconv}
\end{figure}

The definition of $\tauc$ is based on the idea that dynamo action should take place at the base of the convective zone \citep[e.g.,][]{1996ApJ...457..340K}, but where exactly this should be measured is vague. It is thus rather the differential behaviour, from one star to the next, that is important to capture. In that sense, the parameter which affects most the value of $\tauc$ is the depth of the convective zone itself. This is obvious in Figure~\ref{fig:tauconv}, which shows that $\tauc$ is almost uniquely defined by the relative mass of the outer convective zone $m_{\rm cz}$ and is almost insensitive to stellar mass, metallicity and age (once the dependence on $m_{\rm cz}$ has been removed). 

In fact, $\tauc$ can be well fitted for a wide range of pre-main-sequence and main-sequence evolutions by the following relation: 
\begin{eqnarray}
  {\rm Log}(\tau_{\rm c})&=&8.79-2|{\rm Log}(m_{\rm cz})|^{0.349}-0.0194|{\rm Log}(m_{\rm cz})|^2\nonumber\\
  & &-1.62{\rm Min}\left[{\rm Log}(m_{\rm cz})+8.55,0\right]
\label{eq:tauc}
\end{eqnarray}
where the value of $\tauc$ is in seconds and $m_{\rm cz}$ denotes mass of the external convective zone divided by the total stellar mass. 

\subsection{Main-sequence approximation}

\begin{figure}
\centerline{\includegraphics[width=95mm]{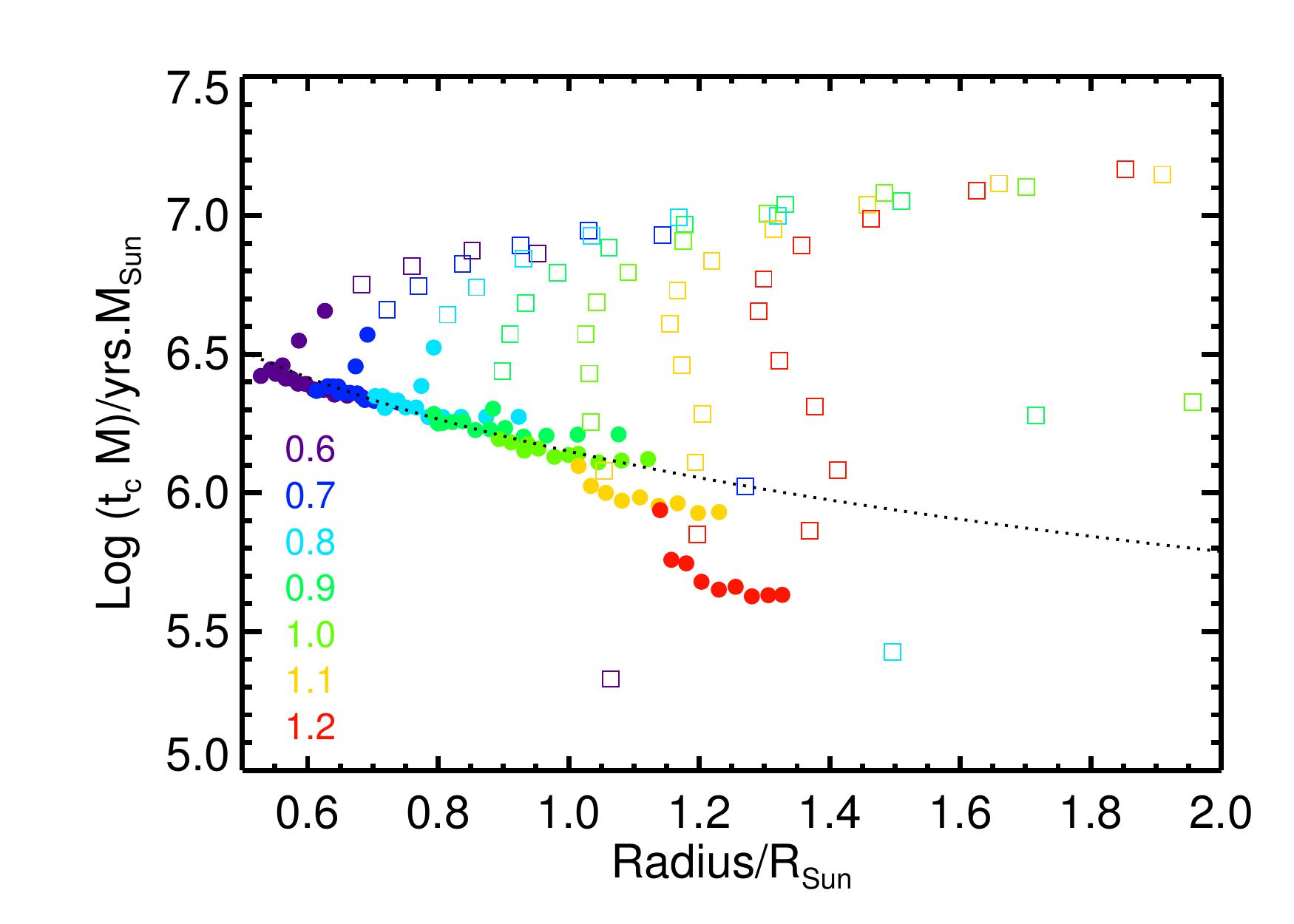}}
\caption{Convective turnover timescale divided by stellar mass as a function of the stellar radius on evolutionary tracks from Landin et al. (2010) for stars with solar composition. Square symbols correspond to age$\,<\,$30\,Myr, circles are for age$\,\ge\,$30\,Myr. The numbers are stellar masses in solar units. The dotted line corresponds to $\tau_{\rm c}M\propto R^{-1.2}$.}
\label{fig:turnover}
\end{figure}

Stars on the pre-main-sequence see their physical properties evolve rapidly and as a consequence, $\tauc$ becomes a strong function of age. However, when they get on to the main-sequence, the time dependence becomes much weaker, and, as shown by Figure~\ref{fig:turnover}, a simple relation then links $\tauc$, stellar mass and stellar radius:
\[
\tau_{\rm c}\propto M^{-1}R^{-1.2}.
\]

This relation holds at least up to 1.0\,M$_\odot$. For higher stellar masses, however, the retreat of the external convective zone implies that $\tauc$ drops, which is at the heart of the existence of the Kraft break, i.e., a much weaker magnetic field and magnetic braking for F-type stars than G-type stars. 

\section{Exploring the magnetic field and mass loss parameter space}
 \label{app:a and d}

\begin{figure}
\hspace*{-0.25in}
  \resizebox{1.12\hsize}{!}{\includegraphics{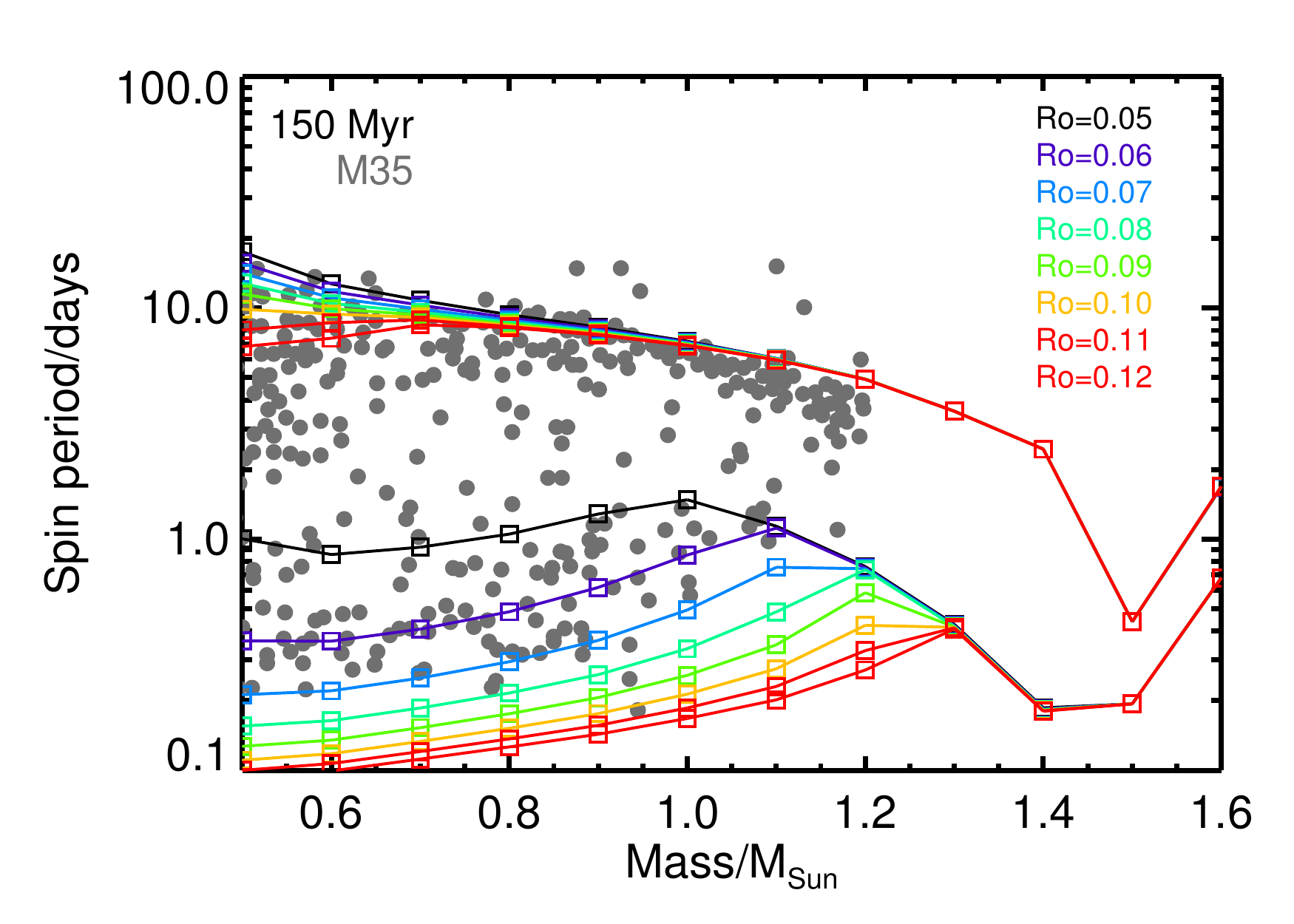}}
\hspace*{-0.25in}
 \resizebox{1.12\hsize}{!}{\includegraphics{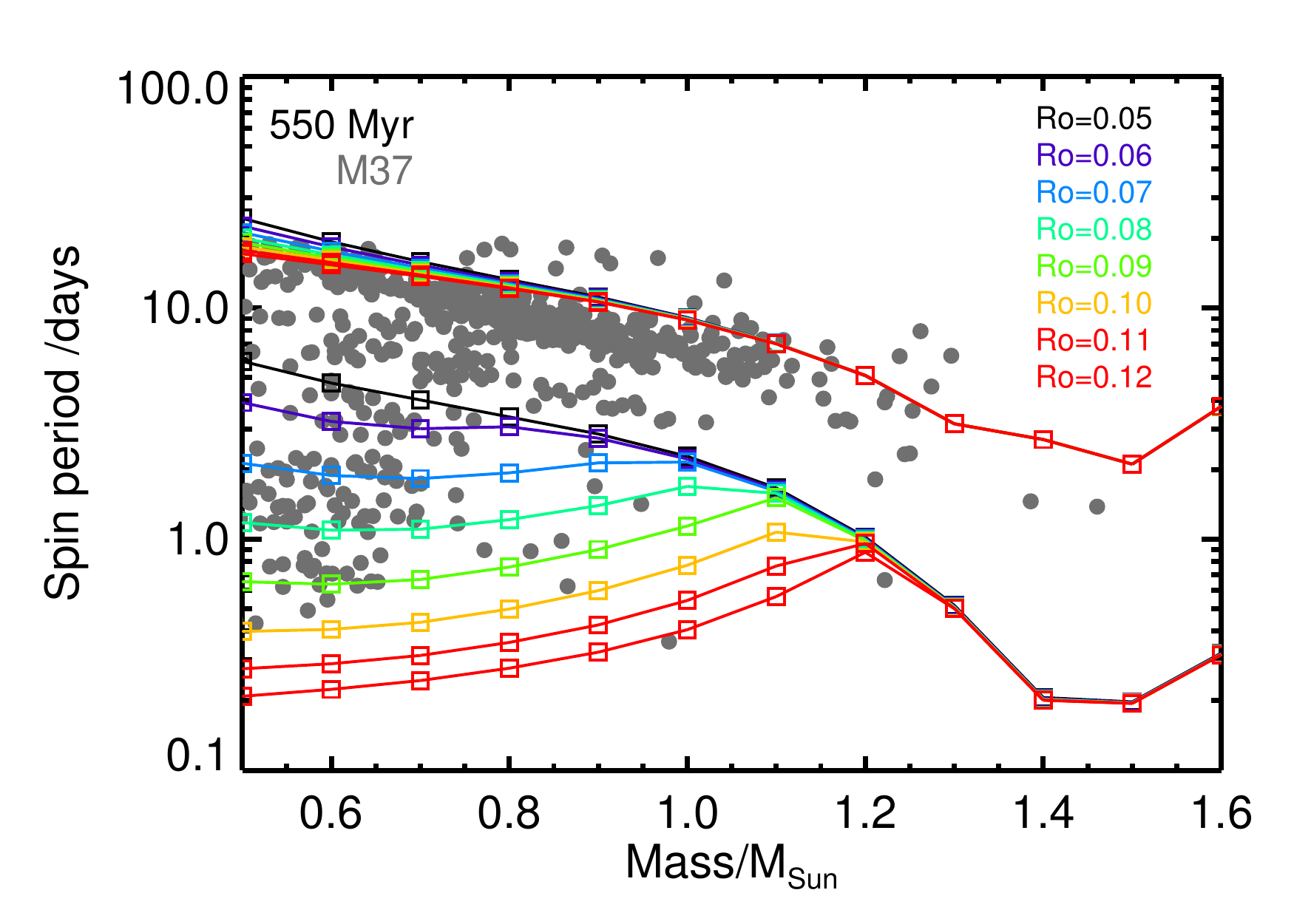}}
\caption{ Test of various $\Rom$ parameter values ($\Ro$ in the figure) ranging from $0.05$ to $0.12$. In these tests $a=1.2$ and $d=1.3$. Models are plotted on top of spin period observations of the two M35 and M37 clusters vs. mass. The numbers in legends and corresponding colours show different values of the critical Rossby number in distinguishing between the slow and fast rotators.}
\label{fig:param_rossby}
\end{figure}
 
\begin{figure}
 \hspace*{-0.25in}
 \resizebox{1.12\hsize}{!}{\includegraphics{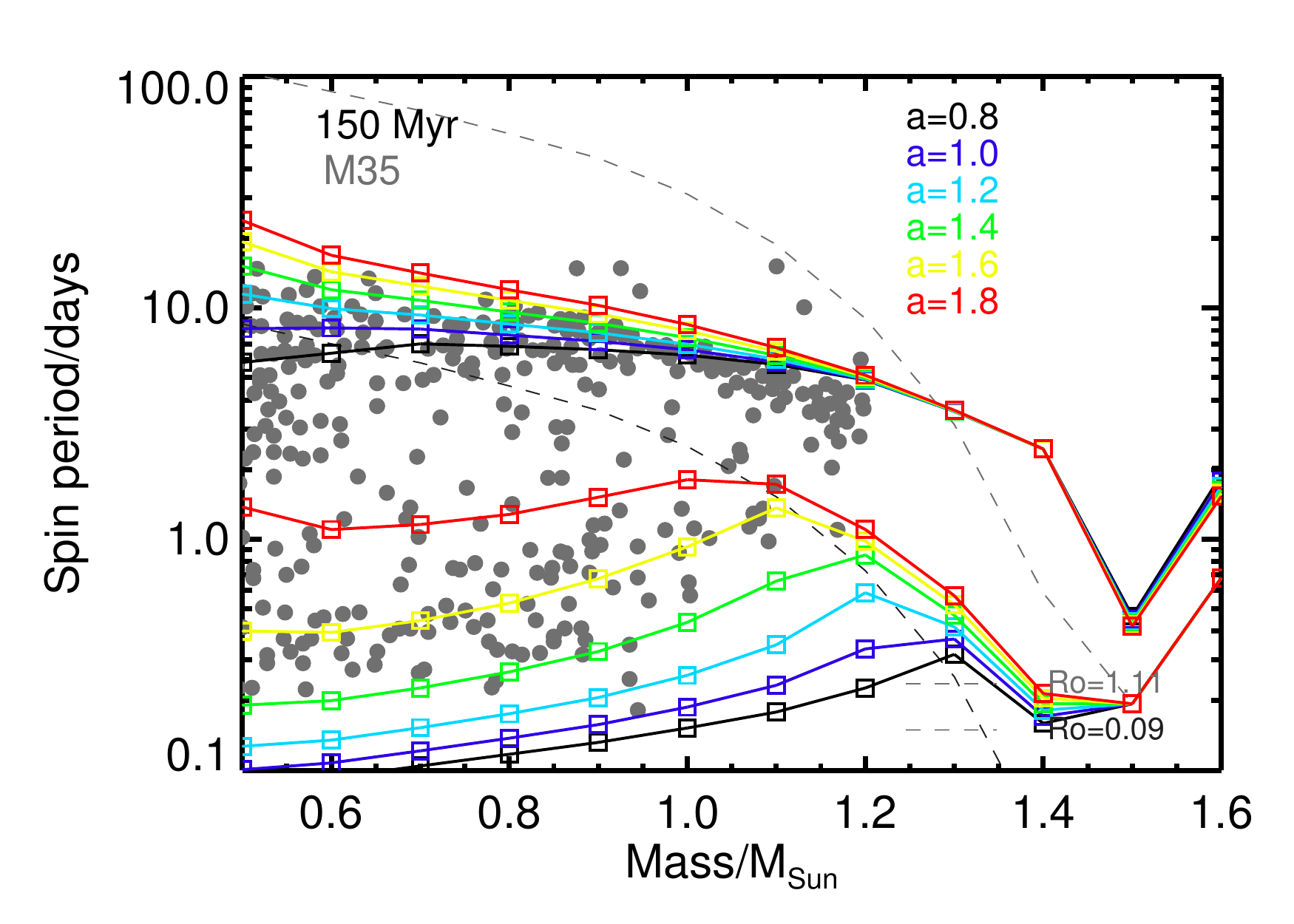}}
 \hspace*{-0.25in}
 \resizebox{1.12\hsize}{!}{\includegraphics{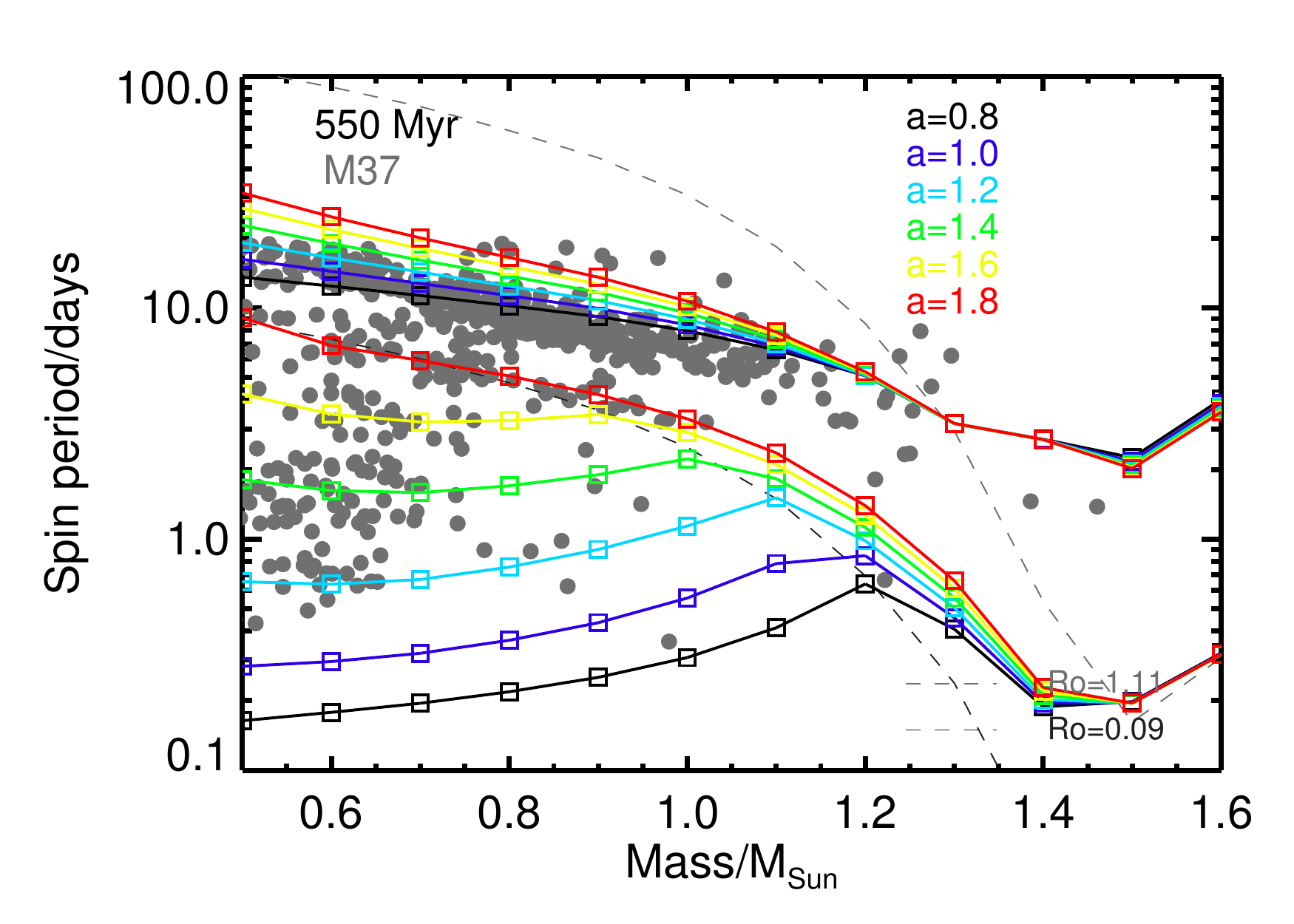}}
\caption{ Tests of the effect of magnetic field parameter $a$ on the spin period of stars as a function of mass against stars in the two M35 and M37 clusters. Different colours correspond to different values of $a$ between 0.8 and 1.8, as labelled. In these tests $\Rom=0.09$ and $d=1.3$. }
\label{fig:param_a}
\end{figure}

\begin{figure}
 \hspace*{-0.25in}
 \resizebox{1.12\hsize}{!}{\includegraphics{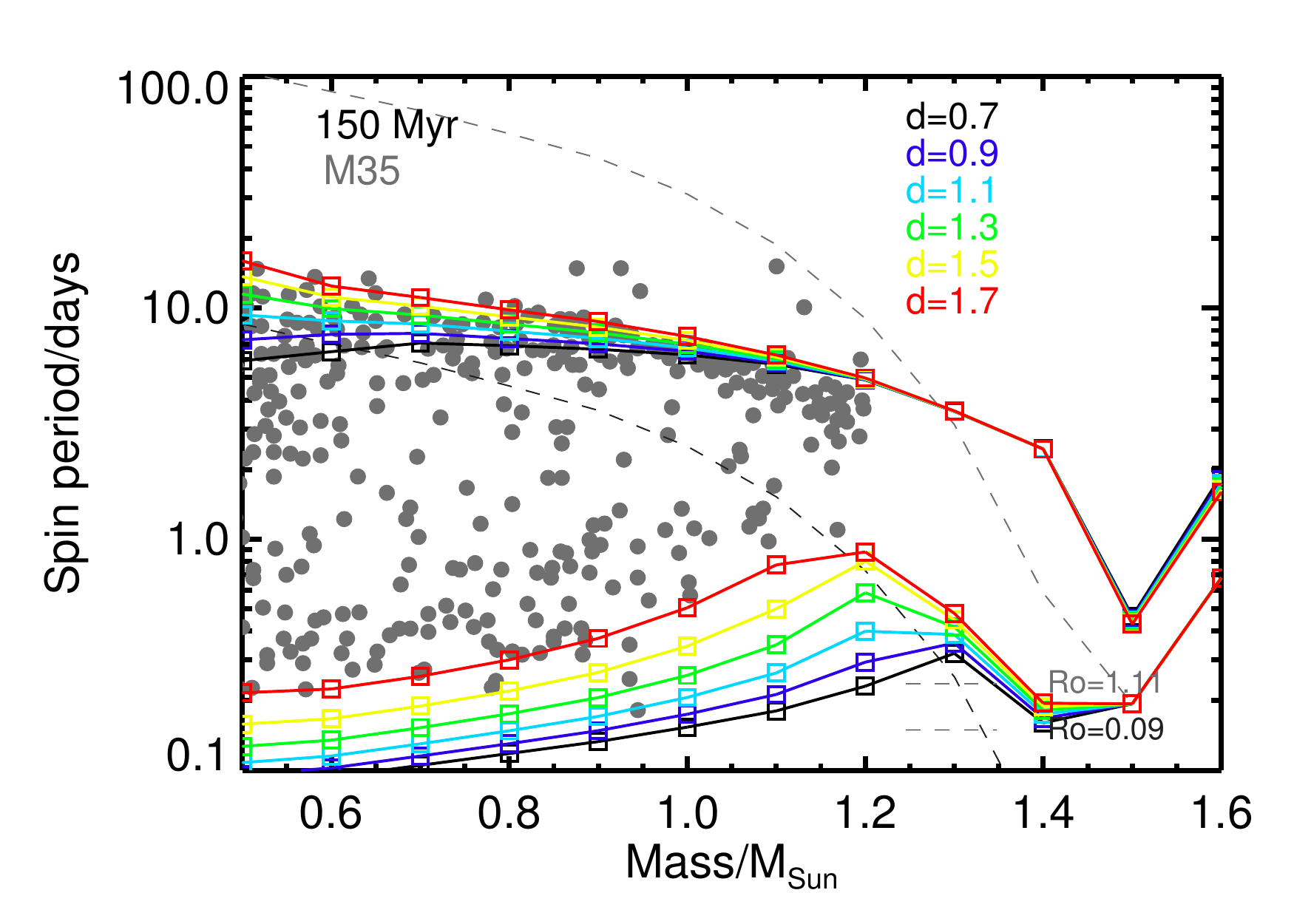}}
  \hspace*{-0.25in}
 \resizebox{1.12\hsize}{!}{\includegraphics{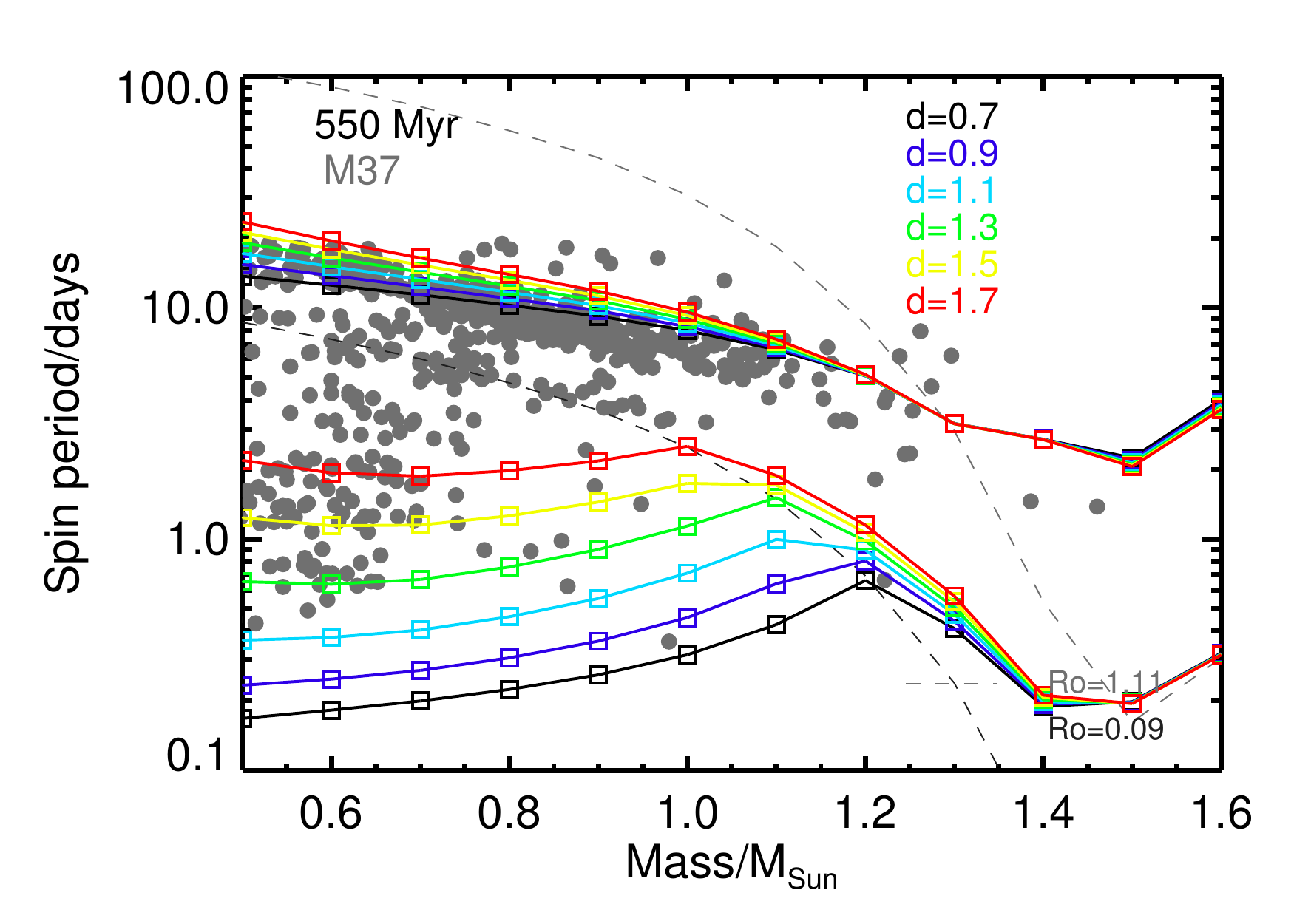}}
\caption{ The effect of changing mass loss parameter; $d$ on the spin period of stars as a function of mass against stars of M35 and M37. Colours correspond to different values of $d$ between $0.7$ and $1.7$. In these tests $a=1.2$ and $\Rom=0.09$.}
\label{fig:param_d}
\end{figure}

In this appendix we discuss how the three parameters~$a$, $d$ \& $\Rom$ affect model predictions. Figures~\ref{fig:param_rossby} through ~\ref{fig:param_d} show how various values for $\Rom$, $a$ and $d$ affect the spin period evolution models against the stars in the M35 (150 Myr) and M37 (550 Myr) clusters.
 
In figure~\ref{fig:param_rossby} we have varied  $\Rom$ from $0.05$ to $0.12$. It should be noted that since values above $\Rom=0.12$ were completely unsuccessful at fitting the observations they have been removed from plots. Small values of $\Rom\approx0.05$ result in a too fast aggregation of the two rotating regimes, while bigger values are slightly more successful at reproducing the slow rotators branch. A value of $\Rom=0.09$ successfully covers both the fast and slow rotators and seems to create the best fit to the  observations. This is in close agreement with the usually adopted value of $\Rom=0.1$ in the literature.
 
In addition, the results of our tests for the $a$ parameter are shown in figure~\ref{fig:param_a}. The results differ quite significantly on the choice of $a$: Small values below 1.2 result in a lower envelope that is detached from the cloud of fast-spinning stars. Large values above 1.6 yield results in contradiction with the existence of low-mass fast spinning stars and also predicts a slow rotation envelope that is detached from the cloud of points for small-mass stars. Based on these results, we choose $a=1.2$ which is between the two values predicted by \cite{Folsom2016} $a=1.0$ and \cite{See2017} who find $a=1.65$.
 
In figure \ref{fig:param_a}, we also show our two critical Rossby numbers $\Rom=0.09$ and $\RoM=1.11$. We can see in the case of these two young clusters (and also for other clusters) that for a majority of stars, the relatively slow rotators, are located between these two lines. The fast rotators are, however, characterised by $\Ro < \Rom$ and are thus in a regime in which the magnetic field saturates (i.e., it is not proportional to the spin rate). This feature is crucial to account for the presence of fast rotating solar-type stars even at relatively old ages. 
 
Finally,  figure~\ref{fig:param_d} shows the results of testing parameter~$d$. It can be seen that  this parameter affects all stellar masses but has a higher effect on fast rotators. Based on this figure we choose $d=1.3$ which is close to the predicted value by \cite{See2017} (d=1.49) and also in agreement with the theoretical discussion of \cite{1992MNRAS.256..269T} who predicted mass loss to be a function of $\sim\Omega$.

\section{Models}
 \label{app:csvs}

 We provide our rotational evolution models for stellar masses $0.5$ to $1.6 \rm M_\odot$ and with five initial rotational periods; $1.0$, $2.0$, $4.0$, $8.0$ and $16.0$ days in electronic format. The models start from zero to 10\,Gyr.  Each model contains the following parameters:

AGE: age of the star [$\rm Myr$], 

H: the total angular momentum  [$\rm g\cdot cm^2/s$], 

PSPIN1: rotation period of the primary [$\rm days$],

M1: mass [$\rm g$],

R1: radius [$\rm cm$],

MTI1: moment of inertia [$\rm g\cdot cm^2/s$],

TEFF1: effective temperature [$\rm K$],

PSPINDEEP1: period of inner radiative layer [$\rm days$],

XRDEEP1: radius of inner radiative layer [$\rm cm$],

XMEXTCZ1: mass of external convective zone [$\rm g$],

MTIDEEP1: moment of inertia of inner radiative layer  [$\rm g\cdot cm^2/s$],

MTICZ1: moment of inertia of convective zone  [$\rm g\cdot cm^2/s$],

BMEAN1: mean magnetic field [$\rm G$],

MDOT1: mass loss rate [$\rm g/\rm yr$],

ROSSBY1: Rossby number,

TCONV1: convective turnover timescale [$\rm s$].


\bsp	
\label{lastpage}
\end{document}